\documentclass[a4paper,usenatbib,useAMS]{mnras}

\pdfoutput=1
\usepackage{mathtools}
\usepackage{amssymb,indentfirst}
\usepackage{color,xcolor}
\usepackage{epsfig,epstopdf}
\usepackage{graphicx}
\usepackage{hhline}
\usepackage{euscript}
\usepackage{amsmath}
\usepackage{fontenc}
\usepackage{multicol}
\usepackage{multirow}
\usepackage{xspace} 
\usepackage{float}
\usepackage[normalem]{ulem}
\usepackage{makecell}

\hypersetup{bookmarksnumbered,bookmarksdepth=3}

\usepackage{enumerate}
\newcommand{\bk}{\mathbf{k}}
\newcommand{\bq}{\mathbf{q}}
\newcommand{\bx}{\mathbf{x}}
\newcommand{\fnl}{f_{\rm NL}}
\newcommand{\fnll}{f_{\mathrm{NL}}^{\mathrm{loc}}}
\newcommand{\fnle}{f_{\mathrm{NL}}^{\mathrm{equil}}}
\newcommand{\fnlo}{f_{\mathrm{NL}}^{\mathrm{orth}}}
\newcommand{\Mpc}{\ensuremath{\text{$h$/Mpc}}\xspace}
\newcommand*\mean[1]{\overline{#1}}
\newcommand*\inte[1]{\int\frac{d^3{#1}}{(2\pi)^3}}
\newcommand*\intee[2]{\int\frac{d^3{#1}}{(2\pi)^3}\frac{d^3{#2}}{(2\pi)^3}}
\newcommand{\eps}{\mathcal{\epsilon}}
\newcommand*{\order} [1] {\ensuremath{\mathcal{O}(#1)}\xspace}
\newcommand*{\scien} [2] {\ensuremath{{#1} \times 10^{#2}}} 
\newcommand*{\ie} {i.\,\!e.\xspace}
\newcommand*{\eg} {e.\,\!g.\xspace}
\newcommand*{\etc} {\emph{etc.}\xspace}
\newcommand*{\veps} {\varepsilon}
\newcommand*{\eref} [1] {Eq.\ (\ref{#1})}
\newcommand*{\esref} [2] {Eqs.\ (\ref{#1})\ and\ (\ref{#2})}

\newcommand*{\tref} [1] {Table \ref{#1}\xspace}
\newcommand*{\sref} [1] {Sec.\ \ref{#1}\xspace}
\newcommand*{\fref} [1] {Fig. \ref{#1}\xspace}
\newcommand*{\aref} [1] {Appendix \ \ref{#1}\xspace}
\newcommand*{\perm} {\;\text{perm}}
\newcommand*{\MPT} {\textsc{MPTbreeze}\xspace}
\newcommand{\Planck} {\textit{Planck }}

\renewcommand{\min}{\text{min}}
\renewcommand{\max}{\text{max}}

\graphicspath{{./plots/}}

\begin{document}

\title[]{Constraining Primordial non-Gaussianity with Bispectrum and Power Spectum from Upcoming Optical and Radio Surveys}
\author[]{Dionysios Karagiannis$^{1}$\thanks{E-mail: {dionysios.karagiannis@pd.infn.it}},
Andrei Lazanu$^{2}$,
Michele Liguori$^{1,2,3}$, 
\newauthor Alvise Raccanelli$^{4,5}$\thanks{Marie Sk\l{}odowska-Curie fellow},
Nicola Bartolo$^{1,2,3}$,
Licia Verde$^{4,6}$
\\
$^{1}$Dipartimento di Fisica e Astronomia ``G. Galilei'',Universit\`a degli Studi di Padova, via Marzolo 8, I-35131, Padova, Italy \\
$^{2}$INFN, Sezione di Padova, via Marzolo 8, I-35131, Padova, Italy \\
$^{3}$INAF-Osservatorio Astronomico di Padova, vicolo dell Osservatorio 5, I-35122 Padova, Italy \\
$^{4}$ICC, University of Barcelona, IEEC-UB, Mart\'i  i Franqu\`es, 1, E08028 Barcelona, Spain \\
$^{5}$Department of Physics \& Astronomy, Johns Hopkins University, Baltimore, MD 21218, USA \\
$^{6}$ICREA, Pg. Llu\' is Companys 23, 08010 Barcelona, Spain}

\maketitle

\begin{abstract}
We forecast constraints on primordial non-Gaussianity (PNG) and bias parameters from measurements of galaxy power spectrum and bispectrum in future radio continuum and optical surveys. 
In the galaxy bispectrum, we consider a comprehensive list of effects, including the bias expansion for non-Gaussian initial conditions up to second order, redshift space distortions, redshift uncertainties and theoretical errors. These effects are all combined in a single PNG forecast for the first time. Moreover, we improve the bispectrum modelling over previous forecasts, by accounting for trispectrum contributions. All effects have an impact on final predicted bounds, which varies with the type of survey. We find that the bispectrum can lead to improvements up to a factor $\sim 5$ over bounds based on the power spectrum alone, leading to significantly better constraints for local-type PNG, with respect to current limits from \textit{Planck}. Future radio and photometric surveys could obtain a measurement error of 
$\sigma(\fnll) \approx 0.2$. In the case of equilateral PNG, galaxy bispectrum can improve upon present bounds only if significant improvements in the redshift determinations of future, 
large volume, photometric or radio surveys could be achieved. For orthogonal non-Gaussianity, expected constraints are generally comparable to current ones.
\end{abstract}


\section{Introduction}

So far, cosmological analyses of Large Scale Structure (LSS) surveys have relied nearly exclusively on matter and galaxy power spectrum estimation. It is however well-known that important extra-information can be extracted via higher-order correlation functions, such as the matter and galaxy {\em bispectrum} (three-point correlation function of Fourier modes), which allow both probing the non-linear regime of structure growth and setting constraints on primordial non-Gaussianity (PNG) \citep[see \eg][ and references therein]{Bernardeau2002}. The latter is in particular an important and general prediction of inflationary theories. It is a direct product of non-linear interactions during the inflationary or reheating stage. The bispectrum of the primordial curvature perturbation field, arising from such interactions, is characterized by a dimensionless, amplitude parameter, $\fnl$, and by a shape function $F(k_1,k_2,k_3)$. While $\fnl$ defines the strength of the PNG signal, the shape describes the functional dependence of the bispectrum on different Fourier space triangles. Both of them are strongly model dependent and provide significant information on the physical mechanisms at work during inflation. Three very important shapes, encompassing a large amount of scenarios, are the so called  {\em local} shape (NG signal peaking on ``squeezed triangles'', \ie $k_1 \ll k_2 \sim k_3$), {\em equilateral shape} ($k_1 \sim k_2 \sim k_3$) and {\em folded-shape} ($k_1 \sim k_2 \sim k_3/2$).

Currently, the tightest experimental $\fnl$ bounds, including  a large number of different shapes, come from \Planck {\em Cosmic Microwave Background} (CMB) bispectrum measurements \citep{Planck_PNG2016}.
Bispectrum measurements of LSS data have been already obtained \citep{Scoccimarro2001b,Feldman2001,Verde2002,Marin2013,GilMarin2014,GilMarin2017}, but the current level of sensitivity is not sufficient to generate useful PNG bounds (current LSS power spectrum constraints on local $\fnl$ are more interesting \citep{Padmanabhan2007,Slosar2008,Xia2010a,Xia2011,Nikoloudakis2012,Agarwal2014,Karagiannis2014,Leistedt2014}, albeit still not competitive with the CMB \citep{Verde1999}).
On the other hand, bispectrum estimates of $\fnl$ with future LSS data do have in principle great potential to improve over CMB bounds, at least for specific shapes. This is because 3D LSS surveys, covering large volumes and probing a wide range of scales, have  access to a much larger amount of modes, with respect to 2D CMB maps. However, LSS measurements will also be very challenging, due to late-time non-linearities, expected to produce much larger non-Gaussian signatures than the primordial 
component. 
These contributions  therefore need to be understood and subtracted with exquisite accuracy. Note that an interesting probe for the local PNG consist of the asymmetric two-point cross-correlation between two different galaxy populations, which is enhanced by non-Gaussian initial conditions \citep[see ][ for details]{Dai2016}.

The issue of theoretical modelling of non-linear effects and of higher order LSS correlators has indeed been long debated in the literature \citep[see \eg][ for a review]{Bernardeau2002} and 
the interest in producing accurate and realistic LSS primordial bispectrum forecasts has been steadily increasing in recent years. Important contributions in this direction include the work of \citet{Scoccimarro2003} and \citet{Sefusatti2007} - where the bispectrum of galaxies is used for the first time to forecast the constraining power of LSS surveys on measuring the amplitude of PNG - and the study of \citet{Song2015}, where information from power spectrum and bispectrum of galaxies is combined - also including redshift space distortion effects - in order to constrain growth parameters and galaxy bias terms. Additional contributions were then made by \citet{Tellarini2016}, who took into account the second order tidal bias term \citep{McDonald2009,Baldauf2012,Chan2012}, as well as the bivariate bias expansion \citep{Giannantonio2010} in the redshift space galaxy bispectrum, in order to constrain the amplitude of local PNG. Finally, the authors of \citet{Baldauf2016} pointed out the importance of including uncertainties in the theoretical modelling of the signal ({\it theoretical errors}) and properly propagating them into the final error bar estimates. 

Many more details on these issues - including a more detailed description of improvements and refinements in bias and redshift space distortion modelling - are provided in sections \ref{sec:mfbias} and \ref{sec:PB_gal_red}. Here  we point out that many of the analysis ingredients mentioned above were considered {\em separately} and independently in previous forecasts, with different works considering the importance of specific new terms, without accounting however for all of them at once (for example, theoretical errors are studied in detail in the real space treatment of \citet{Baldauf2016}, whereas redshift space distortions are accounted for in detail in \citet{Tellarini2016}, without including theoretical errors). Here, for the first time, we consistently include all these terms and produce as complete and realistic as possible PNG
forecasts, in terms of $\fnl$ parameters, combining power spectrum and bispectrum constraints. One advantage of using the bispectrum is that it opens the possibility to explore the full range of primordial shapes, including the equilateral and orthogonal ones. These shapes are very little explored in previous PNG LSS studies.
In addition, we also include -- for the first time in an actual forecast -- trispectrum contributions to $\fnl$ arising from the bias expansion in the galaxy bispectrum, which were pointed out as potentially important \citep{Sefusatti2009,Jeong2009}.  We note that such term could play a significant role in constraining the signal from non-local shapes.

Another important issue in a LSS PNG analysis is of course that of establishing which survey design and which statistical probe provide the best $\fnl$ constraints. Clearly, adding modes by going to smaller scales does in principle improve sensitivity. The obvious caveat is that such approach requires non-linear scales, where the non-PNG contribution gets very large and hard to model. Moreover, late time non-linearities couple different modes. This unavoidably produces a saturation of the available information. To estimate in detail this effect, a full calculation of the bispectrum covariance is needed in the evaluation of the signal-to-noise ratio for $\fnl$. All these issues are still open, and they are currently under a significant amount of scrutiny in the literature \citep{Heavens1998,Crocce2005b,Crocce2005a,Pietroni2008,Bernardeau2008,Wagner2010,Crocce2012,Baumann2012,Carrasco2012,GilMarin2012b,GilMarin2014,Lazanu2015, Lazanu2017,Lazanu2017b}.

In this work we consider the alternative approach: we only investigate large and quasi-linear scales;
 we look at galaxy clustering statistics at high redshift, where non-linearities become important at much smaller scales. Forthcoming radio continuum surveys seem ideally suited to this purpose, as they will probe wide areas of the sky over a large redshift range.  While the power spectrum of radio continuum has been already considered in the literature \citep{Xia2010b,Xia2010c,Xia2011,Raccanelli2011, Camera2015b, Raccanelli2017}, the bispectrum of radio continuum surveys has not been studied so far. We devote particular attention to it in this work. The drawback with radio continuum sources is the lack of a direct determination of their redshifts. Our analysis therefore considers the possibility to extract redshift information via clustering-based estimation methods \citep{Menard2013}. We follow the implementation for forthcoming radio surveys developed in \citet{Kovetz2016}. For comparison, we also present forecasts for optical surveys, both spectroscopic and photometric.
 
The paper is structured as follows: In \sref{sec:basics}, we provide a brief review of the theory behind matter clustering and PNG; in \sref{sec:mfbias} we discuss in more detail the relationship between matter and galaxy statistics; in \sref{sec:galstat} we present the expressions for galaxy  power spectra and bispectra considering all the relevant effects; in \sref{sec:surveys} we describe the survey specifications that we have used for our analysis; in \sref{sec:fisher} we describe the Fisher matrix forecast method; in \sref{sec:theo_errors} we explain how to account for theoretical errors in the forecast. We summarize our theoretical framework in \sref{res_summary}. The main results are then shown in \sref{results}. Finally, we conclude in \sref{sec:conc}. In the Appendix we present a summary of standard perturbation theory (\aref{StanPT}), as well as a small review on the results of the \MPT formalism (\aref{app:MPT}). In addition, we present the derivation of the bias parameters from the adopted peak-background split method (\aref{PBSbias}). Finally in \aref{app:RSD}, we present the kernels used for the redshift space galaxy models.

\section{Galaxy clustering}

\subsection{Matter power spectrum and bispectrum}\label{sec:basics}

In linear theory,  the  matter power spectrum can be related to the primordial gravitational potential power spectrum, via  

 \begin{equation}\label{eq:linmat}
  P_{m}^{L}(k,z)=M^2(k,z)P_{\Phi}(k),
 \end{equation} 
with
 
\begin{equation}
  M(k,z)=\frac{2k^2c^2T(k)D(z)}{3\Omega_{m,0}H_{0}^2} \, .
 \end{equation}

 Here $k\equiv\lvert\bk\rvert$ is the magnitude of the wavevector $\bk$ and $D(z)$ is the linear growth factor, originating from the growing mode of the linear fluid equations, where it is normalized to unity today ($D(0)=1$). $T(k)$ is the matter transfer function normalized to unity on large scales $k \rightarrow 0$ and $c$ is the speed of light. In this work, matter non-linearities will be accounted for in the framework of Standard Perturbation Theory (SPT) {or Renormalized Perturbation Theory (RPT)}. Useful results in SPT are reviewed in \aref{StanPT}.  To be consistent with the SPT approach we will only use the linear power spectrum and the tree-level bispectrum in SPT, since we will restrict our analysis to only mildly non-linear scales (here we consider $k_\max(z)=0.1/D(z)\;h\text{/Mpc}$). There is nevertheless additional information encoded in small, non-linear scales, but modelling this regime is challenging, and will not be considered here. 
  
  Based on the inflationary model considered, primordial gravitational potential bispectra are produced in different shapes. In this work we consider the three most widely used shapes, the local \citep{Salopek1990,Gangui1993,Verde1999,Komatsu2001}, equilateral \citep{Creminelli2005} and orthogonal \citep{Senatore2009} types of PNG. 
  
  The leading, linearly-extrapolated non-Gaussian contribution to the matter density bispectrum is
  \begin{equation} \label{mat:bng}
   B_I(k_1,k_2,k_3,z)=M(k_1,z)M(k_2,z)M(k_3,z)B_{\Phi}(k_1,k_2,k_3) \,,
  \end{equation}
  where $B_\Phi$ is the primordial gravitational potential bispectrum predicted from inflation.
  
The three primordial bispectrum shapes, defining PNG of the local, equilateral and orthogonal 
types, and are defined respectively as
\begin{align}
B_{\Phi}^{\text{loc}}&(k_1,k_2,k_3)=2 \fnll\left[P_{\Phi}(k_1)P_{\Phi}(k_2)+\text{2 perms} \right] \, ,\\
B_{\Phi}^{\text{eq}}&(k_1,k_2,k_3)=6 \fnle\left\{-[P_{\Phi}(k_1)P_{\Phi}(k_2)+\text{2 perms} ]\right. \nonumber \\
&-2[P_{\Phi}(k_1)P_{\Phi}(k_2)P_{\Phi}(k_3)]^{2/3} \nonumber \\
&+[P_{\Phi}^{1/3}(k_1)P_{\Phi}^{2/3}(k_2)P_{\Phi}(k_3)+\text{5 perms}]\left. \right\} \,, \\ 
  B_\Phi^\text{orth}&(k_1,k_2,k_3) = 6\fnlo\big[ 3(P_\Phi^{1/3}(k_1)P_\Phi^{2/3}(k_2)P_\Phi(k_3) \nonumber  \\
  &+5\text{ perms})-3 \left[P_{\Phi}(k_1)P_{\Phi}(k_2)+\text{2 perms} \right] \nonumber \\
  &-8 (P_\Phi(k_1)P_\Phi(k_2)P_\Phi(k_3))^{2/3}
\big] \,.
\end{align} 

Large local PNG can be produced in multi-field scenarios, while sizeable equilateral and orthogonal bispectra can 
be obtained, for example, in single-field scenarios with non-standard kinetic terms and higher derivative interactions. The folded shape, mentioned in the introduction, 
can arise in models with excited initial states and is obtained as a linear combination of equilateral and orthogonal shapes. In fact  the orthogonal shape, while difficult to visualise, is by construction orthogonal to the  equilateral ones. For more details see \eg \citet{Planck_PNG2016}, and references therein.

\subsection{Bias}\label{sec:mfbias}

The bias of tracers  \ie the relation between the statistics of observed objects and the underlying distribution of matter, can be understood as the combination of  two components: the bias of dark matter halos and how the selected tracers populate the halos. Here we first  present our adopted modelling for the bias of halos and then in \sref{sec:hod_bias} we use a halo occupation distribution model to connect this to the bias of selected tracers \citep[see \eg][ for a review]{Desjacques2016}.

\subsubsection{Halo bias}
  
 In order to use the tree-level galaxy bispectrum to forecast the amplitude of PNG coming from future LSS surveys, knowledge of the bias expansion up to second order is needed. A complete set of the bias terms was presented in \citet{Assassi2014}, \citet{Senatore2014} and \citet{Mirbabayi2014}, where the notion of \emph{local-in-matter} bias relation \citep{Coles1993,Fry1993} is generalized to a \emph{local bias} perturbative expansion over operators $O$ representing all possible local gravitational observables along the fluid trajectory. Such operators are built from powers of the tidal tensor $\partial_i\partial_j\Phi$ which can be decomposed into the trace $\nabla^2\Phi\propto\delta$ and the traceless part $s_{ij}$, \ie the tidal field \citep{Catelan2000,McDonald2009,Chan2012,Baldauf2012}.   
  
  \emph{Higher-derivative} terms must also be taken into account in the general local bias expansion \citep{Desjacques2016}. They are constructed from powers of spatial derivatives over each operator $O$ (\ie higher than two derivatives on $\Phi$) and incorporate non-gravitational contributions in the halo formation process (\eg gas heating, feedback processes). These terms introduce a limiting spatial scale,  into the bias expansion, $R_*$, at which they become important. This fundamental cut-off indicates the scale where a perturbative description of bias breaks down \citep[see \eg][ for a discussion]{Desjacques2016}. In the case of dark matter halos it is proportional to the Lagrangian radius $R(M)=(3M/4\pi\mean{\rho}_m)^{1/3}$.

 Including all the above, the most general expansion up to second order in the Eulerian framework with Gaussian initial conditions is \citep{Coles1993,Fry1993,Fry1996,Catelan1997,Catelan2000,McDonald2009,Elia2010,Chan2012,Baldauf2012,Assassi2014,Senatore2014,Mirbabayi2014}
  
  \begin{align}\label{eq:deltaG}
   \delta&_h^{E,(G)}(\bx,\tau)= b_1^E(\tau)\delta(\bx,\tau) +\veps^E(\bx,\tau) \nonumber \\
    &+\frac{b_2^E(\tau)}{2}\delta^2(\bx,\tau) + b_{s^2}^E(\tau)s^2(\bx,\tau)+\veps_{\delta}^E(\bx,\tau)\delta(\bx,\tau) \,,
  \end{align}
  where $\tau$ is the conformal time and $\bx$ are the spatial co-moving coordinates in the Eulerian frame. In addition $s^2=s_{ij}s^{ij}$ is the simplest scalar that can be formed from the tidal field, $\veps^E$ is the leading stochastic bias contribution \citep{Dekel1998,Taruya1998,Matsubara1999} and $\veps_{\delta}^E$ is the stochastic counterpart of the linear bias. The second-order tidal field bias term, following the convention of \citet{Baldauf2012}, is given by $b_{s^2}^E=-2/7(b_1^E-1)$. Note that this expression is not expected to be exact, since it assumes a formation time of $\tau_*\to 0$, as well as a conserved evolution \citep[see \eg][ and references therein for a detailed discussion]{Desjacques2016}.

   Although the subscript $R$ is missing from the fields in the bias expansion, they are all assumed smoothed at a radius $R>R_*$. In the above expansion we have ignored all higher-derivative terms ($\sim\order{R_*^2}$) up to second order, since we consider sufficiently large scales ($k\ll1/R_{*}$) wherein their contribution is suppressed by $(R_*k)^2$. Further, on large scales we can also ignore the velocity bias \citep[see \eg][]{Desjacques2016}.

  The bias coefficients of the general bias expansion can be defined on large scales in the context of the peak-background split \citep{Kaiser1984,Bardeen1986,Cole1989}. Here the mass function proposed by \citet{ShethTormen1999} (ST hereafter) will be used to derive the Eulerian local-in-matter bias terms as a function of the peak height \citep{Scoccimarro2001}. Details on the peak-background split argument can be found in \aref{PBSbias}.

  \subsubsection{Non-Gaussian effects}\label{sec:PNGbias}
  
 In the presence of PNG of the local type, long-wavelength fluctuations of the primordial gravitational potential modulate the initial condition perturbations of small scales, due to the induced coupling between the two. This leads to to a scale-dependent bias correction on large scales \citep{Dalal2008,Slosar2008,Matarrese2008,Afshordi2008,Verde2009,Desjacques2010,Schmidt2010,Desjacques2011a,Schmidt2013}. 
   Additional operators involving the primordial Bardeen potential $\Phi$ (without any derivatives) must be added in the general bias expansion in order to model the scale-dependent corrections (in the same spirit as in the work of \citep{McDonald2008,Giannantonio2010,Baldauf2011}). In the case of an arbitrary quadratic PNG, the full set of operators was derived in \citet{Assassi2015}, where up to second-order in perturbations and linear in $\fnl$ the bias expansion in the Eulerian frame becomes
  
  \begin{align}\label{eq:deltaNG}
   \delta_h^{E,(NG)}(\bx,\tau)&=b_{\Psi}^E(\tau)\Psi(\bq) \nonumber \\ &+b_{\Psi\delta}^E(\tau)\Psi(\bq)\delta(\bx,\tau)+\veps_{\Psi}^E(\bx,\tau)\Psi(\bq)\,,
  \end{align}
  where $\bq$ are the spatial coordinates in the Lagrangian frame and $\veps_{\Psi}^E$ is the stochastic counterpart of the field $\Psi$. The latter is a non-local transformation of the primordial Bardeen potential given by, 
  
  \begin{equation}
   \Psi(\bq)=\int \frac{d^3\bk}{(2\pi)^3}k^{\alpha}\Phi_G(\bk)e^{i\bk\bq}\,,
  \end{equation}
 where $\alpha$ can take real values, which depend on the PNG type. This field originates from the generalization of the local ansatz in the case of a general quadratic PNG \citep[\eg][]{Schmidt2010,Scoccimarro2011,Schmidt2012,Assassi2015}.
   
  By generalizing the peak-background split argument (see \aref{PBSbias}), the values of the additional non-Gaussian bias parameters can be calculated. The leading bias term in the squeezed limit for the case of a general non-local quadratic non-Gaussianity is \citep{Desjacques2011a,Schmidt2013}
  
  \begin{equation}
   b_{\Psi}^E(M,z)=A\fnl^X\left[2\delta_cb_1^L+4\left(\frac{d\ln\sigma_{R,-\alpha}^2}{d\ln\sigma_{R}^2}-1\right)\right]\frac{\sigma_{R,-\alpha}^2}{\sigma_{R}^2},
  \end{equation}
  where $\delta_c$ is the spherical collapse threshold in an Einstein-de Sitter Universe and the superscript $X$ in $\fnl$ denotes one of the three non-Gaussian types considered here. We defined $ \sigma_{R,n}^2=\frac{1}{2\pi^2}\int k^{2+n} P_{R}^{L}(k,z) dk$, and different primordial shapes correspond to different choices of parameters: $\alpha=2$, $A=3$ for the equilateral shape and $\alpha=1$, $A=-3$ for the orthogonal configuration, while the usual local case is described by $A=1$ and $\alpha=0$ \citep[see \eg][and \aref{PBSbias} for more details]{Dalal2008,Slosar2008,Giannantonio2010}. The higher-order bias parameter result is given in the Eulerian framework by
  \begin{align}\label{eq:bpsidE}
   b_{\Psi\delta}^E(M,z)&=2A\fnl^X\bigg[\delta_c\left(b_2^E+\frac{13}{21}(b_1^E-1)\right) \nonumber \\
   &+b_1^E\left(2\frac{d\ln\sigma_{R,-\alpha}^2}{d\ln\sigma_{R}^2}-3\right)+1\bigg]\frac{\sigma_{R,-\alpha}^2}{\sigma_{R}^2}\,.
  \end{align}
  
  \noindent For the local case the above formula was derived in \citet{Giannantonio2010}. Here we extend it to the equilateral and orthogonal cases [\eref{eq:bpsidE}], by using the peak-background split argument  (see \aref{PBSbias} for details of the full derivation). 
  In the case of equilateral PNG, the scale-dependent bias term approaches a constant value on large scales and thus becomes degenerate with the linear bias parameter, excluding the possibility of constraining $\fnle$ through this non-Gaussian correction. It must be noticed that the transfer function inside $M(k)$ in \eref{eq:sc_dep_stand} also introduces a scale dependence, due to PNG, in the bias at {\em small} scales. However, for large $k$ (\ie $k\sim1/R_*$), another strong degeneracy arises, this time between the PNG bias contribution and higher-order derivative bias terms \citep{Assassi2015}. Due to the presence of degeneracies on both large and small scales, the power spectrum does not essentially bear any constraining power on equilateral PNG. Note that, in our analysis, we never consider scales ($k_\max<1/R_*\approx 0.25\;h/\text{Mpc}$ at redshift $z=0$) where higher-order derivative terms produce significant contributions.
    
  The presence of PNG will affect the halo number density and will therefore introduce a small scale-independent correction in the Gaussian bias parameters. In \citet{Desjacques2009} and \citet{Sefusatti2012} expressions for these offsets\footnote{For the analytic results see \aref{PBSbias}.}, up to the quadratic local-in-matter Eulerian bias parameter, were derived,
  
  \begin{align}
   &\delta b_{1,NG}^E(\fnl)=-\frac{1}{\delta_c}\frac{\nu}{R_{NG}}\frac{\partial R_{NG}}{\partial\nu} \,,\\
   &\delta b_{2,NG}^E(\fnl)=\frac{\nu^2}{\delta_c^2R_{NG}}\frac{\partial^2R_{NG}}{\partial\nu^2}+2\nu(b_1^E-\frac{17}{21})\delta b_{1,NG}^E \,,
  \end{align}
  where $\nu(M,z)=\delta_c/\sigma_R(M,z)$ is the height of the peak (see \aref{PBSbias} for a discussion), $\sigma_R(M,z)$ is the variance of the smoothed density field on mass scale $M$ and $R_{NG}$ is the non-Gaussian correction to the mass function, as derived in \citet{LoVerde2007}.

  \subsubsection{Galaxy bias}\label{sec:hod_bias}
  
  In order to connect the halo bias formalism, presented in the previous two sections, with the bias of galaxies, one needs a method for describing the way galaxies populate the dark matter halos. One way to achieve this is to adopt a halo model \citep{Cooray2002}, where a Halo Occupation Distribution (HOD) function is used to provide the mean number density of galaxies ($\left<N\right>_M$) per halo of a given mass $M$. The HOD model will be used in our Fisher matrix analysis only for predicting the fiducial values of the galaxy bias parameters, in the case of the radio continuum surveys. For optical surveys we will not use a HOD. We will instead use specific bias values, taken from the literature describing relevant galaxy populations (see \sref{sec:optical_surveys} for details).
  The galaxy bias coefficients can be obtained from a weighted average of the halo bias over the range of host halo masses corresponding to the desired galaxy type as
  
  \begin{equation}\label{eq:bg}
   b_i(z)=\frac{\int_{M_\min(z)} d\ln M n_h(M,z)b_i^h(M,z)\left<N(M,z)\right>}{\mean{n}_g(z)},
  \end{equation}
  
  \noindent where $n_h(M,z)$ is the mean number of halos above mass $M$, given by the halo mass function, and $\left<N(M,z)\right>$ is the mean number of galaxies per dark matter halo of mass $M$, given by the HOD model. Finally $M_\min$ is the minimum mass a halo must have to contain a central galaxy and $\mean{n}_g(z)$ is the mean galaxy density given by
  \begin{equation}
   \mean{n}_g(z)=\int_{M_\min(z)} d\ln M n_h(M,z)\left<N(M,z)\right>.
  \end{equation} 
   The HOD used here is a simple three-parameter model proposed by \citet{Tinker2005} and immediately adopted in the literature \citep[\eg][]{Conroy2006,Sefusatti2007}.
  \begin{equation}
    \left<N(M,z)\right>= 
  \begin{dcases}
    1+\frac{M}{M_1(z)}\exp\left({-\frac{M_{cut}(z)}{M}}\right), & \text{if } M\geq M_\min(z)\\
    0,              & \text{otherwise},
  \end{dcases}
  \end{equation}
  \noindent where $M_1$ is the mass required for a halo to contain a second satellite galaxy. In \citet{Conroy2006}, a relationship between $M_1$ and $M_{cut}$ has been derived, $\log_{10}(M_{cut})=0.76\log_{10}(M_1)+2.3$, by fitting the HOD free parameters on $N$-body simulations at different redshifts and number densities. In addition, it has been shown that the ratio $\log_{10}(M_1/M_\min)=1.1$ is almost redshift and mean number density independent. We will use both relationships in order to simplify the HOD model, leaving $M_\min$ as the only free parameter. Finally, we set $M_\min$ so that $\mean{n}_g$ matches  the  expected  mean galaxy number density  of  the survey in each redshift bin.

 \subsection{Galaxy statistics}
  \label{sec:galstat}
  
  \subsubsection{Real space}\label{sec:PB_gal_real}
     
  The halo bias expansion discussed in the previous section can be generalized for any kind of dark matter tracer. In the case of galaxies, the bias expansion in the Eulerian framework can be expressed up to second order in Fourier space, for non-Gaussian initial conditions, by \esref{eq:deltaG}{eq:deltaNG} as 
  
    \begin{flalign}\label{eq:dg}
   \delta&_g(\bk)=\delta_g^{(G)}(\bk)+\delta_g^{(NG)}(\bk)&& \nonumber \\
   &=b_1^E\delta(\bk)+b_{\Psi}^E\Psi(\bk)+\veps^E(\bk) +\intee{q_1}{q_2}\delta_D(\bk-\bq_{12}) \nonumber \\
   &\times\bigg[\left(\frac{b_2^E}{2}+b_{s^2}^E S_2(\bq_1,\bq_2)\right)\delta(\bq_1)\delta(\bq_2) \nonumber \\
   &+\frac{1}{2}\left(\veps_{\delta}^E(\bq_1)\delta(\bq_2)+\delta(\bq_1)\veps_{\delta}^E(\bq_2)\right) \nonumber \\
   &+\frac{1}{2}\bigg((b_{\Psi\delta}^E-b_{\Psi}^EN_2(\bq_2,\bq_1))\Psi(\bq_1)\delta(\bq_2) \nonumber \\
   &+(b_{\Psi\delta}^E-b_{\Psi}^EN_2(\bq_1,\bq_2))\delta(\bq_1)\Psi(\bq_2)\nonumber \\
   &+\veps_{\Psi}^E(\bq_1)\Psi(\bq_2)+\Psi(\bq_1)\veps_{\Psi}^E(\bq_2)\bigg)\bigg],
  \end{flalign}  
  where the kernels $S_2(\bk_1,\bk_2)$ and $N_2(\bk_1,\bk_2)$ are defined as:

  \begin{equation}
   S_2(\bk_1,\bk_2) = \frac{(\bk_1\cdot\bk_2)^2}{k_1^2 k_2^2}-\frac{1}{3}\;\;;\;\;N_2(\bk_1,\bk_2) = \frac{\bk_1\cdot\bk_2}{k_1^2}.
  \end{equation}
  The kernel $S_2$ arises from the Fourier transform of the tidal field scalar $s^2$  \citep{McDonald2009,Baldauf2012}, while the kernel $N_2(\bk_1,\bk_2)$ originates from the Fourier transform of the linear displacement field in the mapping between the Eulerian and Lagrangian frames. For non-Gaussian initial conditions the displacement field is coupled to the primordial gravitational potential, introducing additional terms in the tree-level galaxy bispectrum \citep{Giannantonio2010,Baldauf2011}. Henceforth, we drop the superscript \textit{E} from the bias parameters, since we consider galaxy statistics at the time of observation. 
  
  \begin{figure*}
\centering
   \includegraphics[width=0.8\textwidth]{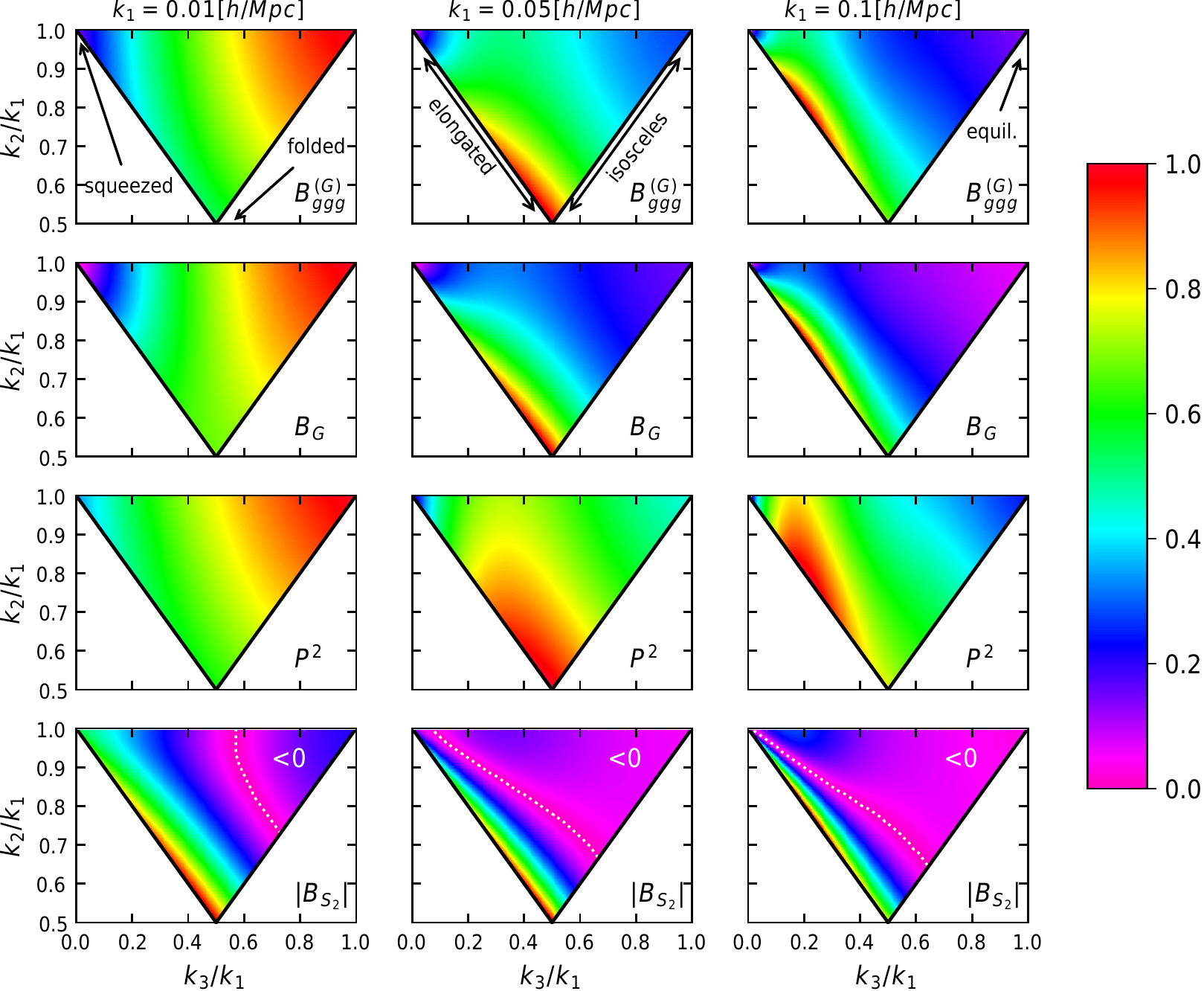}
    \caption{The shape of the galaxy bispectrum for Gaussian initial conditions, together with its highest contributing terms. In each panel the bispectrum, normalized to its absolute maximum value, is plotted as a function of $k_2/k_1$ and $k_3/k_1$ for all configurations in the case of fixed $k_1$. We consider the following three values: $k_1=0.01,\;0.05,\;0.1\;h\text{/Mpc}$, where the triangle sides satisfy $k_3\leq k_2\leq k_1$. In the first row the Gaussian galaxy bispectrum $B^{(G)}$ without the trispectrum contribution is plotted [\eref{eq:Bg_Gauss}], in the second we plot the tree-level matter bispectrum ($B_G$) as predicted by SPT, in the third the quadratic bias term indicated as $P^2$ is plotted (\ie $P_m^L(k_1)P_m^L(k_2)+2\perm$) and finally in the last row we plot the tidal bias term contribution (\ie $B_{S_2}=S_2(k_1,k_2)P_m^L(k_1)P_m^L(k_2)+2\perm$). Note that for the latter the absolute value is plotted, where a white dotted line shows the separation between the positive (left side) and negative (right side) values. For a detailed explanation, see the main text (\sref{sec:PB_gal_real}).
} \label{fig:gauss_bisp}
\end{figure*}
  
 Before presenting the galaxy power spectrum and bispectrum model, we should note that we only consider up to linear terms in $\fnl$, since we assume an $\fnl=0$ fiducial cosmology. In addition we would like to stress that for the bias expansion an implicit smoothing scale $R$ is assumed for the density field and thus $W_{R}(\bk)$ factors are not retained in the expressions; this is allowed since we are working on scales much larger than the smoothing radius ($k_\max \ll 1/R$). Following \citet{Heavens1998}, we can evolve the non-linear density field up to the desired order, smooth it with a filter and then apply the general bias expansion. 
The galaxy power spectrum can be written up to tree level contributions \citep{Dalal2008,Slosar2008,Matarrese2008,Verde2009},

  \begin{equation}\label{eq:Pg_full}
   P_g(k,z)=(b_1+\Delta b(k,\fnl))^2P_m^L(k,z)+P_{\veps} \,,
  \end{equation}
  where $P_m^L$ is the linear matter power spectrum [\eref{eq:linmat}], $P_\veps(k)=\langle \veps(\bk)\veps(\bk')\rangle$ is the large-scale stochastic contribution and $\Delta b(k,\fnl)$ is given by \citet{Agullo2012}
  
  \begin{equation}\label{eq:sc_dep_stand}
   \Delta b(k,\fnl,z)\equiv \frac{b_{\Psi}k^{\alpha}}{M(k,z)}.
  \end{equation}

  For the galaxy bispectrum, by keeping terms  up to $\mathcal{O}(\delta^4)$,  and  for Gaussian initial conditions we obtain
  \begin{align}\label{eq:Bg_Gauss}
   &B_{ggg}^{(G)}(k_1,k_2,k_3,z)=b_1^3B_G(k_1,k_2,k_3,z)+B_{\veps}+2b_1P_{\veps\veps_{\delta}}P_m^L(k_1,z) \nonumber \\  
   &+2b_1^2\left(\frac{b_2}{2}+b_{s^2}S_2(\bk_1,\bk_2)\right)P_m^L(k_1,z)P_m^L(k_2,z) +2\text{ perm.} \nonumber \\  
   &+b_1^2\inte{q}\left(\frac{b_2}{2}+b_{s^2}S_2(\bq,\bk_3-\bq)\right)T_{\delta}(\bk_1,\bk_2,\bq,\bk_3-\bq,z) \nonumber \\
   &+2\text{ perm}. \, ,
  \end{align}
   where $B_G$ is the tree-level gravity-induced matter bispectrum \citep[see \eg][ and references therein]{Verde1998,Verde2000,Bernardeau2002} and $T_\delta$ is the trispectrum of the non-linear matter overdensity (\ie $\delta(\bk)$). Here $B_\veps$ is the bispectrum of the leading stochastic field (\ie $\veps(\bk)$), $P_{\veps\veps_{\delta}}$ is the cross power spectrum between $\veps$ and the next-to-leading order stochastic field (\ie $\veps_\delta$). The full result including non-Gaussian terms up to linear order in $\fnl$ is then given by
  
   \begin{align}\label{eq:Bg_full}
   &B_{ggg}(k_1,k_2,k_3,z)=B_{ggg}^{(G)}(k_1,k_2,k_3,z)+b_1^3B_I(k_1,k_2,k_3,z) \nonumber \\
   &+b_1b_{\Psi}\left(\frac{k_1^{\alpha}}{M(k_1,z)}+\frac{k_2^{\alpha}}{M(k_2,z)}\right)\bigg[2\bigg(b_1F_2^{(s)}(\bk_1,\bk_2) \nonumber \\
   &+\frac{b_2}{2}+b_{s^2}S_2(\bk_1,\bk_2)\bigg)P_m^L(k_1,z)P_m^L(k_2,z) \nonumber \\
   &+\inte{q}\left(\frac{b_2}{2}+b_{s^2}S_2(\bq,\bk_3-\bq)\right)T_{\delta_G\delta}(\bk_1,\bk_2,\bq,\bk_3-\bq)\bigg] \nonumber  \displaybreak[1]  \\
   &+b_1^2\bigg[\bigg(\frac{(b_{\Psi\delta}-b_{\Psi}N_2(\bk_2,\bk_1))k_1^{\alpha} }{M(k_1,z)}+\frac{(b_{\Psi\delta}-b_{\Psi}N_2(\bk_1,\bk_2))k_2^{\alpha} }{M(k_2,z)}\bigg)\nonumber \\
   &\times P_m^L(k_1,z)P_m^L(k_2,z) \nonumber \displaybreak[1] \\
   &+\frac{1}{2}\inte{q}\bigg(\frac{(b_{\Psi\delta}-b_{\Psi}N_2(\bk_3-\bq,\bq))q^{\alpha} }{M(q,z)} \nonumber \displaybreak[1] \\
   &+\frac{(b_{\Psi\delta}-b_{\Psi}N_2(\bq,\bk_3-\bq))|\bk_3-\bq|^{\alpha} }{M(|\bk_3-\bq|,z)}\bigg)T_{\delta_G\delta}(\bk_1,\bk_2,\bq,\bk_3-\bq)\bigg]\nonumber \displaybreak[1] \\
   &+2b_1P_{\veps\veps_{\Psi}}\frac{P_m^L(k_1)k_1^{\alpha}}{M(k_1,z)}+2\text{ perm},
  \end{align}
  where $\delta_D(\sum_i \bk_i)T_{\delta_G\delta}=\langle \delta_G(\bk_1)\delta(\bk_2)\delta(\bk_3)\delta(\bk_4)\rangle$, with $\delta_G$ being the Gaussian part of the density field (\ie the Gaussian part of the primordial curvature perturbation, linearly propagated via Poisson equation), and $P_{\veps\veps_{\Psi}}$ is the cross power spectrum between $\veps$ and $\veps_\Psi$.
  \noindent The fiducial values of the stochastic terms in Eqs. (\ref{eq:Pg_full}), (\ref{eq:Bg_Gauss}) and (\ref{eq:Bg_full}), are taken to be those predicted by Poisson statistics (as $k\ll1/R_*$), and are given by \citep{Schmidt2015,Desjacques2016}:
  
  \begin{equation}
   P_{\veps}=\frac{1}{\mean{n}_g};\;P_{\veps\veps_{\delta}}=\frac{b_1}{2\mean{n}_g};\;P_{\veps\veps_{\Psi}}=\frac{b_{\Psi} }{2\mean{n}_g};\;B_{\veps}=\frac{1}{\mean{n}_g^2}.
  \end{equation}
  
  \noindent The non-Gaussian stochastic bias term in \eref{eq:Bg_full} can also be written as $b_1P_{\veps\veps_{\Psi}}=b_\Psi P_{\veps\veps_\delta}$. 
  
  The Gaussian part of the galaxy bispectrum [\eref{eq:Bg_Gauss}], without the trispectrum contribution, is plotted in \fref{fig:gauss_bisp} for all the triangle configurations generated after keeping $k_1$ fixed and $k_1\geq k_2 \geq k_3$. Three different values are chosen for $k_1$, \ie $k_1=0.01\;h\text{/Mpc}$, $k_1=0.05\;h\text{/Mpc}$ and $k_1=0.1\;h\text{/Mpc}$. In addition, the highest contributing terms to $B_{ggg}^{(G)}$, are also shown in \fref{fig:gauss_bisp}. These include the tree-level gravity-induced matter bispectrum ($B_G$), the quadratic bias term ($P_m^L(k_1,z)P_m^L(k_2,z)+2\perm$), denoted in the plot as $P^2$, and the tidal bias term ($S_2(k_1,k_2)P_m^L(k_1)P_m^L(k_2)+2\perm$), denoted as $B_{S_2}$. In order to show the shape dependence on the triangle configurations, the amplitude of each term is normalized to the maximum value in each panel.
  
  In the second row of \fref{fig:gauss_bisp} the tree-level matter bispectrum signal is plotted, as derived in SPT.
We can see that this term peaks mainly at the elongated ($k_1=k_2+k_3$) and folded ($k_1=2k_2=2k_3$) configurations, while for the squeezed triangles ($k_1\simeq k_2\gg k_3$) its contribution vanishes. This is due to presence of the second order SPT kernel $F_2^{(s)}(\bk_i,\bk_j)$ [\eref{eq:F2kernel}] in $B_G$, which disappears in the squeezed limit and has a maximum at the folded/elongated triangles \citep[see \eg][ for a discussion]{Sefusatti2007,Sefusatti2009,Jeong2009}. As we approach large scales (first column in \fref{fig:gauss_bisp}) we observe that the maximum signal of the matter bispectrum is now at the equilateral triangles ($k_1=k_2=k_3$). This is due to the fact that in this regime the matter power spectrum increases as a function of $k$ and therefore we can get an excess in the signal of $B_G$ when all sides of the triangle are equally large, \ie equilateral configurations \citep[\eg][]{Jeong2009}. The quadratic bias term (shown in the third row of \fref{fig:gauss_bisp}) follows a similar behaviour, where the only difference from $B_G$ is the absence of the $F_2^{(s)}$ kernel. This leads to an enhancement at the squeezed limit and a suppression for the folded/elongated triangles. 
  
The tree-level galaxy bispectrum contribution, proportional to the tidal bias (\ie $B_{S_2}$), peaks on elongated configurations, as we can see from the bottom row of \fref{fig:gauss_bisp}. Note that, due to the presence of the $S_2$ kernel, $B_{S_2}$ can have an amplitude with a negative sign. In order to avoid the saturation of the colour maps in \fref{fig:gauss_bisp}, we show the absolute value of $B_{S_2}$ and we use a white dotted line to separate negative and positive $B_{S_2}$ regions. For most configurations, $B_{S_2}$ is negative on small scales (right column of \fref{fig:gauss_bisp}), while the occurrence of positive values increases on large scales. Note here that, for all equilateral and for most isosceles triangles the tidal bispectrum term is negative, independently of the scale. This behaviour can be explained by the nature of the $S_2(\bk_i,\bk_j)$ kernel, which takes its maximum positive value (for simplicity $\bk_i=\bk_1$ and $\bk_j=\bk_2$) when $\bk_1=a\bk_2$, where $a>1$ (\ie elongated and folded triangles), and its maximum negative value for $\bk_1=\bk_2$ (\ie equilateral triangles). In the folded limit we have, $B_{S_2}\propto [P_m^L(k)]^2+2P_m^L(2k)P_m^L(k)$, and since the matter power spectrum increases towards large scales, the peak of the signal moves towards this configuration. On the other hand, for isosceles configurations ($k_1>k_2=k_3$) the resulting tidal term can be positive or negative depending on the relative size of $k_1$ with respect to the other sides of the triangle. 

                  \begin{figure*}
\centering
   \includegraphics[width=0.7\textwidth]{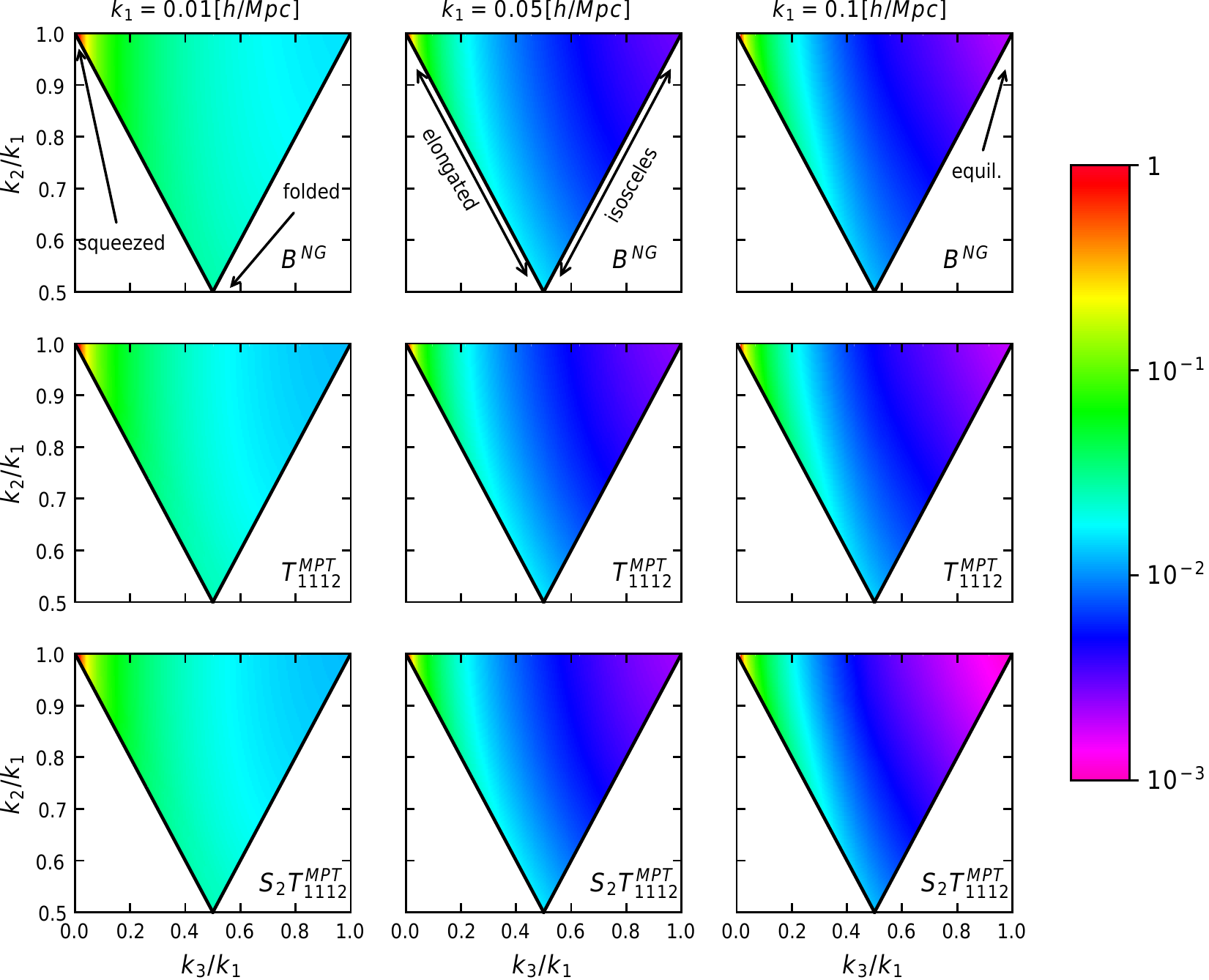}
    \caption{The shape of the non-Gaussian part of the galaxy bispectrum for local-type PNG (top panel) in \eref{eq:Bg_full}, \ie $B^{NG}(k_1,k_2,k_3)=B_{ggg}(k_1,k_2,k_3)-B_{ggg}^{(G)}(k_1,k_2,k_3)$ (cf. \fref{fig:gauss_bisp} for the Gaussian part). The panels display the amplitude of the galaxy bispectrum, normalized to the respective maximum value (note that this implies that a direct comparison of the color scale between different panels is meaningless). The non-linear evolution of the matter field is treated here with the \MPT perturbation theory scheme (see \aref{app:MPT}). In the middle panel the trispectrum loop quadratic bias correction (the $b_2$ trispectrum term in \eref{eq:Bg_Gauss}, \ie $\inte{q}T_\delta^{(5)}(\bk_1,\bk_2,\bq,\bk_3-\bq)+2\perm$), is plotted. Finally, in the bottom panel the tidal bias term trispectrum correction (\ie $\inte{q}S_2(\bq,\bk_3-\bq)T_\delta^{(5)}(\bk_1,\bk_2,\bq,\bk_3-\bq)+2\perm$) is plotted. All the terms plotted here peak at the squeezed limit for this type of PNG. } \label{fig:Bng_loc}
\end{figure*}

         \begin{figure*}
\centering
   \includegraphics[width=0.7\textwidth]{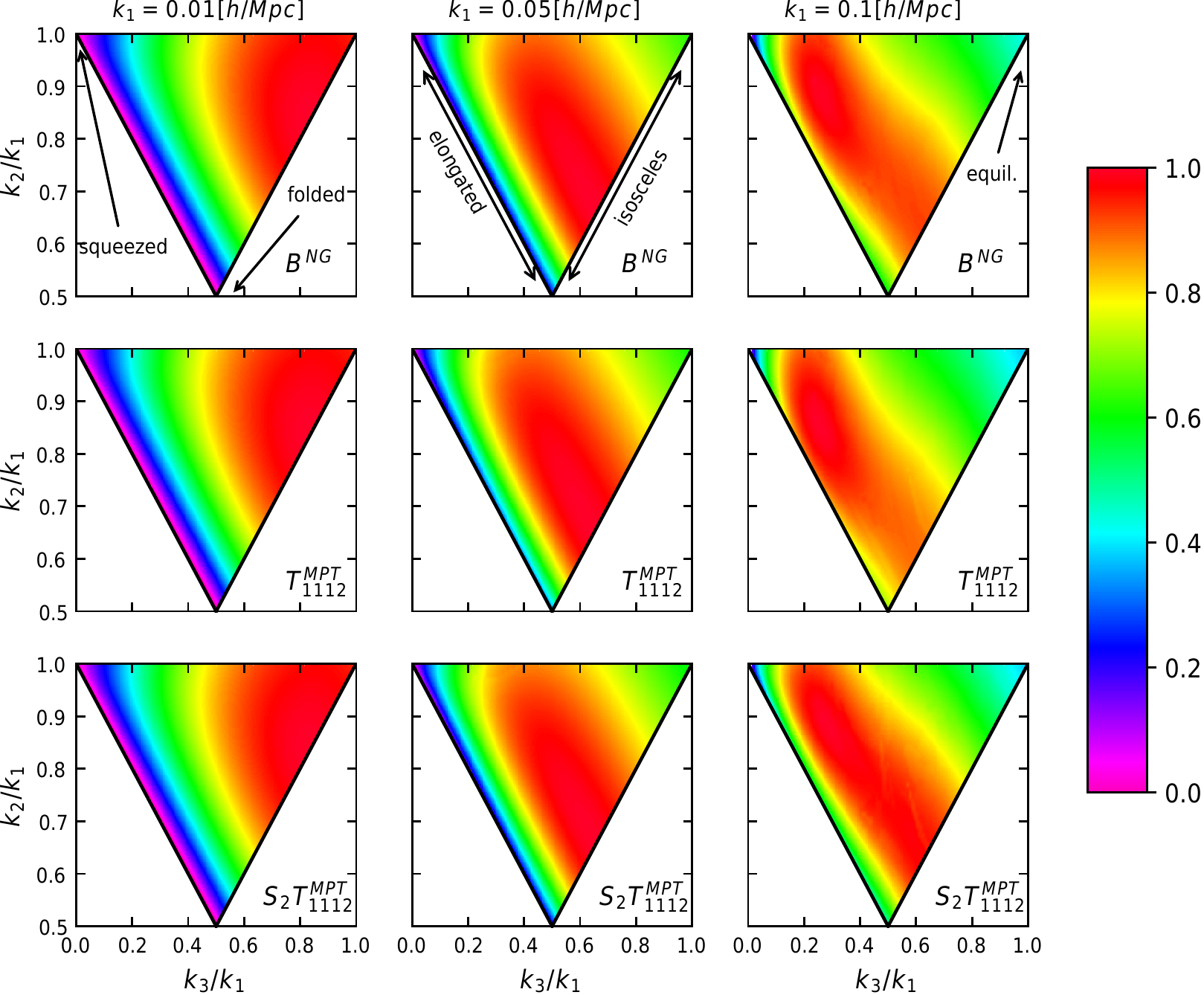}
    \caption{Same as \fref{fig:Bng_loc} but for the equilateral type of PNG. For this type of PNG the galaxy bispectrum is taken to be that of \eref{eq:Bg_Gauss} (\ie $B^{NG}$ is here the sum of the trispectrum bias corrections), since any term proportional to the field $\Psi$ in \eref{eq:Bg_full} (introduced to model the scale dependent bias corrections) is excluded as we discuss in \sref{sec:PNGbias}. } \label{fig:Bng_equil}
\end{figure*}

         \begin{figure*}
\centering
  \includegraphics[width=0.7\textwidth]{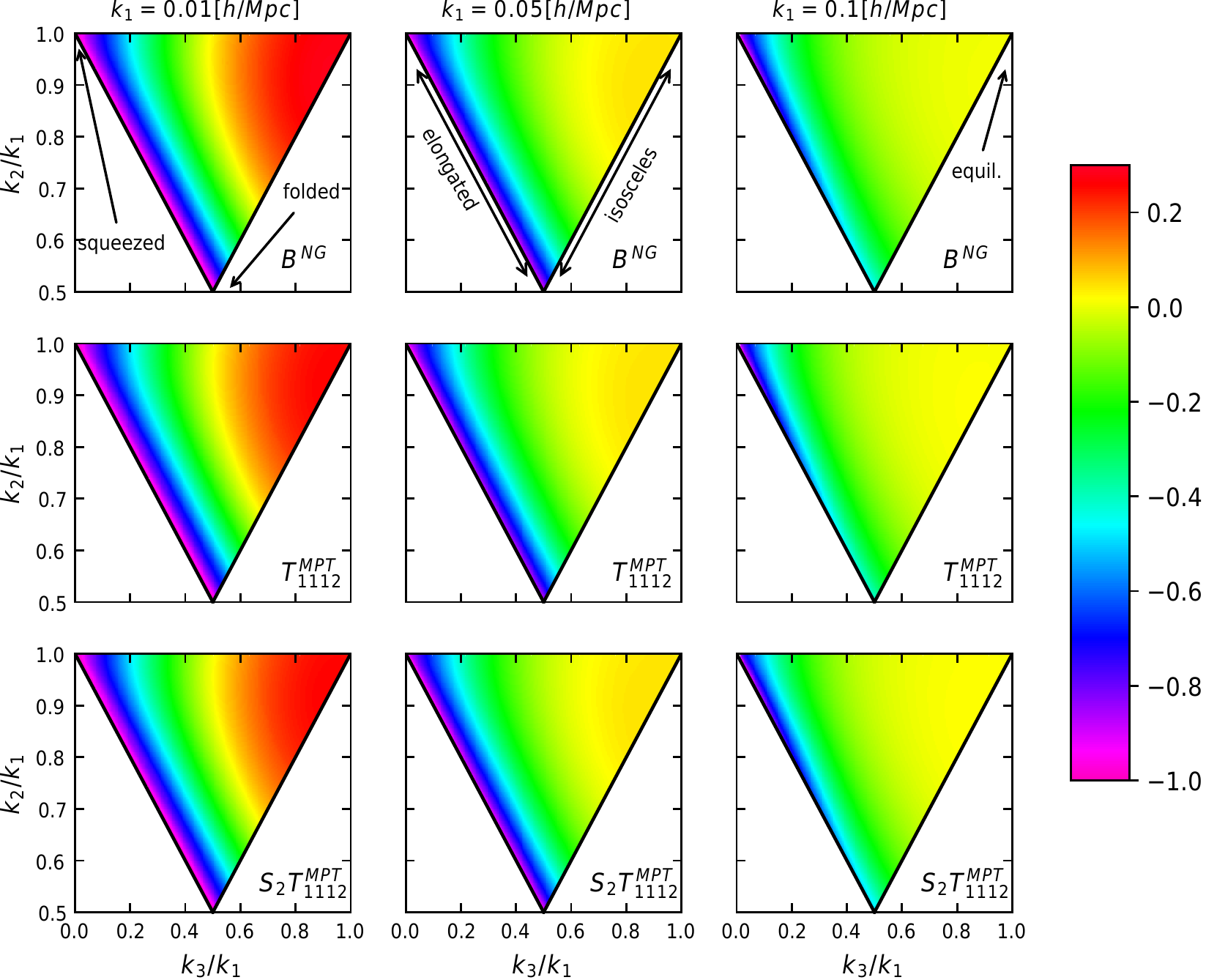}
    \caption{Same as \fref{fig:Bng_loc} but for the orthogonal type of PNG. In this case the equations used are the same as for the local PNG.} \label{fig:Bng_ortho}
\end{figure*}

  Let us now discuss in greater detail the trispectrum terms generated by the bias expansion in the galaxy bispectrum. The importance of the non-linear bias term in \eref{eq:Bg_Gauss} was recognised in the work of \citet{Sefusatti2009} and \citet{Jeong2009} for increasing the sensitivity of galaxy bispectrum to the non-Gaussian initial conditions. 
  Following the notation of \citet{Sefusatti2009}, each individual trispectrum term is expressed as
   \begin{equation}
   \delta^D(\bk_{1234})T_{ijlm}(\bk_1,\bk_2,\bk_3,\bk_4)=\left\langle\delta_{\bk_1}^{(i)}\delta_{\bk_2}^{(j)}\delta_{\bk_3}^{(l)}\delta_{\bk_4}^{(m)}\right\rangle+cyc.
  \end{equation}
Here the upper indices (\ie $i,j,l,m$) denote the expanding order of the non-linear density field. The primordial matter trispectrum (\ie $T_{1111}=T^{(4)}$) for the local case depends on $\fnl^2$ and $g_{\text{NL}}$ and can therefore be neglected in the Fisher analysis performed here. However, we should note that such term has a dominant large scale behaviour for the squeezed configurations and is larger than the non-Gaussian correction to the galaxy power spectrum [\eref{eq:Pg_full}].  The part of the trispectrum generated by the non-linear gravitational coupling (\ie $T_{\delta}^{(6)}$) exhibits no $\fnl$ dependence up to tree-level and therefore it can be ignored in the linear regime considered here. The important contribution for PNG constraints is a coupling term between a non-zero primordial bispectrum and the tree-level gravitational contribution, given by
  
  \begin{flalign}\label{eq:t5}
   T_{\delta}^{(5)}&(\bk_1,\bk_2,\bk_3,\bk_4)=T_{1112}(\bk_1,\bk_2,\bk_3,\bk_4)= && \nonumber \\
   &2[B_I(k_1,k_2,k_{12})F_2^{(s)}(\bk_{12},\bk_3)P_m^L(k_3) \nonumber \\
   &+B_I(k_2,k_3,k_{23})F_2^{(s)}(\bk_{23},\bk_2)P_m^L(k_1) \nonumber \\
   &+B_I(k_3,k_1,k_{31})F_2^{(s)}(\bk_{31},\bk_2)P_m^L(k_2)] \nonumber \\
   &+3\text{ perm} ,
  \end{flalign}
   where $\bk_{ij}=\bk_i+\bk_j$ and $k_{ij}=\lvert\bk_{ij}\rvert$. The above equation has a linear dependence on $\fnl^X$. On large scales it generates a signal that dominates over non-linear terms, for essentially all triangle configurations in the case of local non-Gaussianity \citep{Sefusatti2009,Jeong2009}. Therefore, the constraints on $\fnl^X$ can be significantly improved, as we will show in \sref{results}. When terms proportional to the primordial field $\Psi$ are also considered in the bias expansion, the corresponding trispectrum corrections in \eref{eq:Bg_full} (\ie $T_{\delta_G\delta}$) exhibits one occurrence of $\delta_G$ and therefore will be missing a permutation in \eref{eq:t5}. However, since we only consider $T^{(5)}$ in the tree level matter trispectrum, all the terms in \eref{eq:Bg_full} with $T_{\delta_G\delta}$ will be $\order{\fnl^2}$ and hence they can be ignored. The only remaining $\order{\fnl}$ trispectrum contribution is the one coming from the $T_\delta$ term of \eref{eq:Bg_Gauss}. Its amplitude is shown in the colour maps of \fref{fig:Bng_loc}, \ref{fig:Bng_equil} and \ref{fig:Bng_ortho}, for the three PNG types considered here. Moreover, the non-Gaussian part of the galaxy bispectrum [\eref{eq:Bg_full}] is also shown.
   
   The colouring in the plots shows the shape of the non-Gaussian terms of the galaxy bispectrum, as a function of $k_3/k_1$ and $k_2/k_1$, for three different fixed values of $k_1=0.01,\;0.05,\;0.1\;h\text{/Mpc}$. In the local case, both the primordial bispectrum and the trispectrum correction peak in the squeezed limit. A small difference between the two is originating from the presence of the tidal kernel $S_2$. The PNG contribution can be easily disentangled from the non-primordial part of the galaxy bispectrum, as already pointed out earlier. On larger scales a small increase in the signal is also observed for all configurations, due to the behaviour of the matter power spectrum and of the $F_2^{(s)}$ kernel in this regime (see \sref{sec:PB_gal_real} for a discussion). 
   
    In the equilateral case, the non-Gaussian scale dependence is easily observed. The peak of the signal moves towards large scales, for equilateral configurations. The PNG contribution of the galaxy bispectrum, considered here, includes only trispectrum corrections and $B_I$ (see \sref{sec:PNGbias}), which explains why the panels of \fref{fig:Bng_equil} display a similar behaviour. Note that the observed scale dependence is stronger and more localized, in the equilateral configurations, with respect to the non-primordial bispectrum signal (see first row in \fref{fig:gauss_bisp}), in the large scale regime. This scale dependence can in principle provide a unique signature for measuring $\fnle$. The same general scale dependent behaviour is observed also for orthogonal models and it can also improve the $\fnl$ constraints also this case.
      
   The trispectrum integrals present an ultraviolet divergence, which is automatically cured by adding the smoothing filter with a finite value of $R$. Nevertheless, this introduces a dependence on the smoothing scale in the integration of the trispectrum for the three shapes we consider here\footnote{This was also observed in \citet{Jeong2009} for the local case.}. This makes the results rely upon a non-fundamental quantity, which is unsatisfactory. On large scales this dependence on the smoothing radius goes like $1/\sigma_R^2$, as was also noted in \citet{Jeong2009}. In order to cancel it,  $\sigma_R^2$ is included explicitly in front of the trispectrum integral and later on is reabsorbed by the bias parameters. The ``new'' bias coefficients so obtained are then considered free parameters in the Fisher matrix analysis.

 An alternative approach would be to use a perturbation theory that applies a renormalized technique, like the renormalized perturbation theory (RPT) \citep{Crocce2005b,Crocce2005a}, time renormalized group model \citep{Pietroni2008} and renormalization of bias \citep{McDonald2006,Schmidt2013,Assassi2014,Senatore2014,Mirbabayi2014}. Regardless of the approach taken, the final result for the statistics of galaxies must be the same, therefore here we will use the \MPT formalism \citep{Bernardeau2008,Crocce2012} which simplifies greatly the computational effort of RPT (see \aref{app:MPT}). The reasoning behind this choice is the exponential cut-off that is generated within this formalism and removes the UV divergence of the trispectrum integral by suppressing the small scales contribution. For the physical motivation and the details of the \MPT formalism we refer the reader to \citep{Bernardeau2008,Crocce2012}. We note that, a reduction in the signal originating from intermediate scales is expected in this approach, due to the drop of the matter power spectrum and bispectrum beyond these scales ($k>0.15\;h\text{/Mpc}$ at $z=0$). This is shown in \citet{Lazanu2015}, \citet{Lazanu2017} and \citet{Lazanu2017b}, where a comparison between different perturbation theories is performed using simulations. The resulting power spectrum and bispectrum in the case of \MPT are the same as those defined in \esref{eq:Pg_full}{eq:Bg_full} respectively,   but multiplied by a suppression factor that  accounts for non-linear coupling [\eref{eq:fk}] (see \aref{app:MPT} for a quick review).

  \subsubsection{Redshift Space}\label{sec:PB_gal_red}
  
  In order to model the statistics of galaxies in redshift space, the effect of redshift space distortions (RSD) \citep{Kaiser1987, Hamilton1998} and the Finger-of-God (FOG) must be taken into account \citep{Jackson1972}. The first can be treated perturbatively \citep{Verde1998,Scoccimarro1999}, where the kernel formalism of SPT (see \aref{StanPT}) can be generalised to include the redshift distortions and the bias terms [\eref{eq:dg}] and the second is described phenomenologically  by a damping factor $D_\text{FOG}^P$ and $D_\text{FOG}^B$ for the power spectrum and bispectrum respectively. Since our analysis is restricted to large scales, we only require up to the second order redshift kernels in order to derive the linear power spectrum and tree-level bispectrum in redshift space. For the general, non-local, PNG the power spectrum and bispectrum are given by

   \begin{equation}\label{eq:Pgs}
   P_g^s(\bk,z)=D_\text{FOG}^P(\bk)[Z_1^2(\bk)P_m^L(k,z)+P_{\veps}], 
   \end{equation}
   \begin{align} \label{eq:Bgs}
   &B_g^s(\bk_1,\bk_2,\bk_3,z)=D_\text{FOG}^B(\bk_1,\bk_2,\bk_3) \nonumber \\ 
   &\times\bigg[Z_1(\bk_1)Z_1(\bk_2)Z_1(\bk_3)B_{I}(k_1,k_2,k_3,z) \nonumber \\ 
   &+\bigg(2Z_1(\bk_1)Z_1(\bk_2)Z_2(\bk_1,\bk_2)P_m^L(k_1,z)P_m^L(k_2,z) \nonumber \\
   &+\inte{q}T_{RSD}^{(5)}(\bk_1,\bk_2,\bq,\bk_3-\bq,z)+2\text{ perm}\bigg) \nonumber \\
   &+2P_{\veps\veps_{\delta}}\big(Z_1(\bk_1)P_m^L(k_1,z)+2\text{ perm}\big)+B_{\veps}\bigg], 
  \end{align}
  where the general non-Gaussian redshift kernels up to second order are given explicitly in \aref{app:RSD}. They are reported in a slightly different notation compared to \citet{Tellarini2016}, and we also neglect $\order{\fnl^2}$ contributions.

  The term $T^{(5)}$ in redshift space, after neglecting all $\order{\fnl^2}$ contributions,  derived by using standard PT formalism, the bias expansion of \eref{eq:dg} and RSD  up to second order, is given by
  
  \begin{align}\label{eq:t5_rsd}
   &T_{RSD}^{(5)}(\bk_1,\bk_2,\bk_3,\bk_4,z)=Z_1^{G}(\bk_2)Z_2^{G,b}(\bk_3,\bk_4)[G_{P_1}\mu_1^2f+b_1F_{P_1}] \nonumber \\
   &+Z_1^{G}(\bk_1)Z_2^{G,b}(\bk_3,\bk_4)[G_{P_2}\mu_2^2f+b_1F_{P_2}] \nonumber \\
   &+Z_1^{G}(\bk_1)Z_1^{G}(\bk_2)\bigg[\left(\frac{b_2}{2}+b_{s^2}S_2(\bk_3,\bk_4)\right)F_{P_3} \nonumber \\
   &+\frac{fk_{34}\mu_{34}}{2}\bigg(Z_1^{G}(\bk_4)\frac{\mu_3}{k_3}G_{P_3}+\frac{\mu_4}{k_4}(b_1F_{P_3}+\mu_3^2fG_{P_3})\bigg)\bigg]\nonumber \\
   &+Z_1^{G}(\bk_1)Z_1^{G}(\bk_2)\bigg[\left(\frac{b_2}{2}+b_{s^2}S_2(\bk_3,\bk_4)\right)F_{P_4} \nonumber \\
   &+\frac{fk_{34}\mu_{34}}{2}\bigg(Z_1^{G}(\bk_3)\frac{\mu_4}{k_4}G_{P_4}+\frac{\mu_3}{k_3}(b_1F_{P_4}+\mu_4^2fG_{P_4})\bigg)\bigg]
  \end{align}

  \noindent where $f$ is the linear growth rate. Here $Z_1^{G}(\bk)$ and $Z_2^{G,b}(\bk_1,\bk_2)$ are the Gaussian parts of the redshift kernels $Z_1$ and $Z_2$ respectively [\esref{eq:Z1}{eq:Z2}], while for $Z_2^{G,b}$ we also exclude the two SPT kernel contributions. The terms denoted $F_{P_i}$ and $G_{P_i}$ are the \textit{i}th permutation of \eref{eq:t5}, where the letter \textit{F} and \textit{G} represent the SPT kernel used in the expression at hand (\eg $G_{P_4}=\langle\delta_{\bk_1}^{(1)}\delta_{\bk_2}^{(1)}\delta_{\bk_3}^{(1)}\theta_{\bk_4}^{(2)}\rangle=2G_2(\bk_{12},\bk_3)B_I(k_1,k_2,k_{12},z)P_m^L(k_3,z)+2\perm$). A redshift space model similar to the one in \eref{eq:Bgs} was used in \citet{Tellarini2016} to study expected constraints on the PNG amplitude for a list of future LSS surveys. However, in this reference, only the local case without trispectrum contributions and stochastic bias terms is considered. Besides these new terms, redshift uncertainties are also included in our redshift model, as discussed in the next paragraph.

   The FOG term models the damping effect of the clustering power induced by the FOG effect on linear scales. \citet{Peacock1994} and \citet{Ballinger1996} find that a good model for the power spectrum as an exponential function and  an analogous function can be written for the bispectrum
  \begin{align}
  &D_\text{FOG}^P(\bk)=e^{-(k\mu\sigma_P)^2} \label{eq:DfogP}, \\
  &D_\text{FOG}^B(\bk_1,\bk_2,\bk_3)=e^{-(k_1^2\mu_1^2+k_2^2\mu_2^2+k_3^2\mu_3^2)\sigma_B^2}. \label{eq:DfogB} 
  \end{align}
   Here we consider the fiducial values of $\sigma_P = \sigma_B = \sigma_{\upsilon}(z)$, where $\sigma_{\upsilon}$ is the usual linear, one dimensional velocity dispersion. Besides the FOG effect, the redshift uncertainty of galaxy surveys must be also taken into account. The redshift error, $\sigma_z$, can be translated into a position uncertainty along the line of sight. The treatment of this effect is the same as in the case of FOG \citep[see \eg][]{Seo_2003}, where the only difference is the fiducial value of the $\sigma$ parameter, which will be $\sigma_r=c\sigma_z(z)/H(z)$. Considering both effects gives the final form of the damping factors in \esref{eq:Pgs}{eq:Bgs}, with the $\sigma$ parameters given by $\sigma_\mathrm{v}^2=\sigma_\upsilon^2+\sigma_r^2$. These multiplicative factors introduce a suppression of the signal for all scales with a large component along the line of sight, affecting mostly small scales. In other words, only $k$-modes satisfying $k\mu\sigma_\mathrm{v}\lesssim 1$ are not dominated by noise and can contribute to the power spectrum and bispectrum measurements.

  The redshift space bispectrum is characterized by five variables, three of them defining the triangle shape (\eg the magnitude of the three wavenumbers, $k_1,\;k_2,\;k_3$) and the remaining two the orientation of the triangle with respect to the direction of the line of sight $\hat{z}$, which we consider to be the polar angle $\mu_1$ and the azimuthal angle $\phi$. The bispectrum will now be $B_g^s(\bk_1,\bk_2,\bk_3)=B_g^s(k_1,k_2,k_3,\mu_1,\phi)$. Taking the average over angles (\ie the monopole term in the Legendre expansion) of \eref{eq:Bgs}, in a similar fashion as it was done in \citet{Kaiser1987} for the power spectrum, one can obtain \citep{Sefusatti2006,GilMarin2012b}

  \begin{align}
   &P_g^s(k,z)=\alpha_P(\beta)P_g^r(k,z) \label{eq:Pgsph} \, ,\\ 
   &B_g^s(k_1,k_2,k_3,z)=\alpha_B(\beta)B^r_{ggg}(k_1,k_2,k_3,z) \label{eq:Bgsph} \, ,
  \end{align}
  
  \noindent where $\alpha_P(\beta)=1+2\beta/3+\beta^2/5$,  
   $\alpha_B(\beta)=1+2\beta/3+\beta^2/9$, with $\beta=f/b_1$. The terms $P^r_g$ and $B^r_{ggg}$ are the real space galaxy power spectrum and bispectrum, given by \esref{eq:Pg_full}{eq:Bg_full} respectively. The redshift space bispectrum presented above, as described in \citet{Sefusatti2006}, is derived after an additional average over $\theta_{12}=\mathrm{acos}(\widehat{k_1}\widehat{k_2})$ and by dropping the dependence on the second-order PT velocity kernel [\eref{eq:G2kernel}] and the FOG effect. This is a good approximation on large scales since these two effects partially cancel out.     
  
  In the cases of local and orthogonal PNG the galaxy power spectrum and bispectrum is described by the full model of Eqs. (\ref{eq:Pg_full}), (\ref{eq:Bg_full}) and (\ref{eq:Pgs}), (\ref{eq:Bgs})  for real and redshift space respectively. In \eref{eq:Bgs} we will keep only the $\order{\fnl}$ terms, while the full form was written down for completeness. For the equilateral case, as we discussed before, the scale dependent bias contribution is degenerate and will not be used. Nevertheless we will use the Gaussian power spectrum, excluding its signal contribution from constraining $\fnl$, and the bispectrum without the terms proportional to the non-local primordial field $\Psi$. In this case, the trispectrum bias contribution can compensate for the missing large scale signal and improve the expected constraints on the non-Gaussian amplitude, as we will show in \sref{results}.

\section{Galaxy Surveys}\label{sec:surveys}
In this Section we describe the specifications of the radio continuum and optical galaxy surveys we used in this work. 

\subsection{Future radio surveys}\label{sec:radio_surveys}
We focus on  future radio surveys to investigate whether the high-redshift, full sky nature of those datasets will provide better constraints on non-Gaussianity parameters, despite the lower precision in redshift information.

   \begin{table}
\centering
\resizebox{\hsize}{!}{
\begin{tabular}{llll|llll}
 \hhline{========}\\
 \multicolumn{4}{c}{Radio continuum, 10 $\mu$Jy} & \multicolumn{3}{r}{Radio continuum, 1 $\mu$Jy}\\ \hline   
 $z$ & $\sigma_z$ & $V$ & $\mean{n}_g$ & $z$ & $\sigma_z$ & $V$ & $\mean{n}_g$ \\ \hline
 $0.86$ & $0.18$ & $12.73$ & $2.54$ & $0.41$ & $0.16$ & $4.32$ & $2.15$  \\
 $1.45$ & $0.28$ & $29.63$ & $2.04$ & $1.01$ & $0.23$ & $18.93$ & $5.84$ \\
 $2.3$  & $0.28$ & $33.74$ & $1.27$ & $1.6$ & $0.3$ & $33.22$ & $9.23$ \\
 $3.46$ & $0.46$ & $52.6$ & $0.43$ & $2.56$ & $0.42$ & $50.6$ & $4.57$ \\
 $5.48$ & $0.63$ & $58.2$ & $0.057$ & $4.1$ & $1.63$ & $175.1$ & $1.17$ \\
 \hhline{========}
 \end{tabular}
 }
 \caption{The basic specifications for the two radio surveys considered here for each redshift bin. The shell volume is in units of $(\text{Gpc/}h)^3$ and the mean number density in $10^{-4}(h\text{/Mpc})^3$. The redshift distributions are obtained with the CBR technique (see \sref{sec:radio_surveys} for details). We take the redshift uncertainty of sources as the same value of half the width of the bin it resides in.}
 \label{table:radio_specs}
\end{table}

We forecast results for two types of radio continuum surveys, inspired by the SKA survey \citep{Jarvis2015}, assuming either a $10\mu Jy$ or a a $1 \mu Jy$ flux limit. In both cases we consider $30,000\;\text{deg}^2$.

      \begin{figure}
    \centering
    \resizebox{\hsize}{!}{\includegraphics{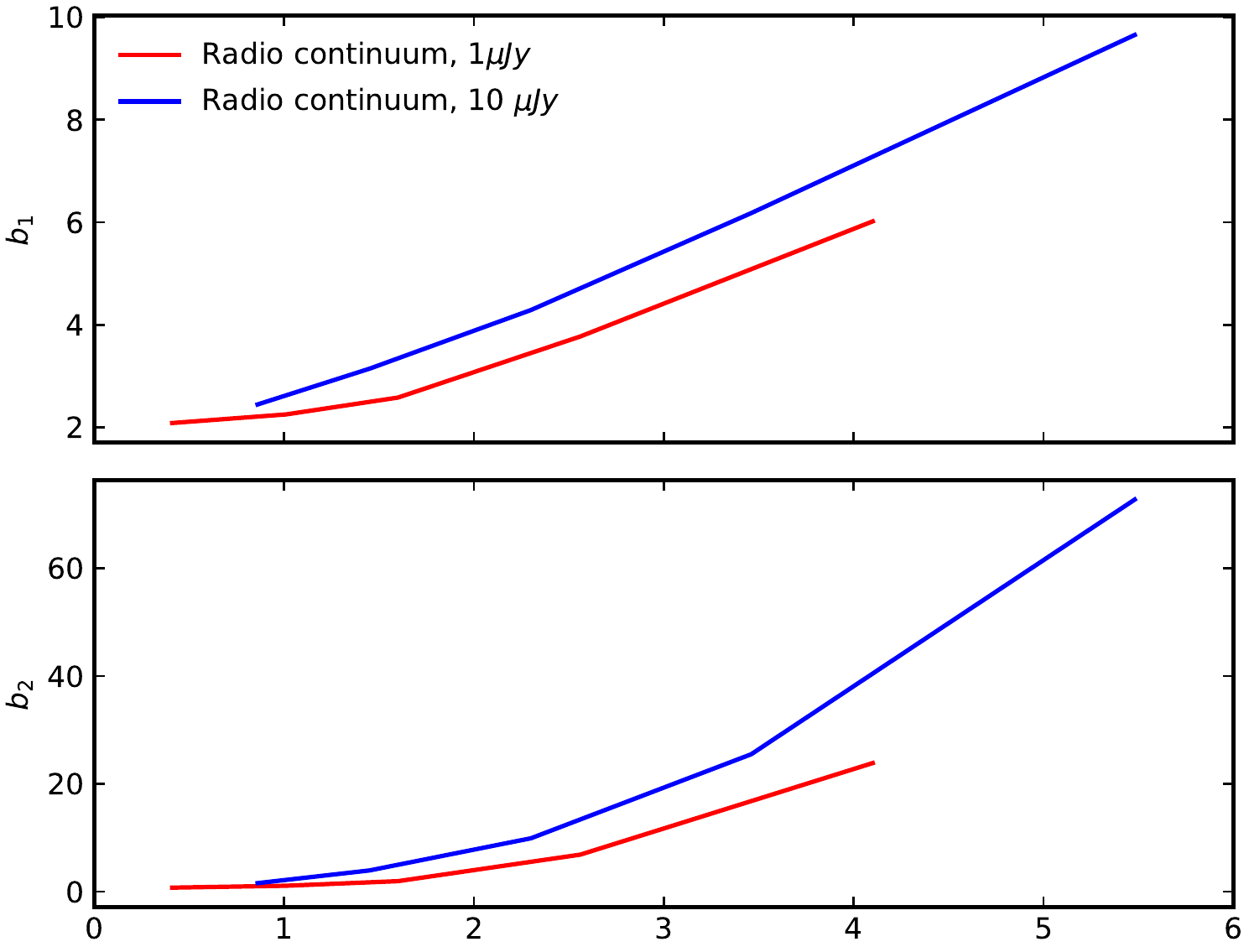}}
    \caption{The galaxy bias, defined from the weighted average over the halo one, as a function of redshift for the two radio surveys considered here, a survey flux limited at $10\mu Jy$ and one with a $1\mu Jy$ flux limit. The details of the surveys are listed in Table \ref{table:radio_specs}.}
\label{fig:bias_radio}
    \end{figure}

Given that we want to forecast the advantages of having both full sky and high-\textit{z} data, we use radio continuum surveys, that are however plagued by the fact that the sources' redshifts are in principle not known. This would allow only the computation of an angular projected power spectrum and bispectrum, in the standard case.

However, recently some techniques have been developed in order to provide such surveys with statistical redshift information; here we follow the Clustering-Based Redshift (CBR) technique, developed in \citet{Schneider2006}, \citet{Newman2008}, \citet{Menard2013} and studied for some cosmological applications, including the SKA in \citet{Kovetz2016}. A disadvantage of this method is that the provided redshift information can have a large uncertainty. In fact it is safe to assume that the redshift error $\sigma_z(z)$ is equal to the width of the bin considered. The resulting predicted redshift distribution and redshift errors for the radio surveys used here are presented in \tref{table:radio_specs}.

The galaxy bias for the two radio galaxy surveys is given by the modelling described in \sref{sec:hod_bias}, while the results for the linear and quadratic bias are plotted in \fref{fig:bias_radio} as a function of redshift.

\subsection{Future optical surveys}\label{sec:optical_surveys}

In the forthcoming years a plethora of large scale structure surveys will provide maps of various galaxy types over large volumes. The large number of modes and high redshifts probed by these surveys will give the possibility to produce tight constraints on the amplitude of PNG. In this section we present the forecasts for spectroscopic and a photometric experiments, covering a variety of redshift ranges and volumes. 

For the spectroscopic survey we consider the redshift range of $0.7\leq z\leq2$ over $15,000\;\text{deg}^2$, while we model the redshift distribution as in \tref{table:Euclid_specs} from the one in \citet{Orsi2010}.

In \tref{table:Euclid_specs} we show the main specifications of the survey used in this work, which is not too dissimilar from an Euclid-class or DESI-class survey  \citep{Laureijs2011,FontRibera2013}. The  adopted size of redshift bins is $\Delta z=0.1$, while the spectroscopic redshift error is given by $\sigma_z(z)\approx0.001(1+z)$.

     \begin{table}
 \centering
 \begin{tabular}{lcc}
 \hhline{===}\\ 
 $z$ & $V$  & $\mean{n}_g$ \\ \hline
 $0.7$ & $2.82$ & $12.95$ \\
 $0.8$ & $3.28$ & $19.95$ \\
 $0.9$ & $3.70$ & $19.13$ \\
 $1.0$ & $4.08$ & $17.7$ \\
 $1.1$ & $4.42$ & $16.0$ \\
 $1.2$ & $4.72$ & $14.31$ \\
 $1.3$ & $4.98$ & $12.85$ \\
 $1.4$ & $5.20$ & $10.72$ \\
 $1.5$ & $5.39$ & $8.64$ \\
 $1.6$ & $5.54$ & $6.24$ \\
 $1.7$ & $5.67$ & $4.07$ \\
 $1.8$ & $5.78$ & $3.82$ \\
 $1.9$ & $5.86$ & $2.28$ \\
 $2.0$ & $5.93$ & $1.26$ \\
 \hhline{===}
 \end{tabular}
 \caption{The basic specifications for a spectroscopic survey for each redshift bin. The shell volume is in units of $(\text{Gpc/}h)^3$ and the mean number density in $10^{-4}(h\text{/Mpc})^3$.}
 \label{table:Euclid_specs}
\end{table}

\begin{figure}
    \centering
    \resizebox{\hsize}{!}{\includegraphics{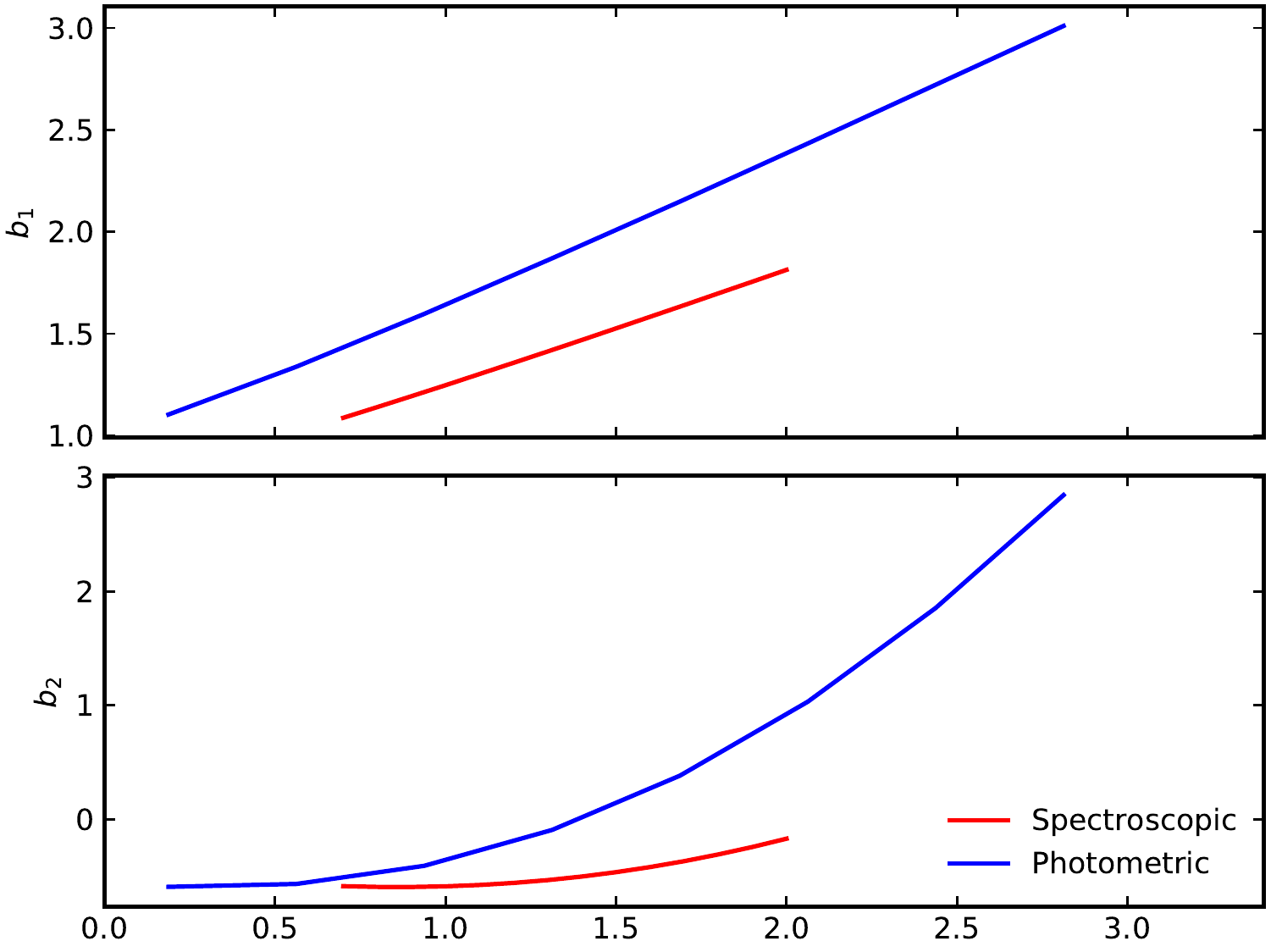}}

    \caption{The linear and quadratic galaxy bias for the optical surveys considered here. The details for the derivation are described in \sref{sec:optical_surveys} for each individual survey.}
    \label{fig:bias_optic}
    \end{figure}

The galaxy sample is assumed to involve a single tracer, whose linear bias is given by $b_1(z)=0.76/D(z)$ \citep{FontRibera2013}. The higher order Eulerian bias coefficients are given from the halo bias predictions (see \aref{PBSbias}), where $\nu$ is determined by $b_1(z)=b_1^h(\nu,z)$ [see \eref{eq:b1hE}]. The results for the linear and quadratic cases can be seen in \fref{fig:bias_optic}.

\begin{figure}
    \centering
    \resizebox{\hsize}{!}{\includegraphics{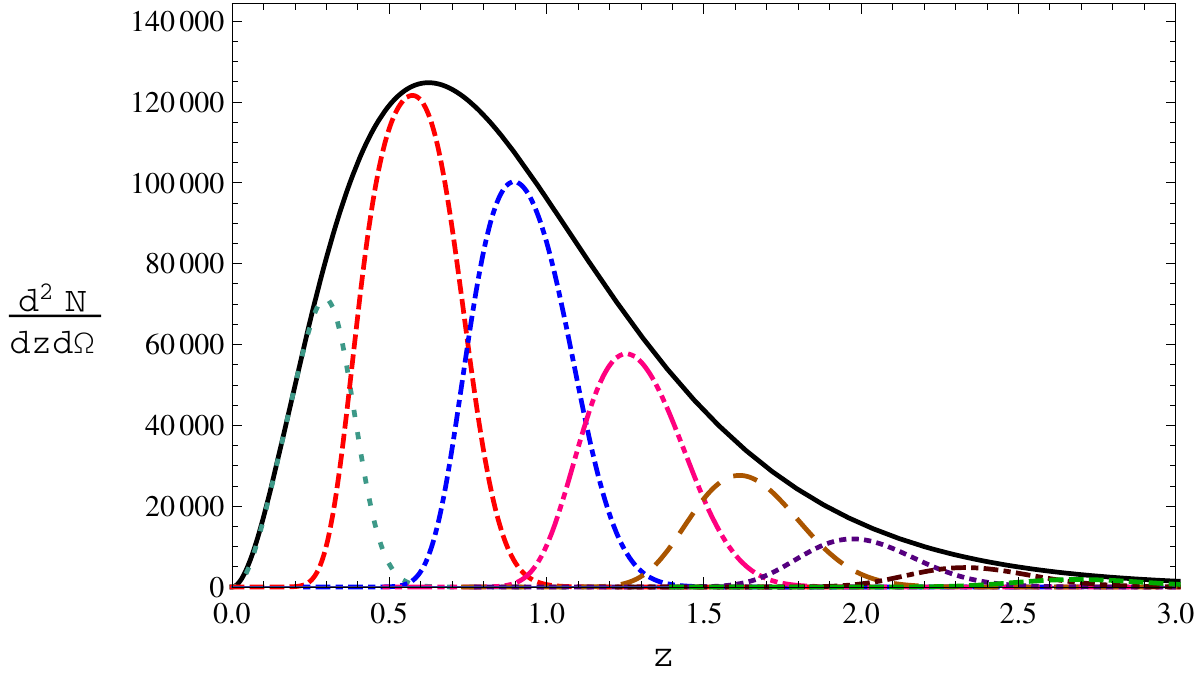}}
    \caption{The overall true redshift distribution $n(z)$ (solid black line) is plotted over the redshift range considered for the photometric survey. The redshift distribution for each tomographic bin, given by \eref{eq:ni_z}, is also shown (dashed, dotted, \etc lines).}
    \label{fig:LSST_nz}
    \end{figure}

In the  photometric survey case, we take inspiration from the future LSST survey~\citep{LSST2009}, and we use a redshift distribution as in \fref{fig:LSST_nz}, where we divide our sample in eight redshift bins covering the range $0\leq z \leq3$ over $18,000\;\text{deg}^2$. The distribution of the true redshifts of galaxies inside each photometric bin $i$ is given by \citep{Ma2006,Zhan2006}

  \begin{equation}\label{eq:ni_z}
   n_i(z)=n(z)\int_{z_\text{ph}^{(i)} }^{z_\text{ph}^{(i+1)} }dz_\text{ph}P(z_\text{ph}|z) \, ,
  \end{equation}
with $z_\text{ph}$ being the photometric redshift, and  where the integration limits define the extent of bin $i$. Following the work of \citet{Ma2006}, we model the photometric redshift error at each redshift by a Gaussian, given by:
  
  \begin{equation}
   P(z_\text{ph}|z)=\frac{1}{\sqrt{2\pi}\sigma_z }\exp[-\frac{(z-z_\text{ph}-z_\text{bias})^2 }{2\sigma_z^2}],
  \end{equation}
  where the photometric redshift bias $z_\text{bias}$ and the photometric rms error $\sigma_z$ are functions of the true redshift. The latter is chosen here to be $\sigma_z(z)=0.05(1+z)$. The fiducial value of the redshift bias is chosen to be $z_\text{bias}(z)=0$. Note that the above probability must be normalised to $\int_0^\infty dz_\text{ph}P(z_\text{ph}|z)$ in order to ensure the positivity of the photometric redshifts. The CBR technique used to acquire the redshift information for the radio continuum sample (\sref{sec:radio_surveys}), can be also adopted for the case of an optical photometric survey (\eg LSST). Given the nature of the clustering redshifts, it would ensure that $z_\text{bias}$ would always be null \citep[see \eg][]{Rahman2016}. 
  
  The overall galaxy true redshift distribution $n(z)=d^2N/dzd\Omega$ is given by the functional form, $n(z)\propto z^\alpha\exp[-(z/z_0)^\beta]$ \citep{Wittman2000}, with $\alpha=2$, $\beta=1$ and $z_0=0.3125$. The normalization of the overall galaxy redshift distribution is fixed by requiring that the total number of galaxies per steradian (\ie $n_\text{tot}=\int_0^\infty dz n(z)$) to be equal to the cumulative galaxy counts of the survey. The overall normalized redshift distribution $n(z)$ is plotted together with the true redshift distribution in each photometric bin in \fref{fig:LSST_nz}. The final number densities for each photometric redshift bin are listed in \tref{table:LSST_specs}. The fiducial value for the linear bias is chosen to be $b_1(z)=1/D(z)$ \citep{LSST2009,FontRibera2013,Passaglia2017}, where the derivation of the higher-order bias parameters follows the prescription described before. The final results for the linear and quadratic biases are plotted in \fref{fig:bias_optic}. We use a surface number density of $n_\text{tot}=40\;\text{gal}/\text{arcmin}^2$. These specifications are chosen to mimic an LSST-like, representative future photometric survey \citep[see \eg][]{LSST2009, Zhan2017}.
  
       \begin{table}
 \centering
 \begin{tabular}{lcc}
 \hhline{===}\\ 
 $z$ & $V$  & $\mean{n}_g$ \\ \hline
 $0.1875$ & $1.95$ & $166.64$ \\
 $0.5625$ & $9.57$ & $82.84$ \\
 $0.9375$ & $17.24$ & $39.62$ \\
 $1.3125$ & $22.46$ & $18.37$ \\
 $1.6875$ & $25.41$ & $8.26$ \\
 $2.0625$ & $26.78$ & $3.60$ \\
 $2.4375$ & $27.14$ & $1.53$ \\
 $2.8125$ & $26.89$ & $0.63$ \\
 \hhline{===}
 \end{tabular}
 \caption{The basic specifications for a photometric survey for each redshift bin. The derivation of these values is described in \sref{sec:optical_surveys}. The shell volume is in units of $(\text{Gpc/}h)^3$ and the mean number density in $10^{-3}(h\text{/Mpc})^3$.}
 \label{table:LSST_specs}
\end{table}

\section{Methodology }

\subsection{Fisher Matrix}\label{sec:fisher}

In order to forecast the precision in measurements of the non-Gaussianity amplitude $\fnl$ and bias parameters, we make use of the Fisher matrix formalism. The Fisher information defines the minimum error on an unknown parameter given an observable. In the case of the galaxy power spectrum, the Fisher matrix is given by
    \begin{equation}\label{eq:fisherPs}
    F_{\alpha\beta}^{Ps}=\sum_{\mu_1=0}^{1}\sum_{k=k_{\rm min} }^{k_\max}\frac{\partial P_g^s}{\partial p_{\alpha}}\frac{\partial P_g^s}{\partial p_{\beta}}\frac{1}{\Delta P^2} \, ,
   \end{equation}
   
   \noindent while for the bispectrum we have \citep{Scoccimarro2003}
   
   \begin{equation}\label{eq:fisherBs}
    F_{\alpha\beta}^{Bs}=\sum_{\mu_1=-1}^{1}\sum_{\phi=-\pi/2}^{\pi/2}\sum_T\frac{\partial B_g^s}{\partial p_{\alpha}}\frac{\partial B_g^s}{\partial p_{\beta}}\frac{1}{\Delta B^2}\, ,
   \end{equation}
   
  \noindent where $p_{\alpha,\beta}$ are the unknown parameters of interest and the derivatives are evaluated at the fiducial value of the parameter vector. We model power spectrum and bispectrum as discussed in \sref{sec:galstat}. The sum over the triangles, which is denoted in \eref{eq:fisherBs} as $\sum_T$, is given by applying the triangle condition on the $k$ modes, while the ordering considered is $k_1\;\geq\; k_2\; \geq k_3$. The wavenumbers are binned with a bin size of $\Delta k$, taken here to be the fundamental frequency of the survey $k_f=2\pi/V^{1/3}$, between a minimum value $k_\min=k_f$ (largest scales probed by survey) and $k_\max$ (smallest scales considered). The angular bin sizes are taken here to be $\Delta\mu_1=0.1$ and $\Delta\phi=\pi/25$. 
   
   The parameter vector considered here per bin consists of the non-Gaussian amplitude, three bias parameters, stochastic power spectrum and bispectrum contributions, the linear growth rate and the velocity dispersion,
   
   \begin{equation}\label{eq:params}
    \mathbf{p}=\{\fnl^X,b_1,b_2,b_{s^2},P_{\veps},P_{\veps\veps_{\delta}},B_{\veps},f,\sigma_{\upsilon}\}.
   \end{equation}      
   The stochastic bias contributions (\ie $P_{\veps},P_{\veps\veps_{\delta}}$ and $B_{\veps}$) are considered here as nuisance parameters and they are marginalised over to acquire the Fisher sub-matrix for the parameters of interest. Cosmological parameters are fixed throughout this work. Since we can determine them with high accuracy, using other probes (\eg CMB, BAO etc.) we do not expect  a significant impact of propagating their errors on $\fnl$, which is instead constrained via power spectrum scale dependence on very large scales, or via the bispectrum. This is actually verified for the CMB primordial bispectrum in \citet{Liguori2008}, and for that case it is also shown that the level of degeneracy between cosmological parameters and $\fnl$ is small\footnote{As this work was being completed, a paper by \citet{Moradinezhad2018} appeared on arXiv, studying the galaxy bispectrum arising from massive spinning particles during Inflation. Their galaxy bispectrum analysis has a different focus and present little overlap with the present one, in terms of bispectrum shapes, range of surveys and many other aspects (e.g. inclusion of theoretical errors, inclusion of trispectrum term in the bispectrum model). However, in \citet{Moradinezhad2018} the cosmological parameter errors are explicitly propagated into the final $\fnl$ errors, for the primordial model they study. They also consider the local shape, and for both cases they show that the impact of parameter marginalisation on $\fnl$ is small, confirming expectations discussed here.} . The final forecasts for the non-Gaussian amplitude are derived after marginalising over the remaining parameters and summing the contribution from all the available redshift bins. Following \eg \citet{Sefusatti2007,Giannantonio2012,Song2015,Tellarini2016}, we neglect cross-correlations between the different redshift bins. This choice is not expected to have a significant effect on the final forecasts, due to the sufficiently large bin width considered here. However, one should perform an explicit check on this, which we leave for a future work. For the spectroscopic survey, the high accuracy of the measured redshifts can provide a redshift distribution function with a negligible overlap, which indicates that the redshift bins' cross-spectra would have a small effect \footnote{In \citet{Bailoni2017} they show that, neglecting cross-correlations between the different redshift bins leads to a few percent effect.} \citep[see \eg][ for a discussion]{Giannantonio2012}. We aim for a complete and conservative analysis, therefore we will restrict the analysis to the linear regime, and exclude non-linear scales. For the redshift evolution of $k_\max(z)$ we consider, $k_\max(z)=0.1/D(z)\;h\text{/Mpc}$ where it slowly varies with redshift while it stays inside the linear and semi-linear regime, ensuring the validity of the bias expansion as well as SPT itself. This limit in the magnitude of the $k$ mode holds also for the bispectrum, where the sides of the triangle satisfy $k_\min\leq k_1, k_2, k_3 \leq k_\max$.

       In our Fisher matrix analysis, only the diagonal part of the covariance matrix (\ie $\Delta P^2$ and $\Delta B^2$) is taken into consideration, neglecting all the cross-correlations between different triangles (bispectrum) and $k$-bins (power spectrum). We adopt a Gaussian approximation for the variance terms:

  \begin{align}
    &\Delta P^2(k,z)=\frac{4\pi^2}{V_{\text{survey}}k^2\Delta k\Delta\mu}P_{\mathrm{tot}}^2\, , \label{eq:deltaP2} \\
    &\Delta B^2(k_1,k_2,k_3,z)=s_{123}\pi V_f\frac{P_{\mathrm{tot}}(k_1)P_{\mathrm{tot}}(k_2)P_{\mathrm{tot}}(k_3)}{k_1k_2k_3\Delta k^3\Delta\mu\Delta\phi}\, ,\label{eq:deltaB2}
   \end{align}

 \noindent where $s_{123}=6,2,1$ for equilateral, isosceles and scalene triangles respectively. The volume of the fundamental shell in Fourier space is  $V_f=k_f^3$. In addition $P_{\mathrm{tot}}(k,z)=P_g^s(k,z)+1/\mean{n}_g$, where the stochastic contribution is excluded from $P_g^s$ and the remaining term accounts for the shot noise. Note that for the Fisher matrix in redshift space, the normalization for the range of the two angles, \ie $N_{\mu}=\mu_{max}-\mu_{min}$ and $N_{\phi}=\phi_{max}-\phi_{min}$, must be applied. The full covariance of the two estimators is outlined in \citet{Sefusatti2006}, where the off-diagonal elements are defined by higher than three point correlators. Although the above results are for redshift space statistics, the reduction to real space is straightforward. 

Recently, the authors of \citet{Chan2017} used dark matter $N$-body simulations, including four halo samples with different number densities, in order to study the full covariance of the power spectrum and bispectrum estimators. They focused on extracting an integrated signal-to-noise ratio for all bins/triangles, checking how this gets degraded when off-diagonal covariance elements and non-Gaussian contributions to the variance are accounted for. While this does not include a specific study of the degradation of error bars for PNG or other cosmological parameters, their results can provide useful guidelines to assess the validity of our diagonal covariance approximation and the error on the parameter forecast we introduce by employing it.
       
       For the dark matter power spectrum, \citet{Chan2017} show that the correlation coefficient between different modes never exceeds $\sim 15\%$ at $z=0$, up to $k_j=0.1\;h\textrm{/Mpc}$. For $z=1$, the correlation reaches at most $\sim 20\%$, up to $k_j\sim 0.15\;h\textrm{/Mpc}$. In all cases, correlations decrease away from the diagonal. On the scale range considered here, the non-Gaussian corrections to the diagonal part of the covariance is negligible at $z=0$, as well as for $z=1$. For halos, these corrections can be up to $\sim 10\%$ for the same scale range, in the case of the small density halo samples, with the exact value depending on the redshift. Furthermore, the results agree with the covariance model predicted by PT up to $k=1\;h\textrm{/Mpc}$ for the abundant halo sample. Therefore we conclude that the exclusion of the off-diagonal part in the galaxy power spectrum covariance will introduce an error of up to $10-15\%$, for the scale range and redshifts considered here, depending on the number density of the sample. In addition, the non-Gaussian corrections to the variance are negligible for the high density samples and scale range considered here.
   
   For the dark matter bispectrum, it is shown that the non-Gaussian corrections to the diagonal Gaussian part (for equilateral configurations) is $\sim 8\%$ at $z=0$ and $k=0.1\;h\textrm{/Mpc}$ and that PT predictions agree with the numerical results up to $k\sim 0.15\;h\textrm{/Mpc}$. For higher redshifts (\ie $z=0.5,\;1$) the corrections are at the level of a few percent, up to $k\sim 0.16\;h\textrm{/Mpc}$, while the PT predictions are in good agreement with the results up to $k\sim 0.2\;h\textrm{/Mpc}$ and $k\sim 0.3\;h\textrm{/Mpc}$ for $z=0.5$ and $z=1$ respectively. In addition, it is shown that the correlation coefficient, used to test couplings between different triangles, is consistent with zero for the large scales and for the redshift slices considered. 

When, instead of dark matter, we consider halos, triangle couplings and non-Gaussian corrections strongly depend on the redshift and density of the sample. Corrections to the diagonal Gaussian part are negligible at $z=0$ and $k=0.1\;h\textrm{/Mpc}$ in the case of a high density sample ($\mean{n}\sim 10^{-3}\;\textrm{[Mpc/}h]^{-3}$). On the other hand, for a low density case ($\mean{n}\lesssim 2\times10^{-5}\;\textrm{[Mpc/}h]^{-3}$) the correction is found to be between few percent and $\sim 10\%$ at low redshift, and increasing  up to $\sim 90\%$ at high redshifts. 
   
The samples considered in this work have a high number density for the majority of the redshift bins (except only for some high redshift slices, where non-linearities are anyway less important), $\mean{n}_g\gtrsim10^4\;(h\text{/Mpc})^3$ (see \tref{table:radio_specs} and \aref{sec:surveys}). Therefore we do not expect the exclusion of the non-Gaussian part in the covariance to overestimate much the S/N ratio and to have a large impact on the final PNG forecasts.  

As explicitly shown in \citet{Chan2017}, on the large scales considered here the full non-Gaussian contribution can be well approximated by including perturbative corrections to the power spectrum appearing in the bispectrum variance expression, obtaining

\begin{align}\label{eq:DB2_NL}
 &\Delta B_\text{NL}^2(k_1,k_2,k_3,z)=\Delta B^2(k_1,k_2,k_3,z)+\frac{s_{123}\pi V_f}{k_1k_2k_3\Delta k^3\Delta\mu\Delta\phi} \nonumber \\
 &\times \left(P_{\mathrm{tot}}(k_1)P_{\mathrm{tot}}(k_2)(P_g^\text{NL}(k_3)-P_g(k_3)+\frac{1}{\mean{n}_g })+2\perm\right) \, ,
\end{align}
where $P_g^\text{NL}(k_3)$ is given by \eref{eq:Pg_full} after replacing the linear matter power spectrum with the non-linear one, as predicted by the \textsc{HALOFIT} algorithm \citep{Smith2003,Takahashi2012}. The reason that \textsc{HALOFIT}, instead of the PT one-loop matter power spectrum, is used, lies in the fact that the latter leads to overestimating the actual variance in the weakly non-linear regime. We will use this expression later on in our analysis, in order to estimate in more detail the effect of neglecting non-Gaussian corrections in our forecasts.

      Another aspect to consider is the covariance between the power spectrum and the bispectrum (PB) when the two are used jointly. This was provided, in the Gaussian case, in \citet{Sefusatti2006} and used in \citet{Song2015}. In \citet{Chan2017}, a comparison between the S/N ratios coming from $N$-body simulations is made in order to test the effect of the $PB$ cross-covariance. In the case of dark matter, they show that, in the redshift range $ 0 \leq z \leq 1$, on linear scales, for abundant samples as those considered here, $PB$ corrections to the overall signal-to-noise amount at most to $\sim 10\%$.  This justifies neglecting the cross-covariance in our forecasts. The combined Fisher matrix will therefore be just obtained as $F_{\alpha\beta}^{P+B}=F_{\alpha\beta}^P+F_{\alpha\beta}^B$.

   \subsection{Theoretical Errors}\label{sec:theo_errors}

   The final ingredient of our analyses is the consideration of theoretical errors in the Fisher matrix formalism. They quantify the uncertainties on the modelling of the matter perturbations and the bias expansion in the statistics of galaxies. In the majority of parameter forecasting analyses they are neglected, while a perfect knowledge of the theoretical model is considered. 
   
   Here we will follow and extend the treatment of \citet{Baldauf2016}, where theoretical errors \textbf{e} are defined as the difference between the true theory and the fiducial theoretical prediction. This formalism considers the true theory to be the model which takes into account at least one more perturbative order than the fiducial one. These errors are bounded by an envelope \textbf{E} and their variation as a function of wavenumbers is characterized by $\Delta  k$ [\eref{eq:cor_coef}]. The value of the correlation length $\Delta k$ is taken to be that of the smallest coherence length of the total power spectrum, that is the scale of the Baryon Acoustic Oscillations (BAO), $\Delta k=\Delta_{BAO}=0.05\;\Mpc$ \citep[see][ for an extensive discussion]{Baldauf2016}. The error covariance matrix is written as:
   
   \begin{equation}\label{eq:err_covar}
    C_{ij}^e=E_i\rho_{ij}E_j
   \end{equation}
   
   \noindent where $i,\;j$ are the indices of the different momentum configurations (\ie number of bins and triangles for the power spectrum and bispectrum respectively). The correlation coefficient $\rho_{ij}$ accounts for the correlations between the momentum configurations, considered to follow a Gaussian distribution, and given by 
   
  \begin{equation}\label{eq:cor_coef}
    \rho_{ij}= 
  \begin{dcases}
    \exp(-(k_i-k_j)^2/2\Delta k^2)	& \text{P},\\
    \prod_{\alpha=1}^3 \exp(-(k_{i,\alpha}-k_{j,\alpha})^2/2\Delta k^2)              & \text{B}.
  \end{dcases}
  \end{equation}   
   For a diagonal error covariance (\ie $\rho_{ii}=1$ and $\rho_{ij}=0$ for $i\neq j$) and a fixed $\Delta k$, the envelope $E(k)$ would be independent of the bin size, contrary to the statistical errors. This means that, for uncorrelated bins, choosing a smaller bin size will reduce the effect of the theoretical errors. The presence of an off-diagonal $\rho_{ij}$ ensures that this does not happen and the relative impact of errors is independent from the size of the $k$ bins. After marginalising over the theoretical errors \textbf{e}, the final covariance that will be used in the Fisher matrix analysis becomes just the sum of the variance of the power spectrum and bispectrum estimators (\esref{eq:deltaP2}{eq:deltaB2} respectively) with the theoretical error covariance [\eref{eq:err_covar}].
    
  One of the goals of this work is to test the effect of theoretical errors on the expected parameter constraints coming from high redshift LSS surveys. The Universe is more linear at high redshifts and hence, for the scales considered in this analysis, we would not expect to see a significant impact on the constraints solely from the theoretical uncertainties attributed to PT. In the formalism proposed by \citet{Baldauf2016}, the envelope is fitted up to two-loops in matter perturbations for both power spectrum and bispectrum while for the bias expansion they consider only the linear bias. Here, we proceed in extending their approach to include the theoretical uncertainties coming from the local-in-matter bias terms (\ie $b_1$, $b_2$, $b_3$, \etc) that appear up to the 1-loop expression of the galaxy power spectrum and bispectrum. This set of terms has been shown to provide an accurate description by comparing with simulations and galaxy catalogues \citep[see \eg][]{Scoccimarro2001b,Feldman2001,Verde2002,Marin2013,GilMarin2014}. Note that the inclusion of all the bias terms at each order (\eg including tidal terms and other operators, see \citet{Desjacques2016} for a review) is potentially important and we will consider their contribution in the theoretical error formalism in the near future (see \sref{sec:radio_error_res} for a discussion).
  
  In order to quantify the theoretical uncertainties, we fit the galaxy power spectrum and bispectrum envelope after including the local-in-matter bias terms up to 1-loop, in addition to the 1-loop matter expressions originating from the description of PT, while assuming Gaussian initial conditions. More precisely, for the galaxy power spectrum we take into account all the terms originating from the bias expansion, that have a dependence on $b_1$, $b_2$ and $b_3$, up to 1-loop (\ie $P_{g,11}$, $P_{g,22}^I$ and $P_{g,13}^I$ of \citet{Sefusatti2009}). For the galaxy bispectrum, the bias terms considered are all those with a dependence on $b_1$, $b_2$, $b_3$ and $b_4$, up to 1-loop, while we exclude those with a dependence on PNG initial conditions \citep[see \eg][ for details]{Sefusatti2009}. For the matter expansion, we consider up to 1-loop terms for both power spectrum and bispectrum in the cases of SPT \citep[see \eg][ for a review]{Bernardeau2002} and \MPT (see \aref{app:MPT} for details). The envelopes for the power spectrum and bispectrum in the the SPT and \MPT schemes are given by

    \begin{equation}\label{eq:pow_theor}
      E_P(k,z)= 
    \begin{dcases}
      D^2(z)P_m^L(k,z)\big[b_1^2(k/0.32)^{1.8}  \\
      \quad+b_2^2(k/0.43)^{1.1}+b_1b_31.13\big]	& \text{SPT},\\
      D^2(z)P_m^L(k,z)\big[b_1^2(k/0.16)^2  \\
      \quad+b_2^2(k/0.43)^{1.1}\big]              & \MPT.
    \end{dcases}
    \end{equation}  
and  
    \begin{align}\label{eq:bis_theor}
   &E_B(k_1,k_2,k_3,z)= \nonumber \\
   &=
  \begin{dcases}
    D^2(z)B_G(k_1,k_2,k_3,z) 
    \big[3b_1^3(\hat{k}/0.32)^{1.8} \\
    \quad+ b_2^3 1.8\hat{k}^{1.25}+b_1b_2b_3 3.2+b_1^2b_4\big]	& \text{SPT},\\
    D^2(z)B_G(k_1,k_2,k_3,z)
    \big[3b_1^3(\hat{k}/0.15)^{1.7}  \\
    \quad+b_2^3 1.8\hat{k}^{1.25}+b_1 b_2b_3 3.2\big]  &  \MPT.
  \end{dcases}
  \end{align}
    
  \noindent where $\hat{k}=(k_1+k_2+k_3)/3$. Note that the fitted coefficients in the above envelopes exhibit a small dependence on the fiducial cosmology, which is negligible for our purposes. The numerical values presented correspond to the cosmological parameters considered in this work. The SPT envelopes are presented here for completeness, since, throughout this work the \MPT description of the matter perturbations will be used.  
  
 A slightly different methodology to quantify the effect of theoretical errors was developed in \citet{Audren2013}. The procedure followed is similar to the one used here. The main difference is that they define theoretical uncertainties by fitting a correction function to the \textsc{HALOFIT}, instead of the desired loop order given by PT. Their technique is applied only to the power spectrum and the relative error to the non-linear $P$ is added to the diagonal part of the covariance matrix. The level of correction is quantified by the precision of \textsc{HALOFIT}, where only the linear bias is taken into account. In order to quantify the difference between the two approaches, we tested this methodology for the power spectrum, since the fitting function is provided only for this case, while a generalisation to the halo bispectrum is not straightforward. To perform the comparison, we adjusted the envelope function of \eref{eq:pow_theor} by removing the higher order bias contributions, as well as by using only the diagonal part of the error covariance [\eref{eq:err_covar}]. In the case of local PNG the difference on the final $\fnl$ constraint turns out to be $\sim 2\%$, while for orthogonal PNG is $\sim 8\%$. Therefore the two approaches produce consistent results, whenever a comparison is possible. We prefer to use the approach of \citet{Baldauf2016} because in this way we take also into account higher-order bias corrections, which are important for the redshift range we consider.

    At low redshifts  using  only the diagonal part of the error covariance underestimates the effect of theoretical uncertainties, but not at  higher redshift, where non-linear corrections are less important. Given that most of the volume and thus the  statistical power is at high redshift we expect the difference between the two methodologies to be at the level of a few percent, in the case where we consider only the linear bias terms for the galaxy power spectrum.

         \begin{figure*}
\centering
   \resizebox{0.8\linewidth}{!}{\includegraphics{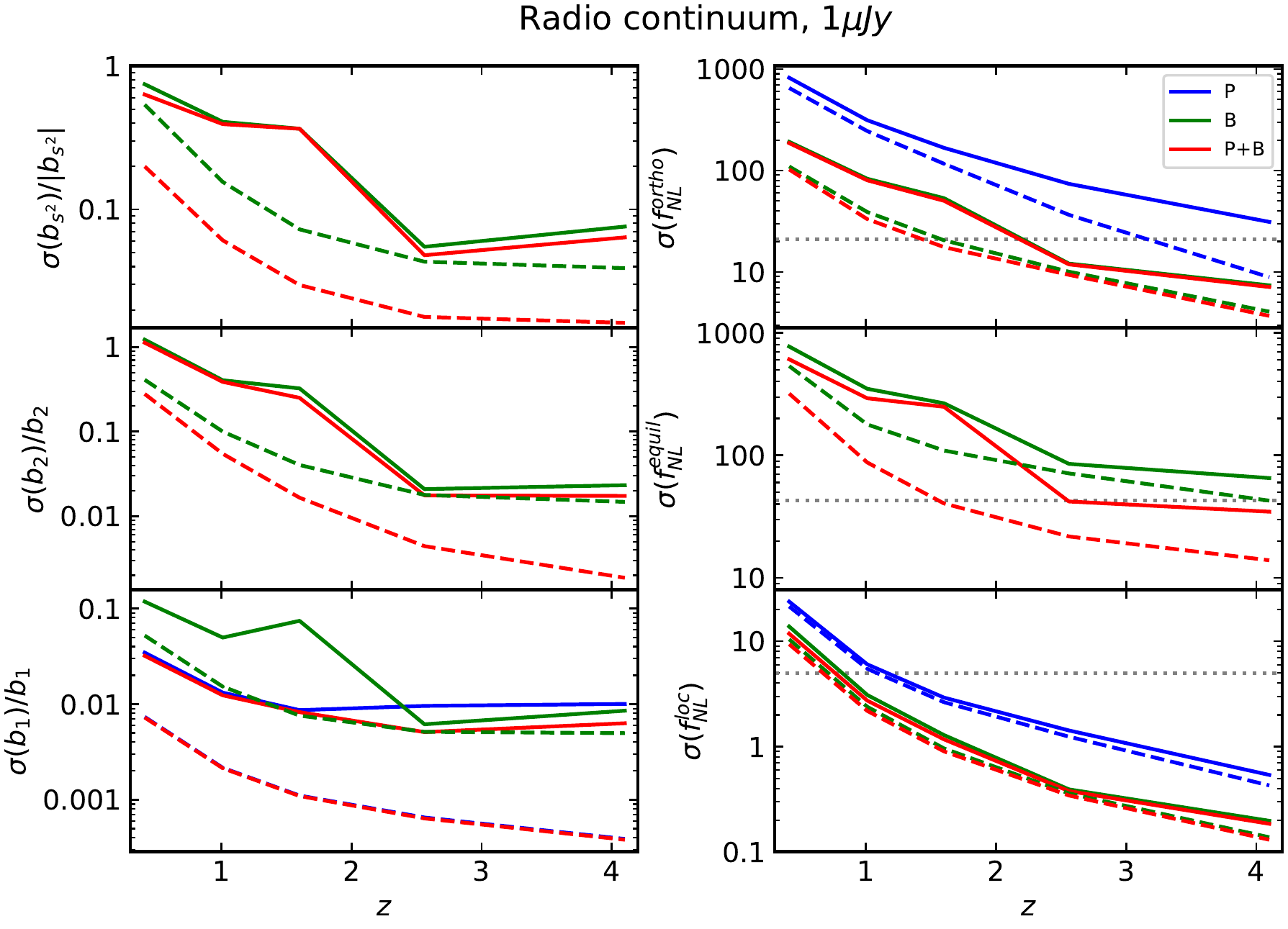}}
    \caption{ The effect of theoretical errors on the bias parameters and on the three $f_{NL}$ parameters per redshift bin (\ie two times the value of $\sigma_z$ in \tref{table:radio_specs}) for the $1\;\mu\text{Jy}$ radio continuum survey. The dashed lines represent our ``idealised model", using a simplified monopole statistic for the bispectrum (see item \ref{it:ideal} in \sref{res_summary}), without theoretical or redshift errors (\ie the starting step of our analysis, as explained in the main text); for the power spectrum we used the linear predictions with PNG [\eref{eq:Pgsph}], while for the bispectrum of galaxies  we used \eref{eq:Bgsph}. The non-linear evolution is treated within the formalism of \MPT. The solid lines represent the same model, but including theoretical errors, as described in \sref{sec:theo_errors}. Forecasts for the power spectrum are plotted in blue, for the bispectrum in green and for both power spectrum and bispectrum in red, without taking into account their cross term in the covariance. The dotted grey line indicates the best constraints on the PNG amplitude, as given by \citet{Planck_PNG2016}.  } \label{fig:comp_theo1}

\end{figure*}

\section{Summary of the analysis method}\label{res_summary}
Here we summarize the modelling used for the galaxy power spectrum and bispectrum, as well as all the steps followed in the next section in order to test the various effects that have an effect on the $\fnl$ forecasts.

\begin{figure*}
\centering
   \resizebox{0.8\linewidth}{!}{\includegraphics{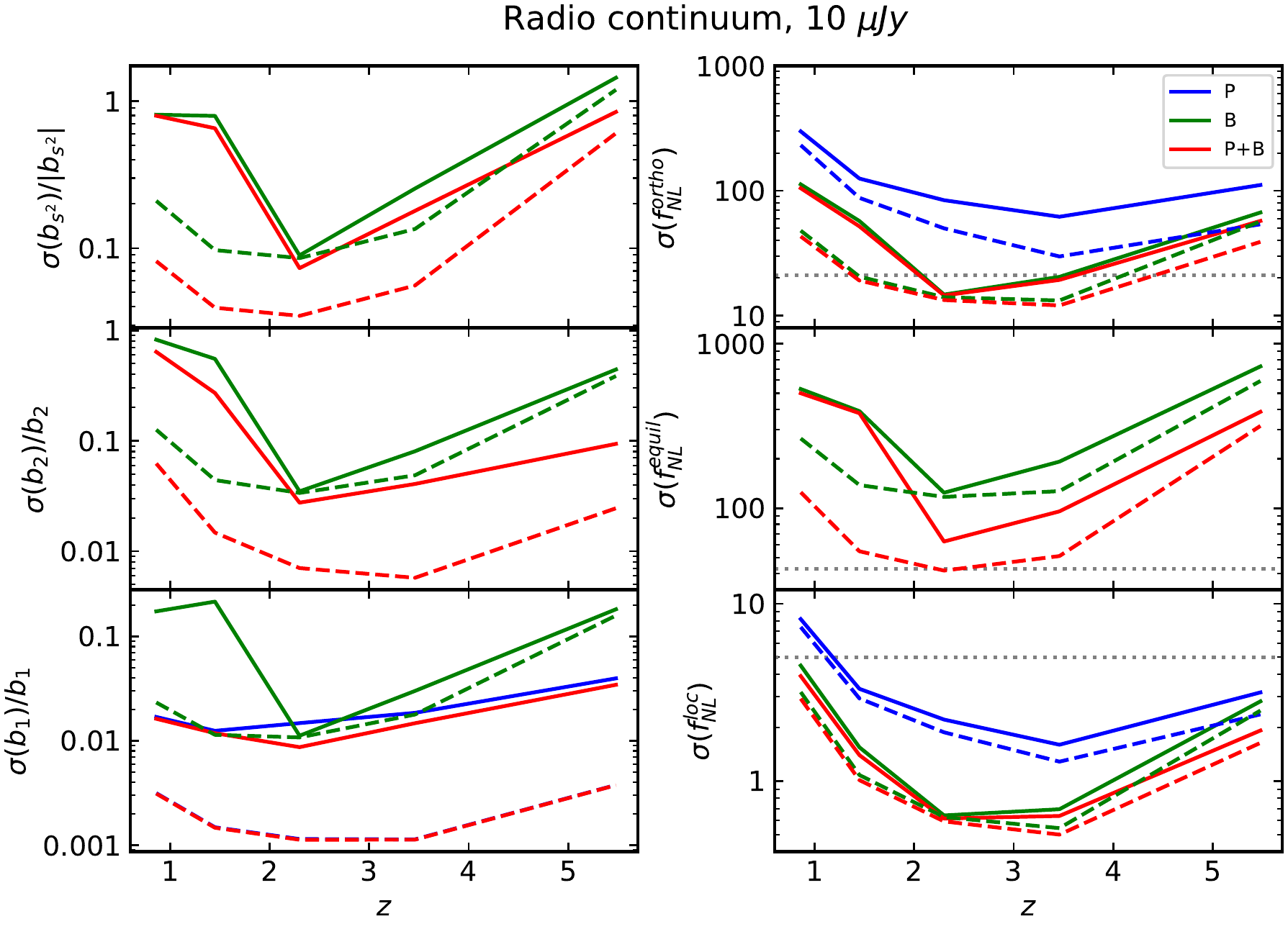}}
    \caption{ Same as in \fref{fig:comp_theo1}, but for the $10\;\mu\text{Jy}$ radio continuum survey. } \label{fig:comp_theo2}

\end{figure*}

\begin{enumerate}[(i)]
\item For the matter power spectrum and bispectrum, the \MPT tree-level description is used throughout. In order to cross-validate our results, we have also considered the scenario of using SPT. The forecast results are in agreement between SPT and \MPT (see \sref{sec:radio_error_res}).
\item Non-Gaussian initial conditions are assumed and a bias expansion up to second order is used [\eref{eq:dg}].
\item The forecasts are performed by following the Fisher matrix formalism (see \sref{sec:fisher}), where a diagonal covariance is used for both power spectrum and bispectrum as described in \esref{eq:deltaP2}{eq:deltaB2} respectively. The analysis is restricted  to scales $k \le k_\max=0.1/D(z)\;h\text{/Mpc}$.
\item \label{it:ideal} We start by considering an idealised scenario, in which we adopt the approximate monopole statistics defined by \esref{eq:Pgsph}{eq:Bgsph}, for the galaxy power spectrum and bispectrum respectively. Here, the use of the term "approximate" stems from the fact that, in Eq. (\ref{eq:Bgsph}), we neglect a number of terms appearing in the full bispectrum monopole expression, as shown \eg in \citet{Tellarini2016}. As explicitly checked, the terms we neglect have a small impact on the large scales included in our analysis, making our bispectrum formulae accurate enough for this preliminary step. See \citet{Sefusatti2006} for a further discussion on this point.The idealisation level is of course much greater for the photometric/radio survey case as in reality  the redshift errors are very large.
Note that, for this warm up exercise, we are also assuming perfect determination of redshifts, neglecting theoretical errors and also excluding the trispectrum term shown in \eref{eq:Bg_full}. 
Despite its lack of realism, it is quite useful to start with this simplified case, for several reasons.
 Firstly, a non-trivial amount of previous literature includes only real space forecasts, or angle averaged statistics in redshift space, without theoretical errors. Therefore, explicitly including this initial step facilitates comparisons.
Secondly, by proceeding in this way we are able to better isolate the relative impact of different effects, such as theoretical and redshift errors, when we add them in subsequent steps. Thirdly, as explained  in more detail in the following, off-diagonal terms in the theoretical error covariance are often neglected in our full redshift space analysis, due to computational limitations. On the other hand, off-diagonal terms can be in most cases fully included at this initial stage. This provides important guidelines to assess the accuracy of the diagonal covariance approximation, used later when needed (we anticipate here that we find this diagonal approximation to be quite good). Finally, by comparison of our preliminary monopole forecasts with the final, full redshift space analysis, we are able to assess the amount of PNG information lost by averaging over angles (\ie the monopole). We will refer to this step of the analysis as to our ``idealised'' case, in the following.
\item The second step is to add the theoretical error covariance given in \eref{eq:err_covar} to the variance of Fisher matrix, in order to account for the uncertainties in the theoretical model (see \sref{sec:theo_errors} for a discussion). Here we account only for the exclusion of 1-loop corrections in the matter perturbations for both power spectrum and bispectrum. For the bias expansion we only quantify the error in excluding the 1-loop contributions that are related to the local-in-matter bias coefficients, \ie $b_1$, $b_2$, $b_3$, \etc 
\item The effect of the theoretical errors on forecasts is shown as a function of redshift in \fref{fig:comp_theo1} and \ref{fig:comp_theo2} for the two radio continuum surveys considered here. The effect on the forecasts coming from the summed signal over all redshift bins is shown in Tables \ref{table:com_ben_werror} and \ref{table:opt_res_mono_theoerr} for the radio and optical surveys respectively.
\item The third step is to move to redshift space and include the full RSD treatment up to second order. The galaxy power spectrum and bispectrum model in redshift space  are given by \esref{eq:Pgs}{eq:Bgs} respectively. Note that the trispectrum term in \eref{eq:Bgs} is still excluded for now. Only the diagonal part of the theoretical error covariance is used in the redshift space models. As also mentioned just above, we argue in \sref{sec:radio_error_res} that the effect of the off-diagonal part on the final $\fnl$ forecasts is small. Note, finally, that we are still not including redshift errors. Therefore our forecast up to this point are still unrealistic for optical photometric and for radio surveys, for which these errors are important. The goal, up to here, remains that of assessing the impact of each ingredient added at each new step of the analysis. Moreover, forecasts with no redshift errors included provide upper limits to the performance of realistic radio continuum surveys, eventually approachable by improving current techniques for redshift determination.
\item In addition to the RSD effect, we consider redshift uncertainties, which are modelled like \esref{eq:DfogP}{eq:DfogB} for the power spectrum and bispectrum respectively (see \sref{sec:PB_gal_red} for a discussion). The effect of RSD, theoretical errors and redshift uncertainties to the $\fnl$ forecasts is shown in Tables \ref{table:RSD_res1}, \ref{table:RSD_res2}  and \ref{table:opt_res_rsd} for the radio and optical surveys respectively.
\item Finally, we take into account the trispectrum term in the galaxy bispectrum for both the idealised and the full RSD models (see \esref{eq:Bg_full}{eq:Bgs}). The effect of this trispectrum correction to the final PNG forecasts is shown, for the idealised case, in Tables \ref{table:tris_res_mono_radio} and \ref{table:opt_res_mono} for the radio and optical surveys respectively. For the final ``full'' model (\ie RSD+theoretical errors+redshift errors+trispectrum), the $\fnl$ forecast are shown in Tables \ref{table:tris_res_rsd_radio} and \ref{table:opt_res_rsd} for the radio continuum and optical surveys respectively.
\item We summarize our final forecast results on the amplitude of PNG coming from future LSS surveys in \tref{table:concl}.
\end{enumerate}

\section{Results}\label{results}
    
    In this section we present the results of our forecasts for radio continuum and optical surveys, obtained with the procedure summarized in \sref{res_summary}. For the local and orthogonal PNG, we consider the power spectrum and bispectrum model that was described in detail in Secs. \ref{sec:PB_gal_real} and \ref{sec:PB_gal_red}, for real and redshift spaces respectively [by Eqs. (\ref{eq:Pg_full}), (\ref{eq:Bg_full}), (\ref{eq:Pgs}) and (\ref{eq:Bgs})].  In the case of the equilateral type of non-Gaussianity, we use the same expressions, but without the corrections from the primordial local gravitational potential $\Psi$ due to degeneracies with the bias parameters (see \sref{sec:PNGbias}). The final results  are derived after marginalising over the nuisance stochastic bias parameters (see \sref{sec:fisher}). The linear matter power spectrum is computed with CAMB \citep{CAMB}, while the cosmological parameters are those determined by \citet{Planck2016_cosmopar}: $h$ = $0.6774$, $\Omega_{c}h^2$  = $0.1188$, $\Omega_{b}h^2$ = $0.0223$, $n_{s}$ = $0.9667$, $\Delta_{\zeta}^2$ = $\scien{2.142}{-9}$, $\tau$ = $0.066$. 

    \subsection{Radio surveys}\label{sec:results_radio}
      
   In this section we present the results of our forecasts for the parameter vector [\eref{eq:params}], considering the two cases described earlier, \ie the $10 \mu Jy$ and the $1 \mu Jy$ flux limit high redshift surveys. The modelling used is described in the previous paragraphs. Complications in the modelling are added in several steps, while the intermediate results are also presented. The reader interested in the results for the ``full'' case can skip to \sref{sec:radio_trisp_res}.

\subsubsection{Theoretical errors effect}\label{sec:radio_error_res}

\begin{figure*}
\centering
\resizebox{0.7\linewidth}{!}{\includegraphics{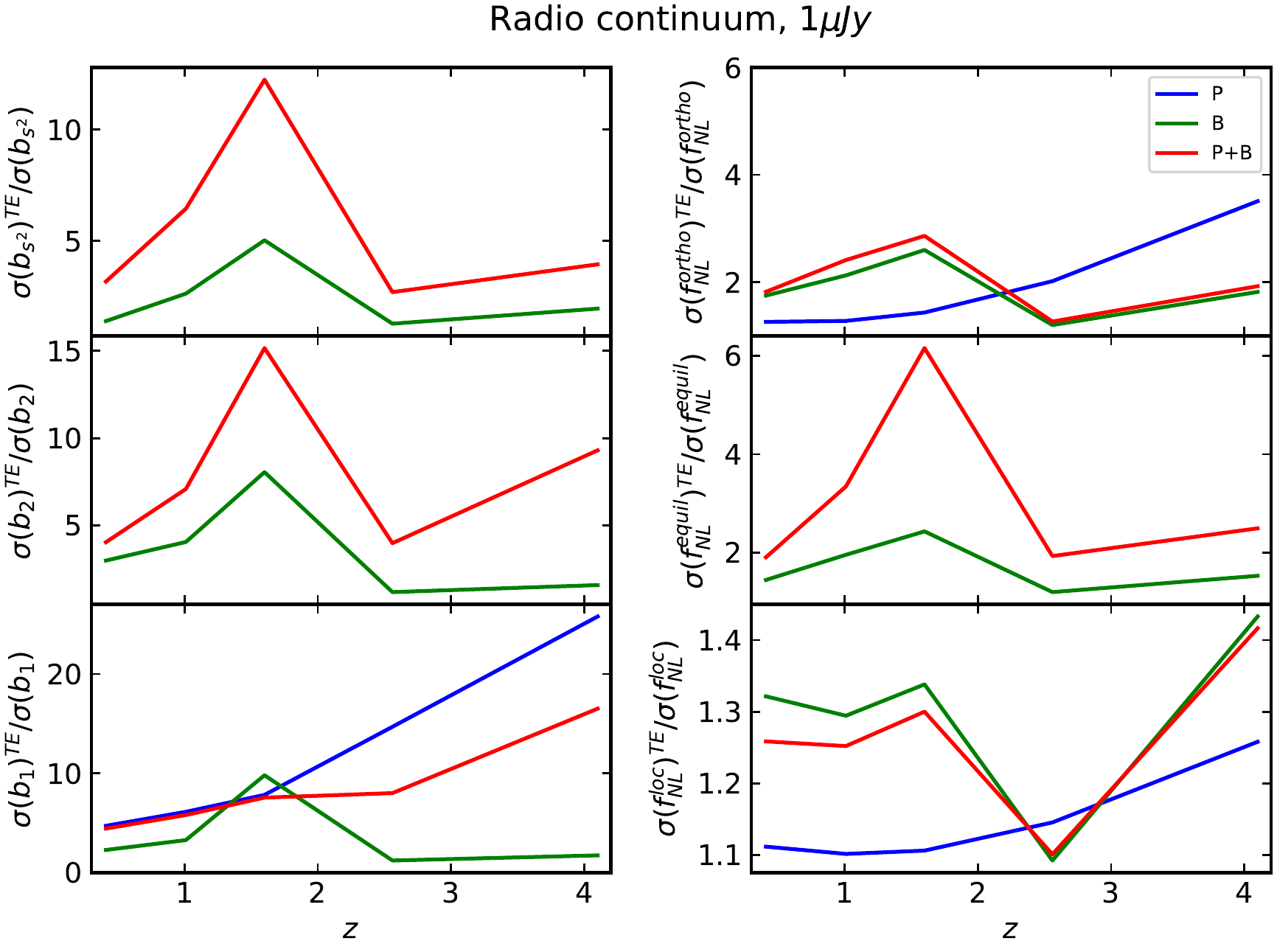}}
    \caption{Same as \fref{fig:comp_theo1}, where here the ratio of the expected 1$\sigma$ errors between the case with theoretical errors (denoted in the plots with the upper index TE) and without is shown. The spikes observed in the expected forecasts, obtained when including the effect of theoretical errors can be attributed to the trade-off between the contributions coming from higher-order terms in the matter and bias expansions (see main text in \sref{sec:radio_error_res} for a discussion). } \label{fig:comp_theo_ratios1}

\end{figure*}

\begin{figure*}
\centering
\resizebox{0.7\linewidth}{!}{\includegraphics{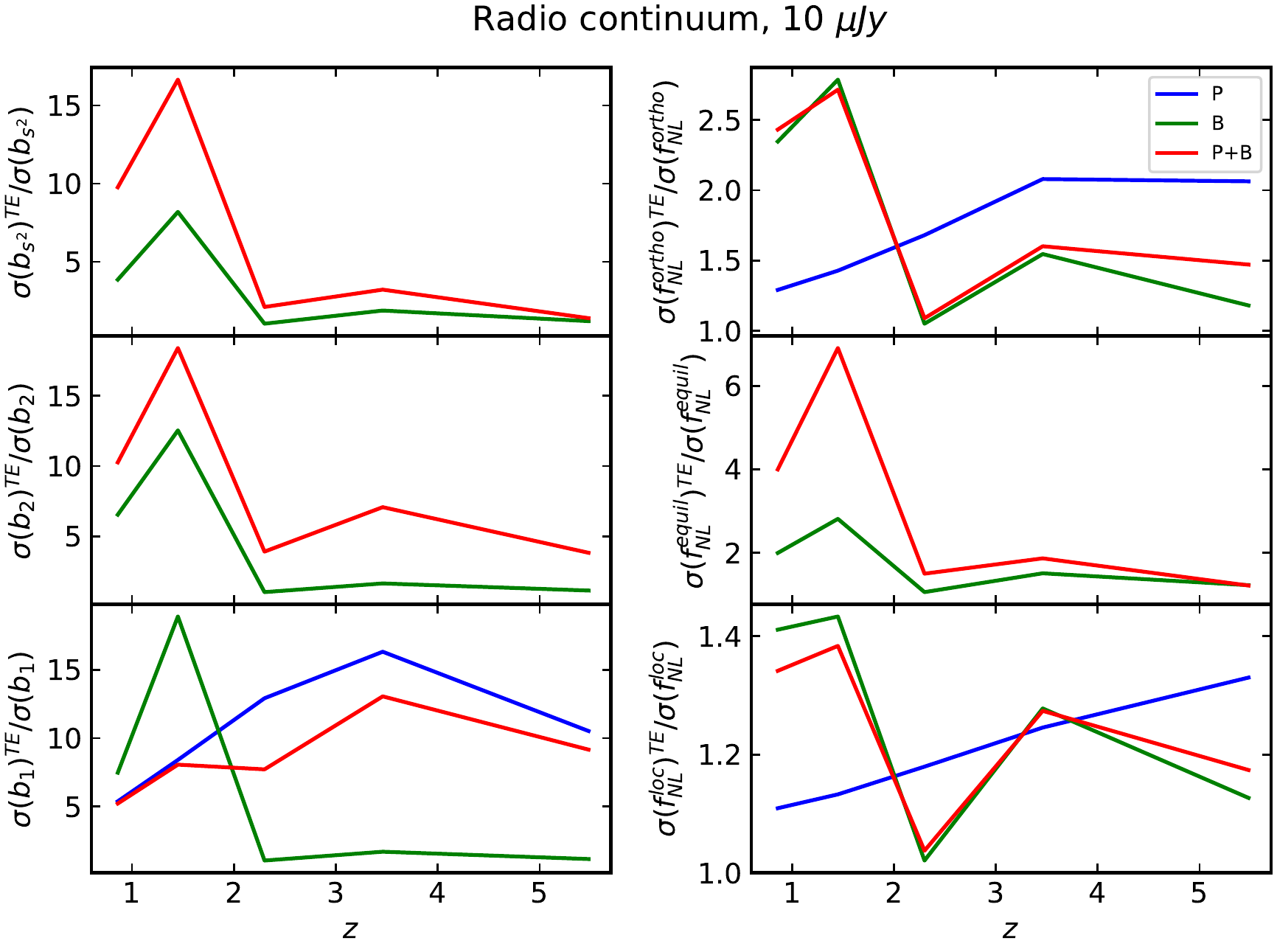}}
    \caption{Same as \fref{fig:comp_theo2}, where here the ratio of the 1$\sigma$ errors between the case with theoretical errors (denoted in the plots with the upper index TE) and without is shown. } \label{fig:comp_theo_ratios2}
\end{figure*}

Our main goal at this stage will be quantifying the effect of theoretical errors, by testing their impact on our initial, idealised forecasts. Since we are not yet including redshift errors and other important effects in the analysis, we immediately warn the reader that our $\fnl$ forecasts are still unrealistic in this section. The purpose here is only the {\em relative} comparison between the two cases, with and without theoretical errors, in order to assess the impact of the latter.  Due to the demanding computational effort needed to perform the inversion of the bispectrum covariance matrix (a $10^6\times10^6$ size matrix for the final redshift bins), which is no longer diagonal when the theoretical errors are included [\eref{eq:err_covar}], we use the full covariance matrix only for the three lowest redshift bins (that's the largest computationally affordable amount; we choose the lowest three redshifts because the effect of off-diagonal terms is largest there), while for the rest we consider only the diagonal contribution.  This means that cross-correlations between modes are excluded. This can in principle affect the impact of the theoretical errors. However, we  performed tests to check the effects of these off-diagonal components, up to the redshift bin allowed by the computational resources available, and we observed that the effect of the off-diagonal terms becomes actually negligible at high-\textit{z}. This is reasonable: the Universe is more linear at higher redshifts, therefore the loop corrections, up to the scales we consider, are expected to be suppressed. Our approximations work therefore very well. Let us note that, in the case of the power spectrum, we always use the full covariance matrix, since no computational issues arise in this case. Let us also note here again that the theoretical modelling was performed using \MPT, which has an embedded cut-off function at high-$k$. This means that higher order contributions, as well as the theoretical error effect will be suppressed on small scales. In order to check whether this has a significant effect, we have performed a similar analysis using SPT. The results were consistent with those presented here, throughout the range of scales chosen for our analysis. The comparison is shown in \fref{fig:comp_theo1} and \ref{fig:comp_theo2} and the forecasts for $\fnl$ are quantitatively reported in \tref{table:com_ben_werror}. In addition, the ratio of the marginal 1$\sigma$ error is presented for the two radio continuum surveys in \fref{fig:comp_theo_ratios1} and \ref{fig:comp_theo_ratios2}, for the cases where theoretical errors are taken into account and when they are omitted.

Note that the effect of theoretical errors doesn't get smaller at high redshifts in all cases, as seen in \fref{fig:comp_theo_ratios1} and \ref{fig:comp_theo_ratios2}, and in particular this is evident for the power spectrum case. The argument of a more linear Universe at high redshifts, which leads to a reduction in the contribution of the higher loop corrections, holds for the case of matter and it is not generally true for the biased tracers. The reason is that at large redshifts, and in the scale range considered here, galaxies become more biased, while the matter loop corrections become less important. The final effect of theoretical errors is determined by adding the monotonic decreasing contribution of matter higher-order terms and the increasing contribution coming from higher-order coefficients in the bias expansion (this also explains the spikes observed in \fref{fig:comp_theo_ratios1} and \ref{fig:comp_theo_ratios2}). Therefore, depending on the redshift evolution of the bias parameters, the depth of the survey and its volume, a different behaviour of the impact of the theoretical errors can be observed, when bias loop corrections are taken into account in the envelope fitting (see \sref{sec:theo_errors} for details). This shows the importance of extending the formalism of \citet{Baldauf2016}, to include theoretical errors attributed to the bias expansion, as we did.

                \begin{table*}
\centering
\resizebox{0.8\linewidth}{!}{
\begin{tabular}{ccc|cc}
\hline
                                & \multicolumn{2}{c|}{Radio continuum, 1 $\mu$Jy} & \multicolumn{2}{c}{Radio continuum, 10 $\mu$Jy} \\ \hline
                                & \thead{Idealised} & \thead{Idealised}\thead{+}\thead{Theoretical \\ Errors} & \thead{Idealised} & \thead{Idealised}\thead{+}\thead{Theoretical \\ Errors}  \\ \hline
\multicolumn{1}{c|}{P(loc)}     & 0.40 & 0.50 & 0.91 & 1.12               \\
\multicolumn{1}{c|}{B(loc)}     & 0.13 & 0.18 & 0.38 & 0.44               \\
\multicolumn{1}{c|}{P+B(loc)}   & 0.12 & 0.17 & 0.35 & 0.41               \\ \hline
\multicolumn{1}{c|}{P(equil)}   & -         & -                  & -         & -                  \\
\multicolumn{1}{c|}{B(equil)}   & 34 & 50 & 70 & 98              \\
\multicolumn{1}{c|}{P+B(equil)} & 11 & 26 & 27 & 51              \\ \hline
\multicolumn{1}{c|}{P(ortho)}   & 9.6 & 28 & 22 & 42              \\
\multicolumn{1}{c|}{B(ortho)}   & 4.3 & 6.2 & 8.5 & 11              \\
\multicolumn{1}{c|}{P+B(ortho)} & 3.8 & 6.0 & 7.8 & 11               \\ \hline
\end{tabular}
}
\caption{Forecasts for the non-Gaussian amplitude $\fnl$ for the three shapes (local, equilateral and orthogonal) over all redshift bins, from the two surveys considered here (the 1 $\mu$Jy -- left and the 10 $\mu$Jy -- right, radio continuum surveys), in the idealised analysis, no theoretical errors case and the case including theoretical errors. The expected constraints were derived from the power spectrum (P), bispectrum (B) and by combining the two (P + B).}
\label{table:com_ben_werror}
\end{table*}

It would seem at this stage that the combined information from the power spectrum and bispectrum of galaxies can provide very tight forecasted constraints on the amplitude of local PNG. For the 1 $\mu$Jy radio continuum survey, an error of $\sigma(\fnll)<0.13$ can be achieved after using the total signal from all redshift bins. In the orthogonal case, this survey can provide $\sigma(\fnlo)\sim3$, while for equilateral PNG, already at this stage the expected constraints are weaker, compared to the other shapes, as we can see from \tref{table:com_ben_werror}. Let us however stress again that these results will deteriorate considerably when considering realistic redshift space measurements and especially when accounting for errors in the determination of the redshift of radio sources (see next subsection). As specified since the beginning of this section, at this stage we do not focus yet on the absolute value of the expected constraint, but on the assessment of the effect of theoretical errors.

   Forecasts on bias parameters are shown in the left panels of \fref{fig:comp_theo1} and \ref{fig:comp_theo2} for each of the surveys considered here. Again, rather than quantitative assessments, which will be refined in the redshift space section later on, it is useful at this level to focus on qualitative behaviours. We see, as expected, that the power spectrum provides the main signal for constraining $b_1$ (blue lines in \fref{fig:comp_theo1} and \ref{fig:comp_theo2}), while the contribution from the bispectrum (green lines) is minimal as we can also see in the results coming from the combination of the two (red lines). For the quadratic bias parameter, as well as the tidal bias, the expected constraints are weaker.  For both these bias coefficients, the constraining signal originates solely from the bispectrum, since these terms appear in loop corrections of the linear galaxy power spectrum, which we do not consider here, and therefore this reduction in the constraining power is justified. In the combined power spectrum and bispectrum case, an improvement is observed in the statistical error of $b_2$ and $b_{s^2}$, due to the tight constraint on the linear bias provided by the power spectrum. In addition, the presence of the tidal bias term breaks the degeneracy between the linear and quadratic terms, improving the predicted errors on $b_2$. For both bias parameters, radio continuum surveys (mainly the high redshift bins) contribute enough signal in order to achieve a few percent precision measurements. However the introduction of theoretical errors deteriorates the expected constraints to a $10-20\%$ precision at high redshift, mainly due to the uncertainty introduced by the exclusion of higher bias terms in the power spectrum and bispectrum of galaxies.

\subsubsection{Redshift Space Distortions and redshift uncertainties.}\label{sec:radio_rsd_res}
           
\begin{figure*}
\centering
\resizebox{0.75\linewidth}{!}{\includegraphics{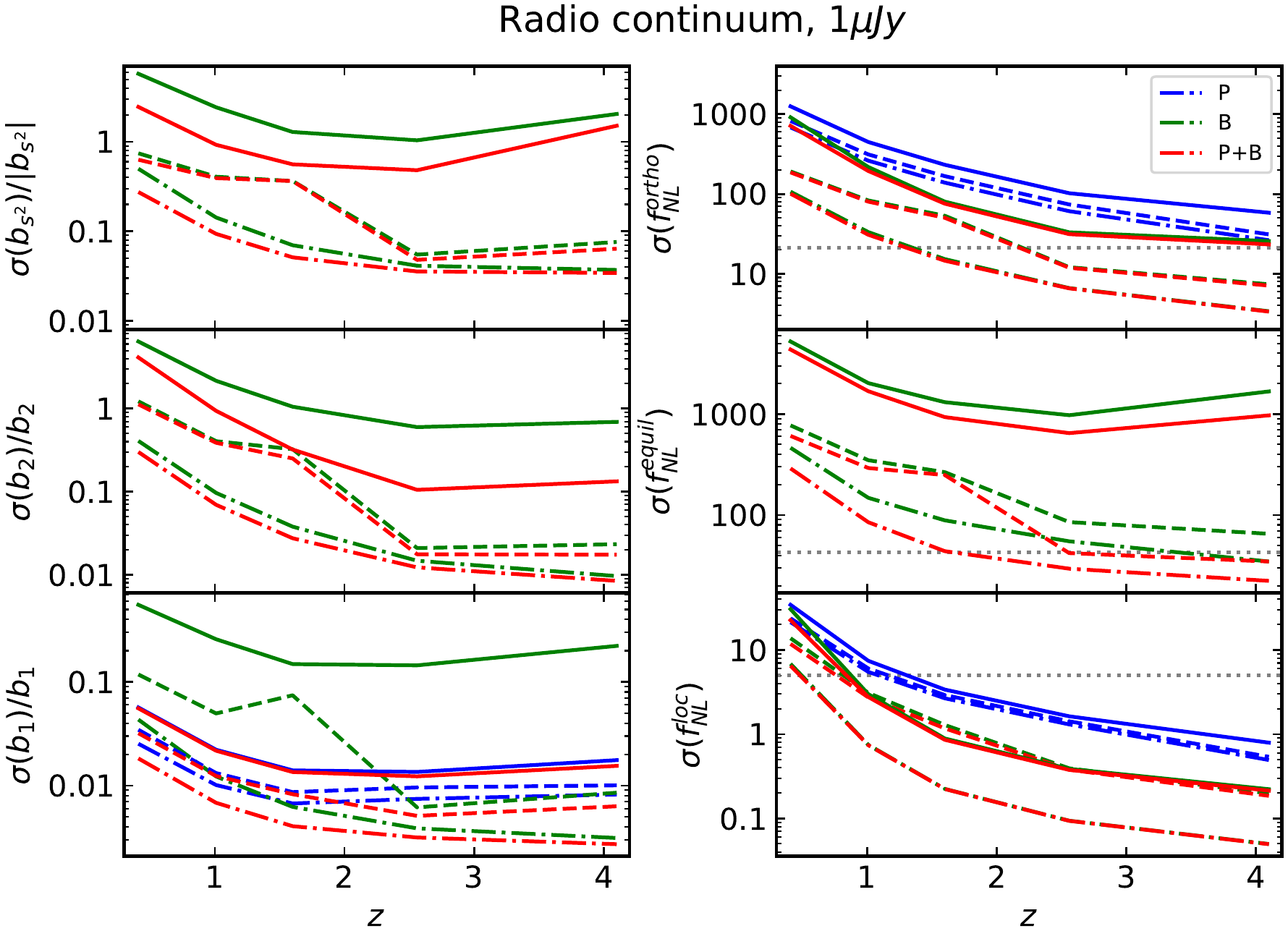}}
    \caption{ Expected constraints on the linear and on the two quadratic bias parameters from the tree-level bias expansion and on the amplitudes of the PNG parameter $\fnl$ for each redshift bin in the case of the 1 $\mu$Jy radio continuum survey when including RSD and FOG effects. The dashed lines represent the idealised case, with theoretical errors included, as described in \sref{sec:radio_error_res}. The dashed-dotted lines represent the model which accounts for RSD and FOG effects, assuming perfect knowledge of redshifts, while the solid lines take also into account the redshift uncertainties. The expected constraints from the power spectrum are plotted in blue, from the bispectrum in green and from both power spectrum and bispectrum in red, without taking into account their cross term in the covariance. The expressions used for the modelling of the power spectrum and bispectrum are given in \esref{eq:Pgs}{eq:Bgs} respectively, while for the latter the trispectrum term is excluded. These results consider only the diagonal part of theoretical error covariance [\eref{eq:err_covar}] for the bispectrum, while for the power spectrum the full error covariance is used. \MPT has been used to derive the matter two- and three-point correlators.}\label{fig:RSD_comp1}
\end{figure*}

\begin{figure*}
\centering
\resizebox{0.75\linewidth}{!}{\includegraphics{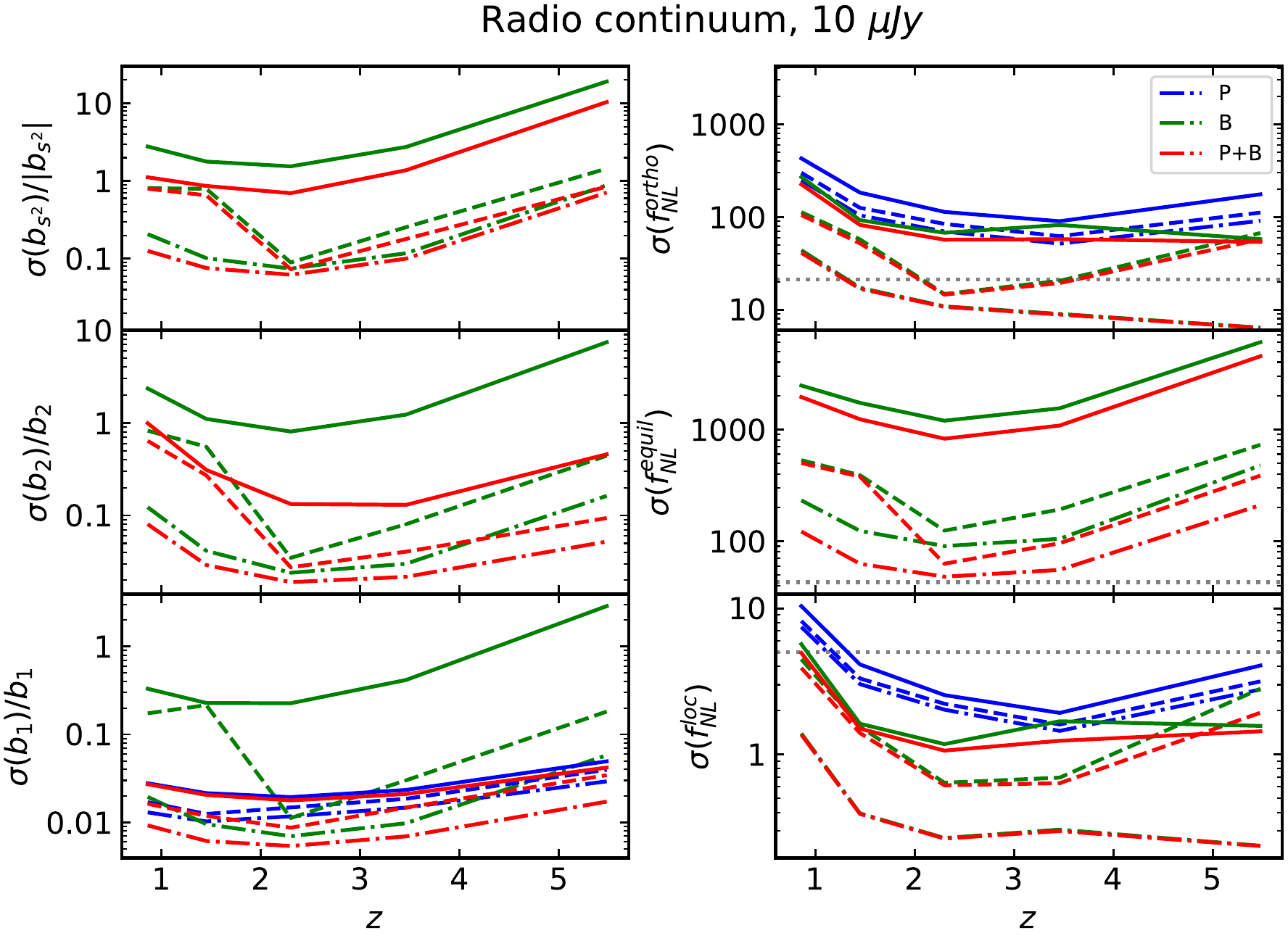}}
    \caption{ Same as \fref{fig:RSD_comp1}, but for the 10 $\mu$Jy flux limited radio continuum survey. }\label{fig:RSD_comp2}
\end{figure*}

   Both the effect of RSD and of uncertainties in the determination of redshifts must be fully taken into account in any realistic galaxy forecast, since galaxies are observed in this coordinate system. In the second step of our analysis, we start by comparing the idealised model (see \sref{sec:PB_gal_red}) with the results derived from including the full RSD, up to second order (\ie for the galaxy power spectrum and bispectrum we use \esref{eq:Pgs}{eq:Bgs} respectively). At this stage, we are still {\em neglecting} redshift errors. We also consider separately the effect of adding RSD and theoretical errors. The goal for this section is to specifically check the impact of including RSD and of going beyond angle averaged statistics, via a relative comparison with previous results. Actual, realistic forecasts, including redshift errors, will finally be presented in the next subsection. The final forecasts use the signal coming from the power spectrum, bispectrum and their combination. Results for bias and PNG parameters, as a function of redshift, are displayed in \fref{fig:RSD_comp1} and \ref{fig:RSD_comp2}. The expected constraints on $\fnl$, coming from the summed signal over all redshift bins, are reported in \tref{table:RSD_res1} and \ref{table:RSD_res2} (column marked by ``RSD (no z-errors)").

   The difference between the model including RSD only and the preliminary, idealised analysis is evident for the bias forecasts, especially the bispectrum derived ones. As we can see in \fref{fig:RSD_comp1} and \ref{fig:RSD_comp2}, the effect of RSD alone is instead overall small for PNG, with the bispectrum again being affected the most. 
   
   As we can see in the two first columns of Tables \ref{table:RSD_res1} and \ref{table:RSD_res2}, the forecasts on $\fnl$ become tighter for all three PNG cases when the full RSD treatment is applied (column denoted as ``RSD (no z-error)'') with respect to the results of the idealised case (column denoted as ``Idealised''). This improvement is expected, since the full RSD model contains additional signal compared to the approximating monopole [\esref{eq:Pgsph}{eq:Bgsph}], used in the idealised case. This indicates that the supplementary information contained in the higher bispectrum multipoles is important and it should be taken into account by future LSS surveys in order to provide tight constraints on the PNG amplitude.

As a further step, in addition to RSD, we include theoretical errors in order to test the combined effect on the $\fnl$ expected constraints for the local, equilateral and orthogonal shapes. A comparison of the $1\sigma$ error bars on the non-Gaussian amplitude obtained respectively considering the idealised models, the RSD model and the RSD $+$ theoretical errors model, without accounting for redshift errors, is displayed in \tref{table:RSD_res1} and \ref{table:RSD_res2}. Here, the off-diagonal part of the theoretical error covariance matrix is ignored in the case of the bispectrum, due to the significant computational cost of inverting this matrix; off-diagonal elements should now be included for each orientation of the triangles [see \eref{eq:fisherBs}]. Excluding the off-diagonal terms underestimates the mode coupling and the overall effect of theoretical errors. In order to quantify this effect, we compare forecasts with and without off-diagonal terms, for the computationally affordable idealised case. Since RSD corrections are not considered in the theoretical error treatment, the effect of the off-diagonal terms in the expected constraints can be transferred to the present forecasts. The marginal error provided by the bispectrum for the PNG amplitude, in the local case, increases in this test by $\sim1\%$, if we include the full theoretical error covariance matrix for the initial redshift bins (see \sref{sec:radio_error_res}), instead of only the diagonal part. For the equilateral and orthogonal cases, we observe $\sim10\%$ and $\sim5\%$ enhancements respectively, after marginalising over the other free parameters, where the degradation of the bias parameters, due to the inclusion of theoretical errors, is propagated on $\fnl$. Note that these quantitative results are valid only for the radio surveys we consider here, since the effect of theoretical errors depends on redshift as well as the fiducial values of the bias parameters. With these caveats in mind, we see that the effect of theoretical errors follows a similar pattern to the one already discussed in the previous section, as expected. In the case of local non-Gaussianity, the impact of theoretical errors is small, while it is very large for the equilateral shape.

  Having considered RSD, we now include redshift uncertainties, modelled like the FOG in \esref{eq:DfogP}{eq:DfogB}. Note that the power spectrum and bispectrum, including the stochastic bias terms, which at large scales resembles the Poisson shot noise, is multiplied by the damping factor, containing the redshift uncertainties, as shown in \esref{eq:Pgs}{eq:Bgs}] respectively.
  
       \begin{table*}
\centering
\resizebox{0.8\textwidth}{!}{
\begin{tabular}{ccccc}
\hline
                                & \multicolumn{4}{c}{Radio continuum, 1 $\mu$Jy}  \\ \hline
                                & \thead{Idealised}  &  RSD (no z-errors)    & \thead{RSD (no z-errors) \\ +Theoretical errors} & \thead{RSD + z-errors \\ +Theoretical errors} \\ \hline 
\multicolumn{1}{c|}{P(loc)}     & 0.40 & 0.38 & 0.47 & 0.70   \\
\multicolumn{1}{c|}{B(loc)}     & 0.13 & 0.039 & 0.053 & 0.19 \\
\multicolumn{1}{c|}{P+B(loc)}   & 0.12 & 0.039 & 0.052 & 0.18 \\ \hline
\multicolumn{1}{c|}{P(equil)}   & -      &    -   &    -     & -     \\
\multicolumn{1}{c|}{B(equil)}   & 34 & 25 & 36 & 664  \\
\multicolumn{1}{c|}{P+B(equil)} & 11 & 8.3 & 20 & 448     \\ \hline
\multicolumn{1}{c|}{P(ortho)}   & 10 & 7.8 & 23 & 49    \\
\multicolumn{1}{c|}{B(ortho)}   & 4.2 & 2.2 & 3.3 & 20  \\
\multicolumn{1}{c|}{P+B(ortho)} & 3.8 & 2.0 & 3.2 & 18    \\ \hline
\end{tabular}}
\caption{Forecast results for the amplitude of PNG ($\fnl$) in the case of three different primordial shapes (local, equilateral and orthogonal), when considering respectively the idealised model, the model taking into account RSD and FOG effects and the redshift space model with theoretical errors, for the 1 $\mu$Jy radio continuum survey. The redshift uncertainties are considered separately and added on top of all the previous effects under the column denoted ``RSD+z-errors+Theoretical errors''. The expected constraints on the bias parameters, up to quadratic order [\eref{eq:dg}], are also presented. The results are derived after marginalising over the unknown parameters. Note that the observed improvement in the P$+$B(equil) results, with respect to B(equil), originate from the tighter forecasts on $b_1$ provided by P(equil). 
}
\label{table:RSD_res1}
\end{table*}

The forecasted constraints on PNG amplitudes and bias, after including the redshift errors, are presented as functions of redshift in \fref{fig:RSD_comp1} and \ref{fig:RSD_comp2}. The final $1\sigma$ error bars on $\fnl$ are displayed in \tref{table:RSD_res1} and \ref{table:RSD_res2} under the column named ``RSD+z-errors+Theoretical errors''. In the case of local PNG, the effect of redshift uncertainties is small and the expected constraints from both radio surveys are still tighter than those originating from \Planck, for more or less the whole redshift range. The final error on $\fnll$ is degraded by a factor of $\sim 1.5$ for a 1 $\mu$Jy survey with respect to the case with a perfect redshift determination. This shows that future radio surveys can tightly constrain PNG of the local type, even with large redshift uncertainties, arising from the statistical nature of redshift determination. For the 1 $\mu$Jy case, an error of $\sigma(\fnll)\sim 0.18$ can be achieved from the combined P $+$ B signal, \ie an expected $\sim30$ times improvement with respect to the constraints of \Planck. In the case of orthogonal PNG the degradation due to redshift uncertainties is much larger, as seen in \fref{fig:RSD_comp1} and \ref{fig:RSD_comp2}. The final power spectrum $+$ bispectrum forecasts for the orthogonal PNG amplitude, adding contributions from all redshift bins, show a $\sim 7$ times deterioration, compared to the case with no $z$-errors, and is slightly worse than the current \Planck constraint. 

     \begin{table*}
\centering
\resizebox{0.8\textwidth}{!}{
\begin{tabular}{ccccc}
\hline
                                & \multicolumn{4}{c}{Radio continuum, 10 $\mu$Jy} \\ \hline
                                & \thead{Idealised}  &  RSD (no z-errors)    & \thead{RSD (no z-errors) \\ +Theoretical errors} & \thead{RSD + z-errors \\ +Theoretical errors} \\ \hline 
\multicolumn{1}{c|}{P(loc)}     & 0.91 & 0.87 & 1.06 & 1.34           \\
\multicolumn{1}{c|}{B(loc)}     & 0.4 & 0.13 & 0.15 & 0.73         \\
\multicolumn{1}{c|}{P+B(loc)}   & 0.3 & 0.13 & 0.15 & 0.63          \\ \hline
\multicolumn{1}{c|}{P(equil)}   & -          & -           & -             & -              \\
\multicolumn{1}{c|}{B(equil)}   & 70 & 55 & 77 & 780           \\
\multicolumn{1}{c|}{P+B(equil)} & 27 & 20 & 38 & 551           \\ \hline
\multicolumn{1}{c|}{P(ortho)}   & 22 & 21 & 39 & 61         \\
\multicolumn{1}{c|}{B(ortho)}   & 8.5  & 3.9  & 5.3 & 35           \\
\multicolumn{1}{c|}{P+B(ortho)} & 7.8  & 3.8  & 5.4 & 30           \\ \hline
\end{tabular}
}
\caption{Same as \tref{table:RSD_res1}, but for the 10 $\mu$Jy flux limited radio continuum survey.}
\label{table:RSD_res2}
\end{table*}

The degradation is even larger for PNG of the equilateral type. In this case, expected constraints provided by both radio surveys are far weaker than those coming from \Planck, for the whole redshift range (dashed-dotted lines in \fref{fig:RSD_comp1} and \ref{fig:RSD_comp2}). The predictions on the $\fnle$ parameter coming from all redshifts and the combined P and B signal get degraded by a factor $\sim 29$, as seen in \tref{table:RSD_res1} and \ref{table:RSD_res2}. The shape-dependence of this error degradation is due to the functional form of the damping factor, which takes a minimum value - and hence has the maximum effect  -  for the equilateral configurations, giving $D_\text{FOG}=\exp[-3(k\mu\sigma_\mathrm{v})^2]$. The impact of the damping factor is instead minimized for squeezed triangles, on intermediate scales, and for folded ones, on large scales. 

Therefore we can conclude that radio continuum samples, in combination with clustering-based redshift estimation, can provide tight constraints for the local PNG amplitude, with an important contribution from bispectrum measurements. In order to achieve tight constraints for other shapes, though, the precision in the determination of redshift should at least match the one achieved by photometric surveys (it is currently estimated to be about an order of magnitude worse). Theoretical errors, as seen in \tref{table:RSD_res1} and \ref{table:RSD_res2}, are less relevant for the $\fnl$ predictions coming from surveys with large redshift uncertainties, since the effect of the latter overshadows completely the impact of the first.

The power spectrum redshift space forecasts presented here (\ie ``RSD+z-errors+Theoretical errors'' columns in \tref{table:RSD_res1} and \ref{table:RSD_res2}) for the two radio continuum surveys are consistent with those presented in \citet{Raccanelli2017}, after taking into account the different flux limits used in the latter. 


The relative errors on bias parameters, shown in \fref{fig:RSD_comp1} and \ref{fig:RSD_comp2}, coming from joint power spectrum and bispectrum estimation, increase by a factor of two when redshift errors are included. For the linear bias, the increase originates mainly from the deterioration of the power spectrum results. For the two quadratic bias terms, the relative errors become larger than unity, since the signal comes only from the bispectrum which is more affected by redshift errors.

Since redshift errors are the main limiting factor in our radio continuum forecasts, it is natural to consider alternative radio datasets in which such issue does not arise. To this purpose, it is interesting to look at HI spectroscopic surveys. The precise redshift information obtained in this case, together with the large volume of the survey, can in principle be very promising for PNG studies. The most interesting aspect is that very tight forecasts could be obtained {\em not only for local-type, but also for equilateral-type NG}, by adding the bispectrum and the trispectrum term to previous PNG power spectrum studies, using HI surveys, such as \citet{Camera2015}. A full bispectrum forecast for the HI SKA galaxy sample requires a dedicated study, and goes beyond the scope of this paper. It is the object of a forthcoming work.

\subsubsection{The effect of the trispectrum term}\label{sec:radio_trisp_res}

    In our final step, we include the trispectrum contribution in the galaxy bispectrum [\eref{eq:Bg_full}], together with theoretical errors. This term will be considered for both the idealised and the full RSD models. The trispectrum term, discussed here, is present only for non-Gaussian initial conditions and exhibits a scale dependence, coming from the primordial bispectrum [\eref{eq:t5}]. It can then be used to tighten the constraints on PNG amplitudes. Interestingly, this is limited not only to the local case, but it also includes the equilateral shape. The importance of the trispectrum contribution was originally investigated by \citet{Sefusatti2009} and \citet{Jeong2009} for local non-Gaussianity, but the term is included in an actual forecast here for the first time. The scale dependence induced in the galaxy bispectrum by the quadratic bias trispectrum correction is analogous to the scale dependent bias of the power spectrum. It can be calculated exactly for any type of PNG, without the need of using any kind of squeezed limit approximation, and  $1/M(k)\propto k^{-2}$. Due to this, the degeneracy of $\fnle$ with the linear bias on large scales can now be broken (see the discussion in \sref{sec:PNGbias}).
          
\begin{figure}
 \centering
     \resizebox{\hsize}{!}{\includegraphics{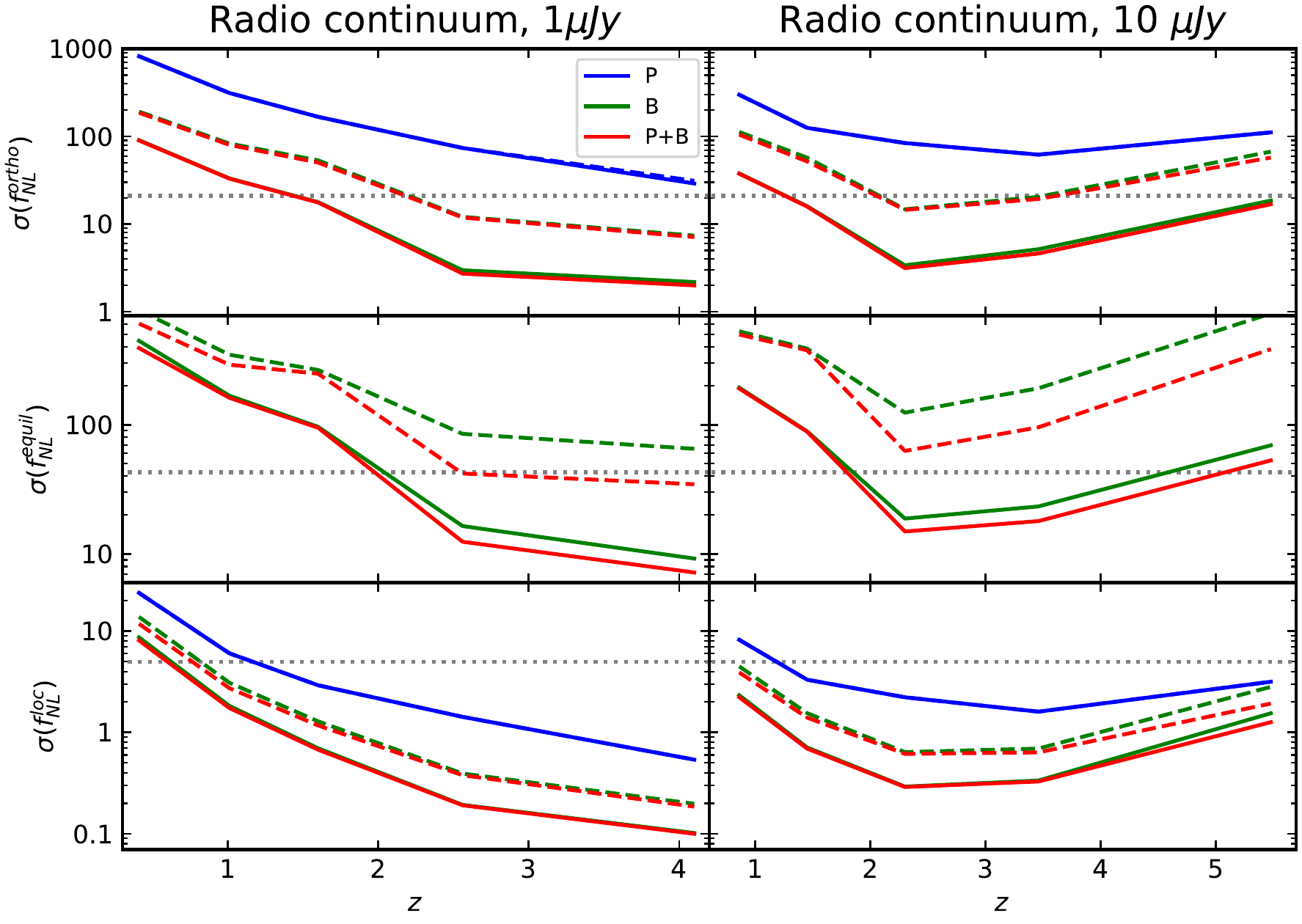}}
    \caption{ Fisher forecasts on $\fnl$ for local, equilateral and orthogonal shapes in the case of a 1 $\mu$Jy survey (left) and a 10 $\mu$Jy  on (right). The dashed lines represent the starting, idealised model, while the solid lines represent the same model when the trispectrum correction is considered [\ie \eref{eq:Pg_full} for the power spectrum and \eref{eq:Bg_full} for the bispectrum]. The full theoretical errors covariance is considered here. The expected constraints from the power spectrum are plotted in blue, from the bispectrum in green and from both power spectrum and bispectrum in red, without taking into account their cross term in the covariance.} \label{fig:wtris_plot}  
 \end{figure}

The results for the PNG amplitude, after including the trispectrum contribution, as well as theoretical errors, are presented in \fref{fig:wtris_plot}, where we compare them to the expected constraints coming from the idealised, preliminary model. An improvement is observed for all PNG amplitudes and especially the expected constraints for the equilateral shape show an impressive enhancement. We have checked that the theoretical errors do not affect significantly $\sigma(\fnle)$ in this case, since the scale dependent signal contribution compensates for the theoretical uncertainties. The bias forecasted constraints are unaffected, since the trispectrum correction disappears for chosen $\fnl=0$ fiducial value. The final predicted errors on the PNG amplitudes for the three cases considered here are presented in \tref{table:tris_res_mono_radio}, after including the trispectrum term from the bias expansion. We see that, in principle, the inclusion of trispectrum corrections allows for significant improvements for all shapes, including equilateral.

 One concern is that the two trispectrum corrections, \ie $T_{1112}^{MPT}$ and $S_2T_{1112}^{MPT}$, depend both on $\fnl$ and on bias parameters $b_1$, $b_2$ and $b_{S_2}$. Therefore, they could generate degeneracies between PNG and bias terms, which is indeed the case. However, we explicitly checked that, for large volume surveys such as the 1 $\mu$Jy radio continuum survey considered here, where very large scales are included, such degeneracies are broken.

\begin{figure}
 \centering
     \resizebox{\hsize}{!}{\includegraphics{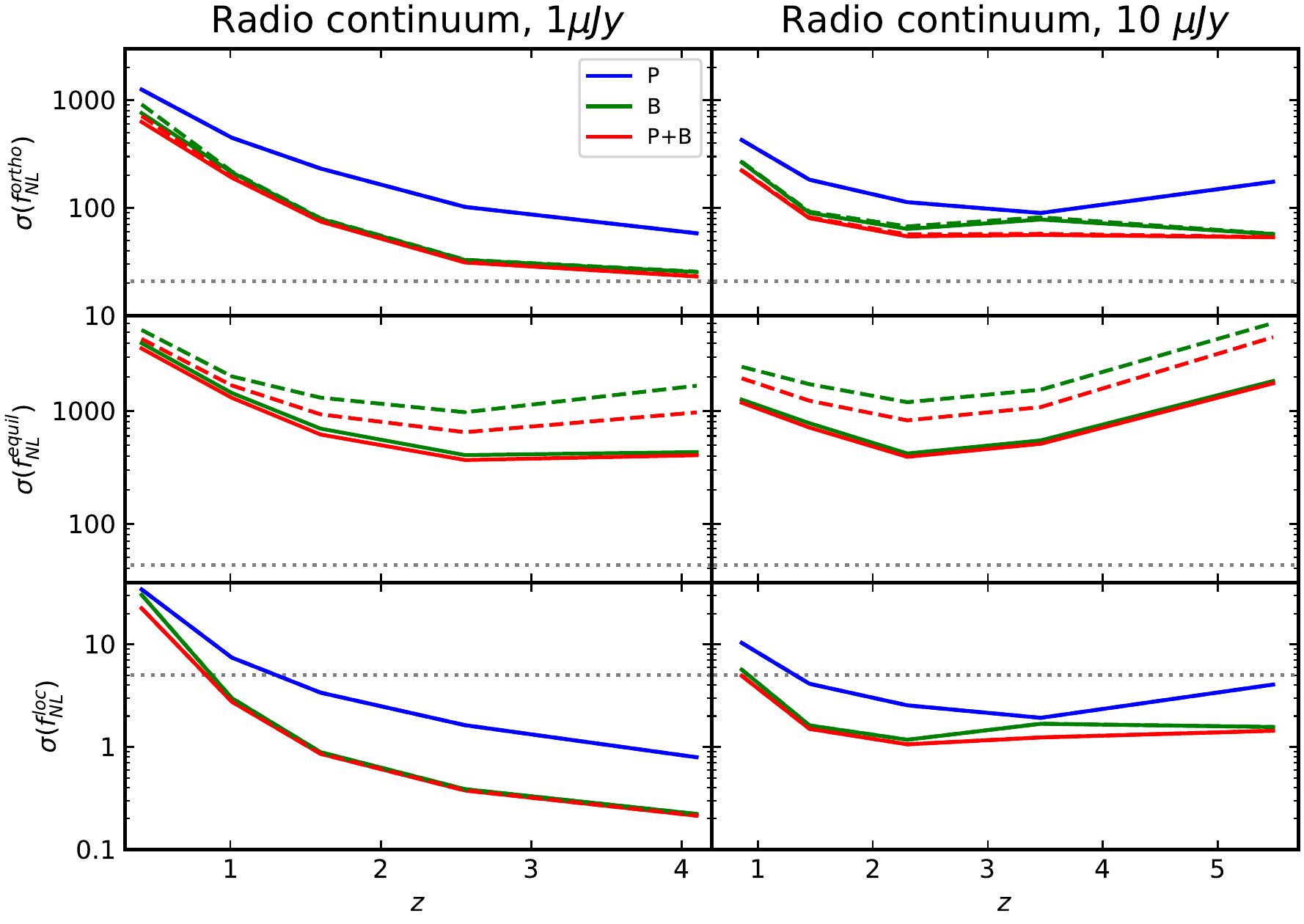}}
    \caption{Fisher forecasts on $\fnl$ for local, equilateral and orthogonal shapes in the case of a 1 $\mu$Jy survey (left) and a 10 $\mu$Jy on (right).  The dashed lines represent the model for the power spectrum and bispectrum in redshift space, given by \esref{eq:Pgs}{eq:Bgs} respectively, while for the latter the trispectrum term is excluded. The solid lines represent the same model when the trispectrum correction [\eref{eq:t5_rsd}] is taken into account. Redshift errors and FOG effects are considered for both cases, while we consider only the diagonal part of theoretical error covariance [\eref{eq:err_covar}]. The expected constraints from the power spectrum are plotted in blue, from the bispectrum in green and from both power spectrum and bispectrum in red, without taking into account their cross term in the covariance.} \label{fig:wtris_rsd_plot}  
 \end{figure}

The extension of this formalism to redshift space, by using the derived expression in \eref{eq:Bgs} with the trispectrum correction [\eref{eq:t5_rsd}], is performed here for the two radio continuum surveys. This model will be considered as our ``full'' treatment case. The resulting forecasts of the PNG amplitude are presented for each redshift bin in \fref{fig:wtris_rsd_plot}. Forecasted constraints for the integrated signal over the whole redshift range are presented in \tref{table:tris_res_rsd_radio}. The modelling for both correlators is described in detail in \sref{sec:radio_rsd_res} (\ie \esref{eq:Pgs}{eq:Bgs} for power spectrum and bispectrum respectively), where the effects of redshift errors and FOG are taken into account. Note that, only the diagonal part of theoretical error covariance is used in these redshift space forecasts (see \sref{sec:radio_rsd_res} for a discussion).

 As seen in \fref{fig:wtris_rsd_plot} and \tref{table:tris_res_rsd_radio}, the improvement provided by the trispectrum term in the forecasts of $\fnll$ is negligible, while for the orthogonal PNG type the contribution is minimal. The constraining power of the trispectrum term is reduced when the RSD effect is taken into account, contrary to the idealised, preliminary analysis (\tref{table:tris_res_mono_radio}). This can be mainly attributed to the presence of redshift errors. On the other hand, the scale dependence provided by the trispectrum term still significantly improves the final forecasted constraints for the equilateral PNG type, in combination with clustering-based redshift estimation. More precisely, the forecasts coming from the redshift space bispectrum are tighter by a factor of $\sim 2.5$, while for the combined signal from the two correlators the improvement can reach up to a factor of $\sim 2$.
 
            \begin{table*}
\centering
\resizebox{0.9\textwidth}{!}{
\begin{tabular}{ccc|cc}
\hline
                                & \multicolumn{2}{c|}{Radio continuum, 1 $\mu$Jy} & \multicolumn{2}{c}{Radio continuum, 10 $\mu$Jy} \\ \hline
                                & \thead{Idealised \\ +Theoretical errors}   & \thead{Idealised \\+Theoretical errors \\+Trispectrum}  & \thead{Idealised \\ +Theoretical errors}   & \thead{Idealised \\+Theoretical errors \\+Trispectrum}     \\ \hline
\multicolumn{1}{c|}{P(loc)}     & 0.50 & 0.50 & 1.12 & 1.12           \\
\multicolumn{1}{c|}{B(loc)}     & 0.18 & 0.089 & 0.44 & 0.21          \\
\multicolumn{1}{c|}{P+B(loc)}   & 0.17 & 0.088 & 0.41 & 0.20          \\ \hline
\multicolumn{1}{c|}{P(equil)}   & -          & -           & -             & -              \\
\multicolumn{1}{c|}{B(equil)}   & 50 & 8 & 98 & 14           \\
\multicolumn{1}{c|}{P+B(equil)} & 26 & 6 & 51 & 11           \\ \hline
\multicolumn{1}{c|}{P(ortho)}   & 28 & 27 & 42 & 42          \\
\multicolumn{1}{c|}{B(ortho)}   & 6.2 & 1.7 & 11 & 2.7           \\
\multicolumn{1}{c|}{P+B(ortho)} & 6.0 & 1.6 & 11 & 2.5           \\ \hline
\end{tabular}
}
\caption{Forecast $1\sigma$ results for the local, equilateral and orthogonal PNG. The calculations are performed by using the approximating monopole [\esref{eq:Pgsph}{eq:Bgsph}], while the full effect of the theoretical errors is taken into account here. The modelling used for the power spectrum is given by \eref{eq:Pg_full} and for the bispectrum by \eref{eq:Bg_full}, where for the latter the trispectrum bias term is taken into account. The idealised model is shown for comparison purposes. }
\label{table:tris_res_mono_radio}
\end{table*}

  \begin{table*}
\centering
\resizebox{0.9\textwidth}{!}{
\begin{tabular}{ccc|cc}
\hline
                                & \multicolumn{2}{c|}{Radio continuum, 1 $\mu$Jy } & \multicolumn{2}{c}{Radio continuum, 10 $\mu$Jy} \\ \hline
                                & \thead{RSD+z-errors \\ +Theoretical errors} & \thead{RSD+z-errors \\ +Theoretical errors \\ +Trispectrum}  & \thead{RSD+z-errors \\ +Theoretical errors} & \thead{RSD+z-errors \\ +Theoretical errors \\ +Trispectrum}     \\ \hline
\multicolumn{1}{c|}{P(loc)}     & 0.70 & 0.70 & 1.35 & 1.35           \\
\multicolumn{1}{c|}{B(loc)}     & 0.19 & 0.19 & 0.73 & 0.73          \\
\multicolumn{1}{c|}{P+B(loc)}   & 0.18 & 0.18 & 0.63 & 0.63          \\ \hline
\multicolumn{1}{c|}{P(equil)}   & -          & -           & -             & -              \\
\multicolumn{1}{c|}{B(equil)}   & 664 & 267 & 780 & 294           \\
\multicolumn{1}{c|}{P+B(equil)} & 448 & 244 & 551 & 274           \\ \hline
\multicolumn{1}{c|}{P(ortho)}   & 49 & 49 & 61 & 61          \\
\multicolumn{1}{c|}{B(ortho)}   & 20 & 20 & 35 & 34           \\
\multicolumn{1}{c|}{P+B(ortho)} & 18 & 18 & 30 & 29           \\ \hline
\end{tabular}
}
\caption{Forecast $1\sigma$ results for the local, equilateral and orthogonal PNG in redshift space. These results consider FOG effects and redshift errors are considered, while only the diagonal part of theoretical error covariance is taken into account here. The modelling used is for the power spectrum \eref{eq:Pgs} and for the bispectrum is given by \eref{eq:Bgs}, where for the latter the trispectrum bias term [\eref{eq:t5_rsd}] is taken into account. The RSD model without the trispectrum contribution (\ie ``RSD'') is shown for comparison purposes.}
\label{table:tris_res_rsd_radio}
\end{table*}

The ``RSD+trispectrum'' results, presented in \tref{table:tris_res_rsd_radio}, correspond to our ``full'' treatment and they are considered as the final forecasts on $\fnl$ for the three PNG types provided by the radio continuum samples. These forecasts show a significant expected improvement with respect to current $\fnll$ \Planck constraints, while for orthogonal PNG the $1\sigma$ forecast results are on the same level as  {\it Planck}. For the equilateral type, the forecasts show worse expected constraining power with respect to what already achieved by {\it Planck}. Nevertheless, we demonstrated the importance of the trispectrum bias term in improving the expected constraints provided by large volume LSS surveys, even in the case of large redshift uncertainties. If large volume surveys (either optical or radio) will allow for accurate redshift measurements, at some stage, not only local models will be measured with high sensitivity, but also all other shapes. This will be further discussed in the optical survey forecasts to follow (\sref{sec:results_opt}).

\subsubsection{Non-Gaussian corrections to the bispectrum variance.}

In this work, as discussed in \sref{sec:fisher}, only the diagonal part of the bispectrum covariance is used in the Fisher matrix formalism. In addition, for the variance we use the predictions of PT up to tree-level. Therefore, at this point, we would like to test the effect of excluding higher-order corrections. Using \eref{eq:DB2_NL}, we test the effect on the expected $\fnl$ constraints coming from the 1 $\mu$Jy radio continuum, and in particular  those originating from the real-space model [\eref{eq:Bg_full}] as well as the redshift space bispectrum [\eref{eq:Bgs}], following the procedure outlined at the end of section \sref{sec:fisher}. For local PNG, expected constraints in real space deteriorate by $\sim 39\%$, while in redshift space the deterioration increases to $\sim 67\%$. For equilateral PNG, the effect seems to be smaller, a degradation of $\sim 12\%$ and $\sim 17\%$ is observed for the real and redshift space case, respectively. This should be kept in mind when quoting final results. A full Fisher matrix analysis, including full covariances, rather than the simplified estimates provided here, will be object of a future study.

 \subsection{Optical surveys}\label{sec:results_opt}  
 
      In this section we show forecasts for a spectroscopic and a photometric survey (see \sref{sec:optical_surveys} for details on the specifications used). These will be ultimately compared with those originating from the two radio continuum surveys (see \sref{sec:radio_surveys}), presented in the previous section. Additionally, these surveys will allow us to test the full effect of theoretical uncertainties, mainly because their smaller volumes reduces significantly the computational time. We will in particular include off-diagonal theoretical error terms in our preliminary, idealised forecasts, up to the highest redshift bins (for nearly all scenarios; few exceptions will be pointed out case by case).

The optical survey analysis will broadly follow the same scheme as adopted in the previous section, based on adding realistic features and higher order corrections step by step, on top of the initial idealised model, in order to check separately their impact. 

 \subsubsection{Idealised treatment}\label{sec:opt_mono_res}

\begin{figure}
 \centering
     \resizebox{\hsize}{!}{\includegraphics{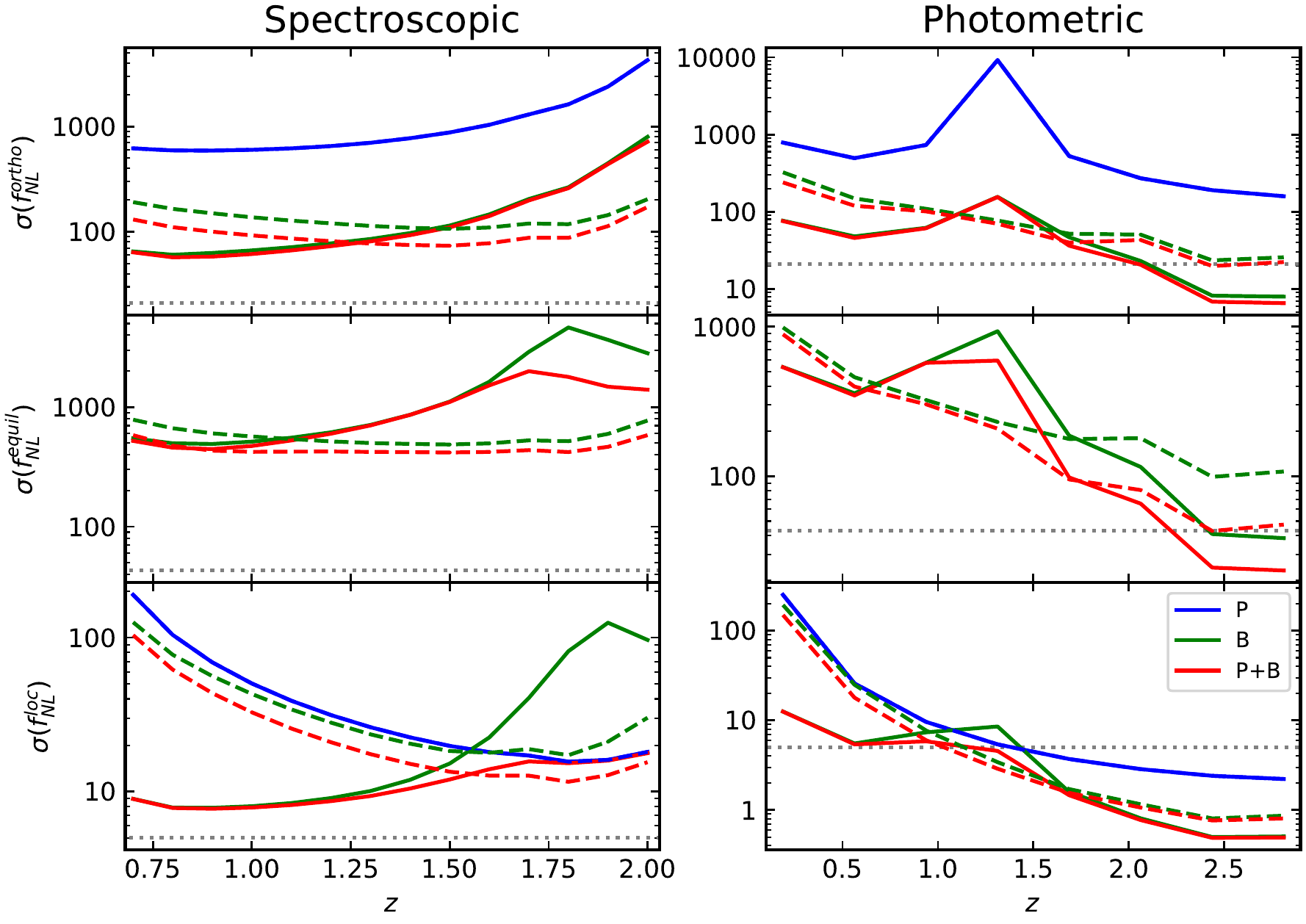}}
    \caption{ Fisher forecasts on $\fnl$ for local, equilateral and orthogonal shapes in the case of a spectroscopic (left) and a photometric survey (right). The dashed lines represent the ``idealised+ theoretical errors" model. The solid lines show the effect of adding the trispectrum correction (\eref{eq:Pg_full} for the power spectrum and \eref{eq:Bg_full} for the bispectrum). The expected constraints from the power spectrum are plotted in blue, from the bispectrum in green and from both power spectrum and bispectrum in red, without taking into account their cross-covariance.} \label{fig:Euclid_Desi_LSST_mono_plot}  
 \end{figure}

 As before, we start with the idealised case and show the effect of adding theoretical errors.  The full error covariance is here used only up to the 6th bin for the photometric case. For the remaining bins, only the diagonal contribution is considered, since for the high redshift bins the off-diagonal terms have a minimum effect on the final PNG amplitude forecasts, as discussed extensively in \sref{sec:radio_error_res}. The full theoretical error covariance will instead always be used over the whole redshift range for the power spectrum, as done before. Besides discussing theoretical errors, we also show the effect of the inclusion of trispectrum corrections. The marginalised $\fnl$ forecasts in both cases are shown in  \fref{fig:Euclid_Desi_LSST_mono_plot}. 
 
 \begin{table}
\centering
\resizebox{\hsize}{!}{
\begin{tabular}{ccc|cc}
\hline
                                & \multicolumn{2}{c|}{\thead{Idealised} } & \multicolumn{2}{c}{\thead{Idealised \\+Theoretical errors}} \\ \hline
                                & Spectroscopic        & Photometric  & Spectroscopic  & Photometric    \\ \hline
\multicolumn{1}{c|}{P(loc)}     & 5.64 & 1.14 & 6.21 & 1.27            \\
\multicolumn{1}{c|}{B(loc)}     & 5.27 & 0.45 & 6.59 & 0.50           \\
\multicolumn{1}{c|}{P+B(loc)}   & 3.85 & 0.42 & 4.52 & 0.46          \\ \hline
\multicolumn{1}{c|}{P(equil)}   & -        & - & - & -             \\
\multicolumn{1}{c|}{B(equil)}   & 109 & 44 & 149 & 59    \\
\multicolumn{1}{c|}{P+B(equil)} & 66 & 16 & 119 & 28    \\ \hline
\multicolumn{1}{c|}{P(ortho)}   & 182 & 61 & 208 & 104   \\
\multicolumn{1}{c|}{B(ortho)}   & 33 & 11 & 35 & 15    \\
\multicolumn{1}{c|}{P+B(ortho)} & 22 & 8 & 24 & 13   \\ \hline
\end{tabular}
}
\caption{Forecast $1\sigma$ results for the three PNG type considered here, originating from a spectroscopic and a photometric survey. The idealised model is as usual specified by \esref{eq:Pgsph}{eq:Bgsph}, and we show the full effect of including theoretical errors.}
\label{table:opt_res_mono_theoerr}
\end{table}
 
 The effect of the full theoretical error covariance is presented in \tref{table:opt_res_mono_theoerr} for the two optical surveys. As done also for the radio continuum surveys, the practicality of studying the idealised case first lies in the fact that a quantification of the effect of the full theoretical error covariance is needed, in order to propagate it to the RSD treatment (next section), where the computational effort is large. It is evident that the behaviour follows the same pattern as in the case of radio continuum. A degradation in the final PNG forecasts is observed ranging between $25-50\%$ depending on the PNG type, as well as the size and redshift range of the survey (see \tref{table:opt_res_mono_theoerr}), where for the equilateral case the effect is maximum. Excluding the off-diagonal elements of the theoretical error covariance introduces an almost negligible error for local PNG for all optical surveys, while for the equilateral and orthogonal forecasts an underestimation of the theoretical uncertainties by $9\%$ and $4\%$ is observed respectively. The forecasts for all thee PNG types coming from the photometric survey are improved with respect to the spectroscopic one.

\begin{figure}
\centering
   \includegraphics[width=0.48\textwidth]{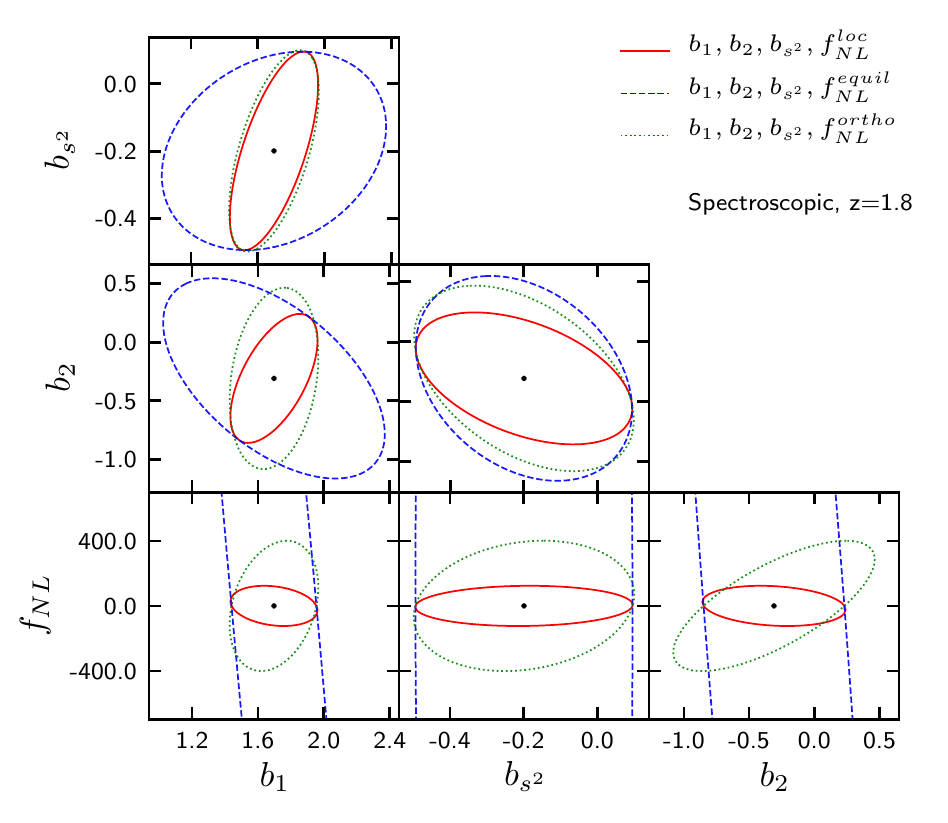}
    \caption{ The joint $1\sigma$ forecasts on the three galaxy bias parameters (\ie $b_1$, $b_2$ and $b_{s^2}$) and PNG from the $z=1.8$ redshift bin of the spectroscopic survey. In each panel the two-dimensional joint forecasts are shown for all the combinations between the parameters, after marginalising over the remaining. The model used is the approximating monopole for both power spectrum and bispectrum [\esref{eq:Pgsph}{eq:Bgsph}], after considering in the galaxy bispectrum the trispectrum bias term shown in \eref{eq:Bg_full}.
    } \label{fig:cont_loc_equil_ortho}

\end{figure}

  Adding trispectrum corrections produces some improvements. For local PNG, low redshift bins produce an order of magnitude improvement, while high redshifts have a much smaller effect. The same behaviour is also observed for the orthogonal shape, but in this case the improvement is much smaller ($\sim 2.5$ times). For equilateral PNG, a small improvement is observed only for the very low redshift bins, while for the large redshifts the expected constraints become even worse. The lack of improvement observed in the small volume spectroscopic survey is due to degeneracies between $\fnl$ and bias parameters, which are enhanced by the presence of the trispectrum term. As seen in \fref{fig:cont_loc_equil_ortho}, $\fnll$ does not show any degeneracies with the three bias terms (\ie $b_1$, $b_2$ and $b_{s^2}$), considered here as free parameters, while $\fnlo$ shows some degeneracy mainly with $b_2$. On the other hand, $\fnle$ shows strong degeneracy with all three and hence this explains the observed degradation in results (see \fref{fig:Euclid_Desi_LSST_mono_plot} and \tref{table:opt_res_mono}) after the inclusion of the trispectrum term. Such degeneracies can be broken if large enough scales are available in the survey (see \fref{fig:wtris_plot} for the radio continuum cases). For example, an overall gain in the forecasts provided by the photometric survey can be observed, since they probe a larger part of the sky and hence grant access to larger scale modes. This fact will  partially break the degeneracies of $\fnl$ with the bias parameters, providing the observed improvement over the results of the idealised model. More specifically, for the local-type PNG, the trispectrum corrections reduce the $1\sigma$ expected constraints on $\fnll$ originating from the photometric survey beyond unity for the high redshift bins.  For the equilateral PNG case, a compelling improvement is observed for both surveys.

 Here we would like to point out the importance of the bispectrum for the PNG amplitude forecasts coming from LSS surveys. As we can see in \fref{fig:Euclid_Desi_LSST_mono_plot}, the gain in the final $\fnl$ expected constraints of the local type is small when the signal from the bispectrum is added on top of the one generated by the power spectrum. However, this changes dramatically for the equilateral and orthogonal types, where the bispectrum is the main source of signal.

 We present the marginalised idealised forecasts for the PNG amplitude in \tref{table:opt_res_mono}. We see that in principle the trispectrum corrections can lead to significant improvements for large volume photometric surveys. In total analogy with the radio continuum case, the introduction of redshift errors at the expected level for next generation surveys will however wash away the enhancement. We explicitly show this in the next section. 
 
         \begin{table}
\centering
\resizebox{\hsize}{!}{
\begin{tabular}{ccc|cc}
\hline
                                & \multicolumn{2}{c|}{\thead{Idealised \\ +Theoretical errors}} & \multicolumn{2}{c}{\thead{Idealised \\+Theoretical errors \\+Trispectrum} } \\ \hline
                                & Spectroscopic   & Photometric   & Spectroscopic    & Photometric        \\ \hline
\multicolumn{1}{c|}{P(loc)}     & 6.21 & 1.27 & 6.21 & 1.27  \\
\multicolumn{1}{c|}{B(loc)}     & 6.60 & 0.50 & 3.00 & 0.32   \\
\multicolumn{1}{c|}{P+B(loc)}   & 4.52 & 0.46 & 2.69 & 0.31    \\ \hline
\multicolumn{1}{c|}{P(equil)}   & -        & -      & -       & -                  \\
\multicolumn{1}{c|}{B(equil)}   & 149 & 59 & 196 & 27    \\
\multicolumn{1}{c|}{P+B(equil)} & 119 & 28 & 181 & 16 \\ \hline
\multicolumn{1}{c|}{P(ortho)}   & 208 & 104 & 208 & 104  \\
\multicolumn{1}{c|}{B(ortho)}   & 35 & 15 & 24 & 5.4   \\
\multicolumn{1}{c|}{P+B(ortho)} & 24 & 13 & 23 & 4.5 \\ \hline
\end{tabular}
}
\caption{Forecast $1\sigma$ results for the three PNG type considered here, originating from a spectroscopic and a photometric survey. Left column: idealised case with $+$ theoretical errors [\esref{eq:Pgsph}{eq:Bgsph}]. Right column: idealised $+$ theoretical errors $+$ trispectrum corrections . The full effect of the theoretical errors is taken into account here, as described at the beginning of \sref{sec:results_opt}.  }
\label{table:opt_res_mono}
\end{table}
 
 All the previous results, assume a Gaussian and diagonal bispectrum covariance matrix. By using \eref{eq:DB2_NL}, proposed in \citet{Chan2017}, we can effectively resum the contributions to the diagonal coming from higher order terms. Doing so will degrade the results presented in \tref{table:opt_res_mono}. In the case of the spectroscopic survey a $\sim 35\%$ and $\sim 16\%$ increase is observed for the $1\sigma$ forecast errors in the case of local and equilateral PNG type respectively. For the photometric survey the deterioration of the forecasted constraints on local and equilateral PNG is $\sim 29\%$ and $\sim 10\%$ respectively. The same trend seen for the radio continuum surveys is also observed here, \ie the local expected constraints are highly affected by the higher order corrections in the bispectrum variance, while a minimum effect is observed for the equilateral PNG case.

  \subsubsection{Full redshift space treatment}\label{sec:opt_rsd_res}

\begin{figure}
 \centering
     \resizebox{\hsize}{!}{\includegraphics{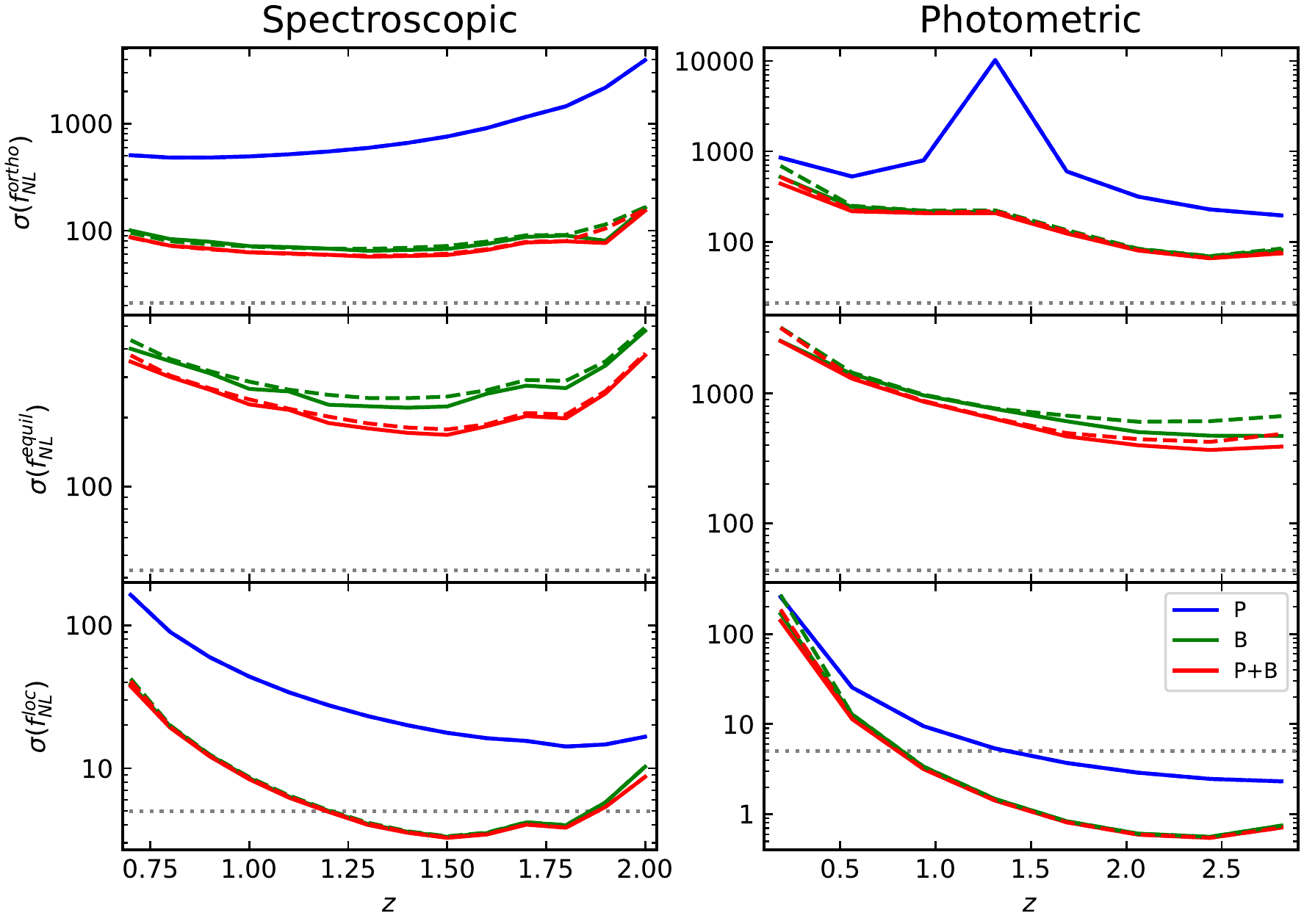}}
    \caption{ Fisher forecasts on the amplitude of the three PNG types considered here in the case of a spectroscopic survey (left) and a photometric one (right). The dashed lines represent the RSD model, given by \esref{eq:Pgs}{eq:Bgs} for the power spectrum and bispectrum respectively, without the presence of the trispectrum term [\eref{eq:t5_rsd}] in the galaxy bispectrum. Solid lines represent the same model but with the trispectrum correction included. The expected constraints from the power spectrum are plotted in blue, from the bispectrum in green and from both power spectrum and bispectrum in red, without taking into account their cross term in the covariance. The FOG effect, as well as the redshift error, is taken into account in both models. Only the diagonal part of the theoretical errors covariance is used here.} \label{fig:Euclid_Desi_LSST_rsd_plot}  
 \end{figure}

  Our next step is to move to redshift space, considering RSD, redshift errors and the FOG smearing effect. As usual, for the sake of comparison, we will use the bispectrum RSD model with and without the trispectrum bias term given in \eref{eq:t5_rsd}. Regarding theoretical errors, only the diagonal part of the error covariance [\eref{eq:err_covar}] will be used in redshift space, due to the high computational cost. The results are presented as a function of redshift bins in \fref{fig:Euclid_Desi_LSST_rsd_plot} for the spectroscopic and photometric surveys.   These surveys can produce improvements on current \Planck bounds for the local shape, but not for the other two, mostly due to theoretical errors. 
  
Adding trispectrum corrections has a minimal effect for spectroscopic surveys, because such corrections need very large scales to be effective and the probed volumes are here too small to make a difference.

\begin{table*}
\centering
\resizebox{\textwidth}{!}{
\begin{tabular}{ccc|cc|cc}
\hline
                                & \multicolumn{2}{c|}{\thead{RSD (no z-errors) \\+Theoretical errors}} & \multicolumn{2}{c|}{ \thead{RSD+z-errors \\Theoretical errors}} & \multicolumn{2}{c}{\thead{RSD + z-errors \\+Theoretical errors \\+trispectrum}} \\ \hline
                                & Spectroscopic    & Photometric &  Spectroscopic   & Photometric  &  Spectroscopic  & Photometric       \\ \hline
\multicolumn{1}{c|}{P(loc)}     & 5.59 & 1.16 & 5.60 & 1.30 & 5.60 & 1.30 \\
\multicolumn{1}{c|}{B(loc)}     & 1.21 & 0.082 & 1.35 & 0.32 & 1.34 & 0.32   \\
\multicolumn{1}{c|}{P+B(loc)}   & 1.19 & 0.081 & 1.31 & 0.31 & 1.30 & 0.31  \\ \hline
\multicolumn{1}{c|}{P(equil)}   & -       & -     &      -    & -     &      - & -      \\
\multicolumn{1}{c|}{B(equil)}   & 74 & 31 & 78 & 276 & 73 & 229 \\
\multicolumn{1}{c|}{P+B(equil)} & 56 & 17 & 60 & 207 & 57 & 184  \\ \hline
\multicolumn{1}{c|}{P(ortho)}   & 173 & 87 & 173 & 124 & 173 & 124   \\
\multicolumn{1}{c|}{B(ortho)}   & 20 & 7.6 & 21 & 40 & 21 & 40  \\
\multicolumn{1}{c|}{P+B(ortho)} & 18 & 7.0 & 19 & 38 & 18 & 38  \\ \hline
\end{tabular}
}
\caption{Forecast marginalised $1\sigma$ results for the three PNG type considered here, originating from  a spectroscopic and a photometric survey. The full RSD model for both power spectrum and bispectrum is considered [\esref{eq:Pgs}{eq:Bgs}]. Three different versions of the model are examined here: without considering the redshift uncertainties (``RSD (no z-errors)+Theoretical errors"), including the redshift errors (``RSD+z-errors+Theoretical errors") and adding to the latter the trispectrum contribution (``RSD+z-errors+Theoretical errors+trispectrum"). Regarding the theoretical errors, only the diagonal part of the covariance is used for all versions.}
\label{table:opt_res_rsd}
\end{table*}
    
  The situation changes a bit for the photometric survey. This survey probes a much larger volume, therefore trispectrum corrections can in principle play a useful role here, as shown explicitly in the previous section. At the same time, however, we are now affected by large photometric redshift errors. Due to the latter, all PNG predicted constraints are now degraded (\fref{fig:Euclid_Desi_LSST_mono_plot}) with respect to the idealised forecasts, where redshift uncertainties are not considered. The full redshift space forecasts, coming from the integrated signal over the entire redshift range, are presented in \tref{table:opt_res_rsd} for the galaxy power spectrum, bispectrum and their combination. Both cases, with and without the trispectrum term, are shown. To better quantify the effect of redshift uncertainties on our final $\fnl$ predicted constraints, we present also the results of the RSD model, taking $\sigma_z(z)=0$. The local PNG case is affected the least by redshift uncertainties and therefore the observed improvements in the $\fnll$ predicted constraints for the photometric survey are justified. Its constraining power is significantly better than that of {\it Planck}, for bins with $z\gtrsim 1$. Adding the trispectrum correction has a minimal impact on the bispectrum forecasts for the orthogonal and local types, while it can have an overall improvement in the equilateral results of the large-volume surveys. Still, large redshift errors make the trispectrum contribution far from enough to improve over current bounds.

As we show in \sref{sec:opt_mono_res}, the effect of neglecting off-diagonal terms from the theoretical error covariance is expected to be small for surveys with wide redshift range. In order to quantify this, we performed a full error covariance analysis for the spectroscopic survey. The survey volume size shrinks in this case the computational effort tremendously, compared to radio continuum cases, and makes a full numerical analysis feasible. The deterioration level in the final PNG expected constraints is consistent with what reported for radio continuum. More specifically, we observe a $<1\%$ and $\sim7\%$ degradation for local and equilateral PNG cases respectively.

   \begin{table*}
\centering
\resizebox{\hsize}{!}{
\begin{tabular}{cccccc}
\hline
                                 & \Planck  & Radio continuum, 1 $\mu$Jy     & Radio continuum, 10 $\mu$Jy  &  Spectroscopic  & Photometric       \\ \hline

\multicolumn{1}{c|}{Local}           &  5.0  & 0.2 & 0.6  & 1.3       & 0.3             \\ 
\multicolumn{1}{c|}{Equilateral}   & 43    & 244  & 274   & 57   & 184 \\ 
\multicolumn{1}{c|}{Orthogonal}  & 21    & 18    & 29  & 18     & 38      \\ \hline
\end{tabular}
}
\caption{Summary of $1\sigma$ limits for the three PNG types considered, from radio continuum and optical surveys  derived from combining the power spectrum and bispectrum and accounting for RSD, the trispectrum term and theoretical errors. See text for more details.}
\label{table:concl}
\end{table*} 
 
  The  ``RSD+trispectrum'' results in \tref{table:opt_res_rsd} are considered as our final forecasts on the three PNG types originating from the optical surveys. Comparing these results with those available in the literature, we see that our power spectrum spectroscopic results, after excluding the theoretical errors, are consistent with the forecasts in \citet{Giannantonio2012} for the local and orthogonal PNG. On the other hand our forecasts on the local type, coming from the spectroscopic survey, are slightly worse than those reported in \citet{Tellarini2016}, where the difference originates from the fact that we take into account the presence of theoretical errors, as well as the effect of redshift uncertainties. The differences are slightly higher for redshift space bispectrum forecasts, owing to the higher impact of theoretical errors and to the inclusion of redshift errors (small, but not completely negligible also for spectroscopic surveys).
  
  As done for the radio analysis, we also tested the effect of neglecting non-Gaussian corrections in the bispectrum {\em variance}, by using \eref{eq:DB2_NL}. In the case of the spectroscopic survey, a degradation of $\sim 42\%$ and $\sim 20\%$ is observed in the forecasts for the local and equilateral PNG types respectively. For the photometric probe we get a deterioration of $\sim 32\%$ and $\sim 10\%$ for the final predicted constraints on local and equilateral PNG respectively. This is not overall a very large effect, in the context of a Fisher forecast. However, it does make it clear again that such NG contributions to the bispectrum variance are not negligible, and future analyses will have to include them to achieve high accuracy.

   \section{Conclusions}
   \label{sec:conc}
  
In this paper we have investigated possible future constraints on the amplitude of the non-Gaussian parameter $\fnl$ for three types of PNG shapes -- local, equilateral and orthogonal -- and on galaxy bias parameters, through galaxy power spectrum and bispectrum measurements on large scales using a Fisher matrix approach.
We thoroughly accounted for a large number of effects in modelling the gravitational non-Gaussian contributions, including a full second order treatment of bias and RSD, going beyond real space angle averaged $\fnl$ forecasts presented in other works. We carefully investigated the propagation of theoretical uncertainties, following the approach introduced in \citet{Baldauf2016}, but extending it to bias loop-corrections. All these effects were to a various extent included in previous literature, but for the first time we accounted for all of them at once in a single forecast analysis.  The cross-correlation between power spectrum and bispectrum was ignored in this work, and the bispectrum covariance approximated as diagonal. However, for the large scales considered here, we discussed how this should have a small impact, based on recent results in the literature \citep{Chan2017}. We also employed, as standardly done, a Gaussian approximation for the bispectrum variance. We presented an explicit estimate of the effect of ignoring non-Gaussian contributions to the variance, by considering leading non-Gaussian corrections. We found that, in the worst case scenario, such corrections can degrade constraints by $\sim 50 \%$. A more detailed study of the bispectrum covariance, including PNG corrections, will be included in a future work. Likewise, it will be important to account in the future for the effect of relativistic corrections on the bispectrum, especially for large volume surveys \citep[see \eg][]{DiDio2016,Raccanelli2016, Umeh2017, Bertacca2017}.

In addition to the previous ingredients, we improved the modelling of the galaxy bispectrum by considering a complete second order bias expansion, which includes for the first time the trispectrum term [\eref{eq:dg}]. We only consider the zeroth order (tree-level) expansions in the matter fields, because we are only interested in the large-scale contributions. For dark matter,  we have used the \MPT perturbation theory, based on Renormalised Perturbation Theory, which provides a natural cut-off in the non-linear regime, such that ultraviolet divergences are automatically removed.
The final bispectrum model used for our forecast is presented in \eref{eq:Bg_full}.

Our results are based on radio continuum (with 10 $\mu$Jy and 1 $\mu$Jy flux limits) and optical (spectroscopic and photometric) surveys.
In the case of radio surveys, we assume a clustering-based estimate of redshifts. 

We have summarised our main results in Table \ref{table:concl}, where we have reported the forecasts derived by combining power spectrum and bispectrum for the three non-Gaussian shapes, together with the \Planck temperature and polarisation constraints \citep{Planck_PNG2016}, as a comparison.

For the local shape,  LSS measurements can provide important improvements over current CMB bounds, even using only the power spectrum, but the bispectrum will allow doing much better, with a further improving factor $\sim 5$.
For equilateral PNG, the bispectrum is by far the most useful statistics, but it is plagued by the fact that theoretical errors peak in the equilateral limit. Therefore, improving upon current forecasts by including smaller scales and more modes, at low redshift, will require exquisitely accurate modelling of late-time non-linearities. On the other hand, large future radio continuum surveys will provide access to much larger volumes and higher redshifts. Note that the constraining power of the galaxy bispectrum on probing PNG can be improved by using the multitracer technique \citep{Seljak2009,Hamaus2011} as was in \citet{Yamauchi2017}.

The trade-off for these surveys is however represented by large errors in the determination of redshifts, which become dominant and now overshadow the effect of theoretical errors.
In a very idealised case where galaxy redshifts could be accurately known  for all objects,  even in presence of significant theoretical errors, large improvements (\eg up to a factor $\sim 5$ for radio) with respect to \Planck equilateral constraints are obtained.  This constraining power mostly comes from very large scales and trispectrum contributions, which display a $\sim k^{-2}$ scale-dependence in the equilateral case. Such contributions therefore deserve further attention. In a more realistic case, redshift errors affect these forecasts significantly. The gain obtained by including the trispectrum term is actually still large when z-errors are included, but it becomes insufficient to improve over current equilateral $\Planck$ bounds. The final forecasts are in the first two columns of Table \ref{table:concl}. 

We note that the issue above cannot be circumvented simply by looking at the trispectrum term in forthcoming optical spectroscopic surveys, because they have too small a volume to make such term significant. We argued that looking instead at HI spectroscopic surveys is another interesting approach, which we will pursue in a forthcoming work. The other possible way forward would be finding  strategies for better determination of redshifts in either future photometric or
radio continuum surveys.

\[\]{\bf Acknowledgments:}\\
We are indebted to Emiliano Sefusatti and Sabino Matarrese for illuminating discussions throughout the elaboration of the manuscript; to Ely Kovetz and Mubdi Rahman for contributions at the beginning of the development of the project. The authors also thank Guido d'Amico for useful discussions.
AR has received funding from the People Programme (Marie Curie Actions) of the European Union H2020 Programme
under REA grant agreement number 706896 (COSMOFLAGS) and the Templeton
Foundation. Funding for this work was partially provided by the
Spanish MINECO under MDM-2014-0369 of ICCUB (Unidad de Excelencia ``Maria de Maeztu").
LV acknowledges support  by Spanish Mineco via AYA2014-58747-P AEI/FEDER UE and  MDM-2014-0369 of ICCUB (Unidad de Excelencia Maria de Maeztu) and   European UnionÕs Horizon 2020 research and innovation programme ERC (BePreSySe, grant agreement 725327).

\bibliography{bibliographypaper}

\begin{thebibliography}{}
\makeatletter
\relax
\def\mn@urlcharsother{\let\do\@makeother \do\$\do\&\do\#\do\^\do\_\do\%\do\~}
\def\mn@doi{\begingroup\mn@urlcharsother \@ifnextchar [ {\mn@doi@}
  {\mn@doi@[]}}
\def\mn@doi@[#1]#2{\def\@tempa{#1}\ifx\@tempa\@empty \href
  {http://dx.doi.org/#2} {doi:#2}\else \href {http://dx.doi.org/#2} {#1}\fi
  \endgroup}
\def\mn@eprint#1#2{\mn@eprint@#1:#2::\@nil}
\def\mn@eprint@arXiv#1{\href {http://arxiv.org/abs/#1} {{\tt arXiv:#1}}}
\def\mn@eprint@dblp#1{\href {http://dblp.uni-trier.de/rec/bibtex/#1.xml}
  {dblp:#1}}
\def\mn@eprint@#1:#2:#3:#4\@nil{\def\@tempa {#1}\def\@tempb {#2}\def\@tempc
  {#3}\ifx \@tempc \@empty \let \@tempc \@tempb \let \@tempb \@tempa \fi \ifx
  \@tempb \@empty \def\@tempb {arXiv}\fi \@ifundefined
  {mn@eprint@\@tempb}{\@tempb:\@tempc}{\expandafter \expandafter \csname
  mn@eprint@\@tempb\endcsname \expandafter{\@tempc}}}

\bibitem[\protect\citeauthoryear{{Afshordi} \& {Tolley}}{{Afshordi} \&
  {Tolley}}{2008}]{Afshordi2008}
{Afshordi} N.,  {Tolley} A.~J.,  2008, \mn@doi [\prd]
  {10.1103/PhysRevD.78.123507}, \href
  {http://adsabs.harvard.edu/abs/2008PhRvD..78l3507A} {78, 123507}

\bibitem[\protect\citeauthoryear{{Agarwal}, {Ho}  \& {Shandera}}{{Agarwal}
  et~al.}{2014}]{Agarwal2014}
{Agarwal} N.,  {Ho} S.,   {Shandera} S.,  2014, \mn@doi [\jcap]
  {10.1088/1475-7516/2014/02/038}, \href
  {http://adsabs.harvard.edu/abs/2014JCAP...02..038A} {2, 038}

\bibitem[\protect\citeauthoryear{{Agullo} \& {Shandera}}{{Agullo} \&
  {Shandera}}{2012}]{Agullo2012}
{Agullo} I.,  {Shandera} S.,  2012, \mn@doi [\jcap]
  {10.1088/1475-7516/2012/09/007}, \href
  {http://adsabs.harvard.edu/abs/2012JCAP...09..007A} {9, 007}

\bibitem[\protect\citeauthoryear{{Assassi}, {Baumann}, {Green}  \&
  {Zaldarriaga}}{{Assassi} et~al.}{2014}]{Assassi2014}
{Assassi} V.,  {Baumann} D.,  {Green} D.,   {Zaldarriaga} M.,  2014, \mn@doi
  [\jcap] {10.1088/1475-7516/2014/08/056}, \href
  {http://adsabs.harvard.edu/abs/2014JCAP...08..056A} {8, 056}

\bibitem[\protect\citeauthoryear{{Assassi}, {Baumann}  \& {Schmidt}}{{Assassi}
  et~al.}{2015}]{Assassi2015}
{Assassi} V.,  {Baumann} D.,   {Schmidt} F.,  2015, \mn@doi [\jcap]
  {10.1088/1475-7516/2015/12/043}, \href
  {http://adsabs.harvard.edu/abs/2015JCAP...12..043A} {12, 043}

\bibitem[\protect\citeauthoryear{{Audren}, {Lesgourgues}, {Bird}, {Haehnelt}
  \& {Viel}}{{Audren} et~al.}{2013}]{Audren2013}
{Audren} B.,  {Lesgourgues} J.,  {Bird} S.,  {Haehnelt} M.~G.,   {Viel} M.,
  2013, \mn@doi [\jcap] {10.1088/1475-7516/2013/01/026}, \href
  {http://adsabs.harvard.edu/abs/2013JCAP...01..026A} {1, 026}

\bibitem[\protect\citeauthoryear{{Bailoni}, {Spurio Mancini}  \&
  {Amendola}}{{Bailoni} et~al.}{2017}]{Bailoni2017}
{Bailoni} A.,  {Spurio Mancini} A.,   {Amendola} L.,  2017, \mn@doi [\mnras]
  {10.1093/mnras/stx1209}, \href
  {http://adsabs.harvard.edu/abs/2017MNRAS.470..688B} {470, 688}

\bibitem[\protect\citeauthoryear{{Baldauf}, {Seljak}  \& {Senatore}}{{Baldauf}
  et~al.}{2011}]{Baldauf2011}
{Baldauf} T.,  {Seljak} U.,   {Senatore} L.,  2011, \mn@doi [\jcap]
  {10.1088/1475-7516/2011/04/006}, \href
  {http://adsabs.harvard.edu/abs/2011JCAP...04..006B} {4, 006}

\bibitem[\protect\citeauthoryear{{Baldauf}, {Seljak}, {Desjacques}  \&
  {McDonald}}{{Baldauf} et~al.}{2012}]{Baldauf2012}
{Baldauf} T.,  {Seljak} U.,  {Desjacques} V.,   {McDonald} P.,  2012, \mn@doi
  [\prd] {10.1103/PhysRevD.86.083540}, \href
  {http://adsabs.harvard.edu/abs/2012PhRvD..86h3540B} {86, 083540}

\bibitem[\protect\citeauthoryear{{Baldauf}, {Mirbabayi}, {Simonovi{\'c}}  \&
  {Zaldarriaga}}{{Baldauf} et~al.}{2016}]{Baldauf2016}
{Baldauf} T.,  {Mirbabayi} M.,  {Simonovi{\'c}} M.,   {Zaldarriaga} M.,  2016,
  preprint, \href {http://adsabs.harvard.edu/abs/2016arXiv160200674B} {}
  (\mn@eprint {arXiv} {1602.00674})

\bibitem[\protect\citeauthoryear{{Ballinger}, {Peacock}  \&
  {Heavens}}{{Ballinger} et~al.}{1996}]{Ballinger1996}
{Ballinger} W.~E.,  {Peacock} J.~A.,   {Heavens} A.~F.,  1996, \mn@doi [\mnras]
  {10.1093/mnras/282.3.877}, \href
  {http://adsabs.harvard.edu/abs/1996MNRAS.282..877B} {282, 877}

\bibitem[\protect\citeauthoryear{{Bardeen}, {Bond}, {Kaiser}  \&
  {Szalay}}{{Bardeen} et~al.}{1986}]{Bardeen1986}
{Bardeen} J.~M.,  {Bond} J.~R.,  {Kaiser} N.,   {Szalay} A.~S.,  1986, \mn@doi
  [\apj] {10.1086/164143}, \href
  {http://adsabs.harvard.edu/abs/1986ApJ...304...15B} {304, 15}

\bibitem[\protect\citeauthoryear{{Baumann}, {Nicolis}, {Senatore}  \&
  {Zaldarriaga}}{{Baumann} et~al.}{2012}]{Baumann2012}
{Baumann} D.,  {Nicolis} A.,  {Senatore} L.,   {Zaldarriaga} M.,  2012, \mn@doi
  [\jcap] {10.1088/1475-7516/2012/07/051}, \href
  {http://adsabs.harvard.edu/abs/2012JCAP...07..051B} {7, 051}

\bibitem[\protect\citeauthoryear{{Bernardeau}, {Colombi}, {Gazta{\~n}aga}  \&
  {Scoccimarro}}{{Bernardeau} et~al.}{2002}]{Bernardeau2002}
{Bernardeau} F.,  {Colombi} S.,  {Gazta{\~n}aga} E.,   {Scoccimarro} R.,  2002,
  \mn@doi [\physrep] {10.1016/S0370-1573(02)00135-7}, \href
  {http://adsabs.harvard.edu/abs/2002PhR...367....1B} {367, 1}

\bibitem[\protect\citeauthoryear{{Bernardeau}, {Crocce}  \&
  {Scoccimarro}}{{Bernardeau} et~al.}{2008}]{Bernardeau2008}
{Bernardeau} F.,  {Crocce} M.,   {Scoccimarro} R.,  2008, \mn@doi [\prd]
  {10.1103/PhysRevD.78.103521}, \href
  {http://adsabs.harvard.edu/abs/2008PhRvD..78j3521B} {78, 103521}

\bibitem[\protect\citeauthoryear{{Bernardeau}, {Crocce}  \&
  {Scoccimarro}}{{Bernardeau} et~al.}{2012}]{2012PhRvD..85l3519B}
{Bernardeau} F.,  {Crocce} M.,   {Scoccimarro} R.,  2012, \mn@doi [\prd]
  {10.1103/PhysRevD.85.123519}, \href
  {http://adsabs.harvard.edu/abs/2012PhRvD..85l3519B} {85, 123519}

\bibitem[\protect\citeauthoryear{Bertacca, Raccanelli, Bartolo, Liguori,
  Matarrese  \& Verde}{Bertacca et~al.}{2018}]{Bertacca2017}
Bertacca D.,  Raccanelli A.,  Bartolo N.,  Liguori M.,  Matarrese S.,   Verde
  L.,  2018, \mn@doi [Phys. Rev. D] {10.1103/PhysRevD.97.023531}, 97, 023531

\bibitem[\protect\citeauthoryear{{Camera} et~al.,}{{Camera}
  et~al.}{2015a}]{Camera2015b}
{Camera} S.,  et~al., 2015a, Advancing Astrophysics with the Square Kilometre
  Array (AASKA14), \href {http://adsabs.harvard.edu/abs/2015aska.confE..25C}
  {p.~25}

\bibitem[\protect\citeauthoryear{{Camera}, {Santos}  \& {Maartens}}{{Camera}
  et~al.}{2015b}]{Camera2015}
{Camera} S.,  {Santos} M.~G.,   {Maartens} R.,  2015b, \mn@doi [\mnras]
  {10.1093/mnras/stv040}, \href
  {http://adsabs.harvard.edu/abs/2015MNRAS.448.1035C} {448, 1035}

\bibitem[\protect\citeauthoryear{{Carrasco}, {Hertzberg}  \&
  {Senatore}}{{Carrasco} et~al.}{2012}]{Carrasco2012}
{Carrasco} J.~J.~M.,  {Hertzberg} M.~P.,   {Senatore} L.,  2012, \mn@doi
  [Journal of High Energy Physics] {10.1007/JHEP09(2012)082}, \href
  {http://adsabs.harvard.edu/abs/2012JHEP...09..082C} {9, 82}

\bibitem[\protect\citeauthoryear{{Catelan}, {Lucchin}, {Matarrese}  \&
  {Porciani}}{{Catelan} et~al.}{1998}]{Catelan1997}
{Catelan} P.,  {Lucchin} F.,  {Matarrese} S.,   {Porciani} C.,  1998, \mn@doi
  [\mnras] {10.1046/j.1365-8711.1998.01455.x}, \href
  {http://adsabs.harvard.edu/abs/1998MNRAS.297..692C} {297, 692}

\bibitem[\protect\citeauthoryear{{Catelan}, {Porciani}  \&
  {Kamionkowski}}{{Catelan} et~al.}{2000}]{Catelan2000}
{Catelan} P.,  {Porciani} C.,   {Kamionkowski} M.,  2000, \mn@doi [\mnras]
  {10.1046/j.1365-8711.2000.04023.x}, \href
  {http://adsabs.harvard.edu/abs/2000MNRAS.318L..39C} {318, L39}

\bibitem[\protect\citeauthoryear{{Chan} \& {Blot}}{{Chan} \&
  {Blot}}{2017}]{Chan2017}
{Chan} K.~C.,  {Blot} L.,  2017, \mn@doi [\prd] {10.1103/PhysRevD.96.023528},
  \href {http://adsabs.harvard.edu/abs/2017PhRvD..96b3528C} {96, 023528}

\bibitem[\protect\citeauthoryear{{Chan}, {Scoccimarro}  \& {Sheth}}{{Chan}
  et~al.}{2012}]{Chan2012}
{Chan} K.~C.,  {Scoccimarro} R.,   {Sheth} R.~K.,  2012, \mn@doi [\prd]
  {10.1103/PhysRevD.85.083509}, \href
  {http://adsabs.harvard.edu/abs/2012PhRvD..85h3509C} {85, 083509}

\bibitem[\protect\citeauthoryear{{Cole} \& {Kaiser}}{{Cole} \&
  {Kaiser}}{1989}]{Cole1989}
{Cole} S.,  {Kaiser} N.,  1989, \mn@doi [\mnras] {10.1093/mnras/237.4.1127},
  \href {http://adsabs.harvard.edu/abs/1989MNRAS.237.1127C} {237, 1127}

\bibitem[\protect\citeauthoryear{{Coles}}{{Coles}}{1993}]{Coles1993}
{Coles} P.,  1993, \mn@doi [\mnras] {10.1093/mnras/262.4.1065}, \href
  {http://adsabs.harvard.edu/abs/1993MNRAS.262.1065C} {262, 1065}

\bibitem[\protect\citeauthoryear{{Conroy}, {Wechsler}  \& {Kravtsov}}{{Conroy}
  et~al.}{2006}]{Conroy2006}
{Conroy} C.,  {Wechsler} R.~H.,   {Kravtsov} A.~V.,  2006, \mn@doi [\apj]
  {10.1086/503602}, \href {http://adsabs.harvard.edu/abs/2006ApJ...647..201C}
  {647, 201}

\bibitem[\protect\citeauthoryear{{Cooray} \& {Sheth}}{{Cooray} \&
  {Sheth}}{2002}]{Cooray2002}
{Cooray} A.,  {Sheth} R.,  2002, \mn@doi [\physrep]
  {10.1016/S0370-1573(02)00276-4}, \href
  {http://adsabs.harvard.edu/abs/2002PhR...372....1C} {372, 1}

\bibitem[\protect\citeauthoryear{{Creminelli}, {Nicolis}, {Senatore}, {Tegmark}
   \& {Zaldarriaga}}{{Creminelli} et~al.}{2006}]{Creminelli2005}
{Creminelli} P.,  {Nicolis} A.,  {Senatore} L.,  {Tegmark} M.,   {Zaldarriaga}
  M.,  2006, \mn@doi [\jcap] {10.1088/1475-7516/2006/05/004}, \href
  {http://adsabs.harvard.edu/abs/2006JCAP...05..004C} {5, 004}

\bibitem[\protect\citeauthoryear{{Crocce} \& {Scoccimarro}}{{Crocce} \&
  {Scoccimarro}}{2006a}]{Crocce2005a}
{Crocce} M.,  {Scoccimarro} R.,  2006a, \mn@doi [\prd]
  {10.1103/PhysRevD.73.063519}, \href
  {http://adsabs.harvard.edu/abs/2006PhRvD..73f3519C} {73, 063519}

\bibitem[\protect\citeauthoryear{{Crocce} \& {Scoccimarro}}{{Crocce} \&
  {Scoccimarro}}{2006b}]{Crocce2005b}
{Crocce} M.,  {Scoccimarro} R.,  2006b, \mn@doi [\prd]
  {10.1103/PhysRevD.73.063520}, \href
  {http://adsabs.harvard.edu/abs/2006PhRvD..73f3520C} {73, 063520}

\bibitem[\protect\citeauthoryear{{Crocce}, {Scoccimarro}  \&
  {Bernardeau}}{{Crocce} et~al.}{2012}]{Crocce2012}
{Crocce} M.,  {Scoccimarro} R.,   {Bernardeau} F.,  2012, \mn@doi [\mnras]
  {10.1111/j.1365-2966.2012.22127.x}, \href
  {http://adsabs.harvard.edu/abs/2012MNRAS.427.2537C} {427, 2537}

\bibitem[\protect\citeauthoryear{{Dai}, {Kamionkowski}, {Kovetz}, {Raccanelli}
  \& {Shiraishi}}{{Dai} et~al.}{2016}]{Dai2016}
{Dai} L.,  {Kamionkowski} M.,  {Kovetz} E.~D.,  {Raccanelli} A.,   {Shiraishi}
  M.,  2016, \mn@doi [\prd] {10.1103/PhysRevD.93.023507}, \href
  {http://adsabs.harvard.edu/abs/2016PhRvD..93b3507D} {93, 023507}

\bibitem[\protect\citeauthoryear{{Dalal}, {Dor{\'e}}, {Huterer}  \&
  {Shirokov}}{{Dalal} et~al.}{2008}]{Dalal2008}
{Dalal} N.,  {Dor{\'e}} O.,  {Huterer} D.,   {Shirokov} A.,  2008, \mn@doi
  [\prd] {10.1103/PhysRevD.77.123514}, \href
  {http://adsabs.harvard.edu/abs/2008PhRvD..77l3514D} {77, 123514}

\bibitem[\protect\citeauthoryear{{Dekel} \& {Lahav}}{{Dekel} \&
  {Lahav}}{1999}]{Dekel1998}
{Dekel} A.,  {Lahav} O.,  1999, \mn@doi [\apj] {10.1086/307428}, \href
  {http://adsabs.harvard.edu/abs/1999ApJ...520...24D} {520, 24}

\bibitem[\protect\citeauthoryear{{Desjacques} \& {Seljak}}{{Desjacques} \&
  {Seljak}}{2010}]{Desjacques2010}
{Desjacques} V.,  {Seljak} U.,  2010, \mn@doi [Advances in Astronomy]
  {10.1155/2010/908640}, \href
  {http://adsabs.harvard.edu/abs/2010AdAst2010E..89D} {2010, 908640}

\bibitem[\protect\citeauthoryear{{Desjacques}, {Seljak}  \&
  {Iliev}}{{Desjacques} et~al.}{2009}]{Desjacques2009}
{Desjacques} V.,  {Seljak} U.,   {Iliev} I.~T.,  2009, \mn@doi [\mnras]
  {10.1111/j.1365-2966.2009.14721.x}, \href
  {http://adsabs.harvard.edu/abs/2009MNRAS.396...85D} {396, 85}

\bibitem[\protect\citeauthoryear{{Desjacques}, {Jeong}  \&
  {Schmidt}}{{Desjacques} et~al.}{2011a}]{Desjacques2011b}
{Desjacques} V.,  {Jeong} D.,   {Schmidt} F.,  2011a, \mn@doi [\prd]
  {10.1103/PhysRevD.84.061301}, \href
  {http://adsabs.harvard.edu/abs/2011PhRvD..84f1301D} {84, 061301}

\bibitem[\protect\citeauthoryear{{Desjacques}, {Jeong}  \&
  {Schmidt}}{{Desjacques} et~al.}{2011b}]{Desjacques2011a}
{Desjacques} V.,  {Jeong} D.,   {Schmidt} F.,  2011b, \mn@doi [\prd]
  {10.1103/PhysRevD.84.063512}, \href
  {http://adsabs.harvard.edu/abs/2011PhRvD..84f3512D} {84, 063512}

\bibitem[\protect\citeauthoryear{{Desjacques}, {Jeong}  \&
  {Schmidt}}{{Desjacques} et~al.}{2016}]{Desjacques2016}
{Desjacques} V.,  {Jeong} D.,   {Schmidt} F.,  2016, preprint, \href
  {http://adsabs.harvard.edu/abs/2016arXiv161109787D} {} (\mn@eprint {arXiv}
  {1611.09787})

\bibitem[\protect\citeauthoryear{{Di Dio}, {Durrer}, {Marozzi}  \&
  {Montanari}}{{Di Dio} et~al.}{2016}]{DiDio2016}
{Di Dio} E.,  {Durrer} R.,  {Marozzi} G.,   {Montanari} F.,  2016, \mn@doi
  [\jcap] {10.1088/1475-7516/2016/01/016}, \href
  {http://adsabs.harvard.edu/abs/2016JCAP...01..016D} {1, 016}

\bibitem[\protect\citeauthoryear{{Elia}, {Kulkarni}, {Porciani}, {Pietroni}  \&
  {Matarrese}}{{Elia} et~al.}{2011}]{Elia2010}
{Elia} A.,  {Kulkarni} S.,  {Porciani} C.,  {Pietroni} M.,   {Matarrese} S.,
  2011, \mn@doi [\mnras] {10.1111/j.1365-2966.2011.18761.x}, \href
  {http://adsabs.harvard.edu/abs/2011MNRAS.416.1703E} {416, 1703}

\bibitem[\protect\citeauthoryear{{Feldman}, {Frieman}, {Fry}  \&
  {Scoccimarro}}{{Feldman} et~al.}{2001}]{Feldman2001}
{Feldman} H.~A.,  {Frieman} J.~A.,  {Fry} J.~N.,   {Scoccimarro} R.,  2001,
  \mn@doi [Physical Review Letters] {10.1103/PhysRevLett.86.1434}, \href
  {http://adsabs.harvard.edu/abs/2001PhRvL..86.1434F} {86, 1434}

\bibitem[\protect\citeauthoryear{{Font-Ribera}, {McDonald}, {Mostek}, {Reid},
  {Seo}  \& {Slosar}}{{Font-Ribera} et~al.}{2014}]{FontRibera2013}
{Font-Ribera} A.,  {McDonald} P.,  {Mostek} N.,  {Reid} B.~A.,  {Seo} H.-J.,
  {Slosar} A.,  2014, \mn@doi [\jcap] {10.1088/1475-7516/2014/05/023}, \href
  {http://adsabs.harvard.edu/abs/2014JCAP...05..023F} {5, 023}

\bibitem[\protect\citeauthoryear{{Fry}}{{Fry}}{1996}]{Fry1996}
{Fry} J.~N.,  1996, \mn@doi [\apjl] {10.1086/310006}, \href
  {http://adsabs.harvard.edu/abs/1996ApJ...461L..65F} {461, L65}

\bibitem[\protect\citeauthoryear{{Fry} \& {Gaztanaga}}{{Fry} \&
  {Gaztanaga}}{1993}]{Fry1993}
{Fry} J.~N.,  {Gaztanaga} E.,  1993, \mn@doi [\apj] {10.1086/173015}, \href
  {http://adsabs.harvard.edu/abs/1993ApJ...413..447F} {413, 447}

\bibitem[\protect\citeauthoryear{{Gangui}, {Lucchin}, {Matarrese}  \&
  {Mollerach}}{{Gangui} et~al.}{1994}]{Gangui1993}
{Gangui} A.,  {Lucchin} F.,  {Matarrese} S.,   {Mollerach} S.,  1994, \mn@doi
  [\apj] {10.1086/174421}, \href
  {http://adsabs.harvard.edu/abs/1994ApJ...430..447G} {430, 447}

\bibitem[\protect\citeauthoryear{{Giannantonio} \& {Porciani}}{{Giannantonio}
  \& {Porciani}}{2010}]{Giannantonio2010}
{Giannantonio} T.,  {Porciani} C.,  2010, \mn@doi [\prd]
  {10.1103/PhysRevD.81.063530}, \href
  {http://adsabs.harvard.edu/abs/2010PhRvD..81f3530G} {81, 063530}

\bibitem[\protect\citeauthoryear{{Giannantonio}, {Porciani}, {Carron}, {Amara}
  \& {Pillepich}}{{Giannantonio} et~al.}{2012}]{Giannantonio2012}
{Giannantonio} T.,  {Porciani} C.,  {Carron} J.,  {Amara} A.,   {Pillepich} A.,
   2012, \mn@doi [\mnras] {10.1111/j.1365-2966.2012.20604.x}, \href
  {http://adsabs.harvard.edu/abs/2012MNRAS.422.2854G} {422, 2854}

\bibitem[\protect\citeauthoryear{{Gil-Mar{\'{\i}}n}, {Wagner}, {Verde},
  {Porciani}  \& {Jimenez}}{{Gil-Mar{\'{\i}}n} et~al.}{2012}]{GilMarin2012b}
{Gil-Mar{\'{\i}}n} H.,  {Wagner} C.,  {Verde} L.,  {Porciani} C.,   {Jimenez}
  R.,  2012, \mn@doi [\jcap] {10.1088/1475-7516/2012/11/029}, \href
  {http://adsabs.harvard.edu/abs/2012JCAP...11..029G} {11, 029}

\bibitem[\protect\citeauthoryear{{Gil-Mar{\'{\i}}n}, {Wagner}, {Nore{\~n}a},
  {Verde}  \& {Percival}}{{Gil-Mar{\'{\i}}n} et~al.}{2014}]{GilMarin2014}
{Gil-Mar{\'{\i}}n} H.,  {Wagner} C.,  {Nore{\~n}a} J.,  {Verde} L.,
  {Percival} W.,  2014, \mn@doi [\jcap] {10.1088/1475-7516/2014/12/029}, \href
  {http://adsabs.harvard.edu/abs/2014JCAP...12..029G} {12, 029}

\bibitem[\protect\citeauthoryear{{Gil-Mar{\'{\i}}n}, {Percival}, {Verde},
  {Brownstein}, {Chuang}, {Kitaura}, {Rodr{\'{\i}}guez-Torres}  \&
  {Olmstead}}{{Gil-Mar{\'{\i}}n} et~al.}{2017}]{GilMarin2017}
{Gil-Mar{\'{\i}}n} H.,  {Percival} W.~J.,  {Verde} L.,  {Brownstein} J.~R.,
  {Chuang} C.-H.,  {Kitaura} F.-S.,  {Rodr{\'{\i}}guez-Torres} S.~A.,
  {Olmstead} M.~D.,  2017, \mn@doi [\mnras] {10.1093/mnras/stw2679}, \href
  {http://adsabs.harvard.edu/abs/2017MNRAS.465.1757G} {465, 1757}

\bibitem[\protect\citeauthoryear{{Hamaus}, {Seljak}  \& {Desjacques}}{{Hamaus}
  et~al.}{2011}]{Hamaus2011}
{Hamaus} N.,  {Seljak} U.,   {Desjacques} V.,  2011, \mn@doi [\prd]
  {10.1103/PhysRevD.84.083509}, \href
  {http://adsabs.harvard.edu/abs/2011PhRvD..84h3509H} {84, 083509}

\bibitem[\protect\citeauthoryear{{Hamilton}}{{Hamilton}}{1998}]{Hamilton1998}
{Hamilton} A.~J.~S.,  1998, in {Hamilton} D.,  ed.,  Astrophysics and Space
  Science Library Vol. 231, The Evolving Universe. p.~185 (\mn@eprint {}
  {astro-ph/9708102}), \mn@doi{10.1007/978-94-011-4960-0_17}

\bibitem[\protect\citeauthoryear{{Heavens}, {Matarrese}  \& {Verde}}{{Heavens}
  et~al.}{1998}]{Heavens1998}
{Heavens} A.~F.,  {Matarrese} S.,   {Verde} L.,  1998, \mn@doi [\mnras]
  {10.1046/j.1365-8711.1998.02052.x}, \href
  {http://adsabs.harvard.edu/abs/1998MNRAS.301..797H} {301, 797}

\bibitem[\protect\citeauthoryear{{Jackson}}{{Jackson}}{1972}]{Jackson1972}
{Jackson} J.~C.,  1972, \mn@doi [\mnras] {10.1093/mnras/156.1.1P}, \href
  {http://adsabs.harvard.edu/abs/1972MNRAS.156P...1J} {156, 1P}

\bibitem[\protect\citeauthoryear{{Jarvis}, {Bacon}, {Blake}, {Brown},
  {Lindsay}, {Raccanelli}, {Santos}  \& {Schwarz}}{{Jarvis}
  et~al.}{2015}]{Jarvis2015}
{Jarvis} M.,  {Bacon} D.,  {Blake} C.,  {Brown} M.,  {Lindsay} S.,
  {Raccanelli} A.,  {Santos} M.,   {Schwarz} D.~J.,  2015, Advancing
  Astrophysics with the Square Kilometre Array (AASKA14), \href
  {http://adsabs.harvard.edu/abs/2015aska.confE..18J} {p.~18}

\bibitem[\protect\citeauthoryear{{Jeong} \& {Komatsu}}{{Jeong} \&
  {Komatsu}}{2009}]{Jeong2009}
{Jeong} D.,  {Komatsu} E.,  2009, \mn@doi [\apj]
  {10.1088/0004-637X/703/2/1230}, \href
  {http://adsabs.harvard.edu/abs/2009ApJ...703.1230J} {703, 1230}

\bibitem[\protect\citeauthoryear{{Kaiser}}{{Kaiser}}{1984}]{Kaiser1984}
{Kaiser} N.,  1984, \mn@doi [\apjl] {10.1086/184341}, \href
  {http://adsabs.harvard.edu/abs/1984ApJ...284L...9K} {284, L9}

\bibitem[\protect\citeauthoryear{{Kaiser}}{{Kaiser}}{1987}]{Kaiser1987}
{Kaiser} N.,  1987, \mn@doi [\mnras] {10.1093/mnras/227.1.1}, \href
  {http://adsabs.harvard.edu/abs/1987MNRAS.227....1K} {227, 1}

\bibitem[\protect\citeauthoryear{{Karagiannis}, {Shanks}  \&
  {Ross}}{{Karagiannis} et~al.}{2014}]{Karagiannis2014}
{Karagiannis} D.,  {Shanks} T.,   {Ross} N.~P.,  2014, \mn@doi [\mnras]
  {10.1093/mnras/stu590}, \href
  {http://adsabs.harvard.edu/abs/2014MNRAS.441..486K} {441, 486}

\bibitem[\protect\citeauthoryear{{Komatsu} \& {Spergel}}{{Komatsu} \&
  {Spergel}}{2001}]{Komatsu2001}
{Komatsu} E.,  {Spergel} D.~N.,  2001, \mn@doi [\prd]
  {10.1103/PhysRevD.63.063002}, \href
  {http://adsabs.harvard.edu/abs/2001PhRvD..63f3002K} {63, 063002}

\bibitem[\protect\citeauthoryear{{Kovetz}, {Raccanelli}  \& {Rahman}}{{Kovetz}
  et~al.}{2017}]{Kovetz2016}
{Kovetz} E.~D.,  {Raccanelli} A.,   {Rahman} M.,  2017, \mn@doi [\mnras]
  {10.1093/mnras/stx691}, \href
  {http://adsabs.harvard.edu/abs/2017MNRAS.468.3650K} {468, 3650}

\bibitem[\protect\citeauthoryear{{LSST Science Collaboration} et~al.,}{{LSST
  Science Collaboration} et~al.}{2009}]{LSST2009}
{LSST Science Collaboration} et~al., 2009, preprint, \href
  {http://adsabs.harvard.edu/abs/2009arXiv0912.0201L} {} (\mn@eprint {arXiv}
  {0912.0201})

\bibitem[\protect\citeauthoryear{{Laureijs} et~al.,}{{Laureijs}
  et~al.}{2011}]{Laureijs2011}
{Laureijs} R.,  et~al., 2011, preprint, \href
  {http://adsabs.harvard.edu/abs/2011arXiv1110.3193L} {} (\mn@eprint {arXiv}
  {1110.3193})

\bibitem[\protect\citeauthoryear{{Lazanu}}{{Lazanu}}{2017}]{Lazanu2017b}
{Lazanu} A.,  2017, preprint, \href
  {http://adsabs.harvard.edu/abs/2017arXiv170909425L} {} (\mn@eprint {arXiv}
  {1709.09425})

\bibitem[\protect\citeauthoryear{{Lazanu}, {Giannantonio}, {Schmittfull}  \&
  {Shellard}}{{Lazanu} et~al.}{2016}]{Lazanu2015}
{Lazanu} A.,  {Giannantonio} T.,  {Schmittfull} M.,   {Shellard} E.~P.~S.,
  2016, \mn@doi [\prd] {10.1103/PhysRevD.93.083517}, \href
  {http://adsabs.harvard.edu/abs/2016PhRvD..93h3517L} {93, 083517}

\bibitem[\protect\citeauthoryear{{Lazanu}, {Giannantonio}, {Schmittfull}  \&
  {Shellard}}{{Lazanu} et~al.}{2017}]{Lazanu2017}
{Lazanu} A.,  {Giannantonio} T.,  {Schmittfull} M.,   {Shellard} E.~P.~S.,
  2017, \mn@doi [\prd] {10.1103/PhysRevD.95.083511}, \href
  {http://adsabs.harvard.edu/abs/2017PhRvD..95h3511L} {95, 083511}

\bibitem[\protect\citeauthoryear{{Leistedt}, {Peiris}  \& {Roth}}{{Leistedt}
  et~al.}{2014}]{Leistedt2014}
{Leistedt} B.,  {Peiris} H.~V.,   {Roth} N.,  2014, \mn@doi [Physical Review
  Letters] {10.1103/PhysRevLett.113.221301}, \href
  {http://adsabs.harvard.edu/abs/2014PhRvL.113v1301L} {113, 221301}

\bibitem[\protect\citeauthoryear{{Lewis}, {Challinor}  \& {Lasenby}}{{Lewis}
  et~al.}{2000}]{CAMB}
{Lewis} A.,  {Challinor} A.,   {Lasenby} A.,  2000, \mn@doi [\apj]
  {10.1086/309179}, \href {http://adsabs.harvard.edu/abs/2000ApJ...538..473L}
  {538, 473}

\bibitem[\protect\citeauthoryear{{Liguori} \& {Riotto}}{{Liguori} \&
  {Riotto}}{2008}]{Liguori2008}
{Liguori} M.,  {Riotto} A.,  2008, \mn@doi [\prd] {10.1103/PhysRevD.78.123004},
  \href {http://adsabs.harvard.edu/abs/2008PhRvD..78l3004L} {78, 123004}

\bibitem[\protect\citeauthoryear{{LoVerde}, {Miller}, {Shandera}  \&
  {Verde}}{{LoVerde} et~al.}{2008}]{LoVerde2007}
{LoVerde} M.,  {Miller} A.,  {Shandera} S.,   {Verde} L.,  2008, \mn@doi
  [\jcap] {10.1088/1475-7516/2008/04/014}, \href
  {http://adsabs.harvard.edu/abs/2008JCAP...04..014L} {4, 014}

\bibitem[\protect\citeauthoryear{{Ma}, {Hu}  \& {Huterer}}{{Ma}
  et~al.}{2006}]{Ma2006}
{Ma} Z.,  {Hu} W.,   {Huterer} D.,  2006, \mn@doi [\apj] {10.1086/497068},
  \href {http://adsabs.harvard.edu/abs/2006ApJ...636...21M} {636, 21}

\bibitem[\protect\citeauthoryear{{Mar{\'{\i}}n} et~al.,}{{Mar{\'{\i}}n}
  et~al.}{2013}]{Marin2013}
{Mar{\'{\i}}n} F.~A.,  et~al., 2013, \mn@doi [\mnras] {10.1093/mnras/stt520},
  \href {http://adsabs.harvard.edu/abs/2013MNRAS.432.2654M} {432, 2654}

\bibitem[\protect\citeauthoryear{{Matarrese} \& {Verde}}{{Matarrese} \&
  {Verde}}{2008}]{Matarrese2008}
{Matarrese} S.,  {Verde} L.,  2008, \mn@doi [\apjl] {10.1086/587840}, \href
  {http://adsabs.harvard.edu/abs/2008ApJ...677L..77M} {677, L77}

\bibitem[\protect\citeauthoryear{{Matsubara}}{{Matsubara}}{1999}]{Matsubara1999}
{Matsubara} T.,  1999, \mn@doi [\apj] {10.1086/307931}, \href
  {http://adsabs.harvard.edu/abs/1999ApJ...525..543M} {525, 543}

\bibitem[\protect\citeauthoryear{{McDonald}}{{McDonald}}{2006}]{McDonald2006}
{McDonald} P.,  2006, \mn@doi [\prd] {10.1103/PhysRevD.74.103512}, \href
  {http://adsabs.harvard.edu/abs/2006PhRvD..74j3512M} {74, 103512}

\bibitem[\protect\citeauthoryear{{McDonald}}{{McDonald}}{2008}]{McDonald2008}
{McDonald} P.,  2008, \mn@doi [\prd] {10.1103/PhysRevD.78.123519}, \href
  {http://adsabs.harvard.edu/abs/2008PhRvD..78l3519M} {78, 123519}

\bibitem[\protect\citeauthoryear{{McDonald} \& {Roy}}{{McDonald} \&
  {Roy}}{2009}]{McDonald2009}
{McDonald} P.,  {Roy} A.,  2009, \mn@doi [\jcap]
  {10.1088/1475-7516/2009/08/020}, \href
  {http://adsabs.harvard.edu/abs/2009JCAP...08..020M} {8, 020}

\bibitem[\protect\citeauthoryear{{M{\'e}nard}, {Scranton}, {Schmidt},
  {Morrison}, {Jeong}, {Budavari}  \& {Rahman}}{{M{\'e}nard}
  et~al.}{2013}]{Menard2013}
{M{\'e}nard} B.,  {Scranton} R.,  {Schmidt} S.,  {Morrison} C.,  {Jeong} D.,
  {Budavari} T.,   {Rahman} M.,  2013, preprint, \href
  {http://adsabs.harvard.edu/abs/2013arXiv1303.4722M} {} (\mn@eprint {arXiv}
  {1303.4722})

\bibitem[\protect\citeauthoryear{{Mirbabayi}, {Schmidt}  \&
  {Zaldarriaga}}{{Mirbabayi} et~al.}{2015}]{Mirbabayi2014}
{Mirbabayi} M.,  {Schmidt} F.,   {Zaldarriaga} M.,  2015, \mn@doi [\jcap]
  {10.1088/1475-7516/2015/07/030}, \href
  {http://adsabs.harvard.edu/abs/2015JCAP...07..030M} {7, 030}

\bibitem[\protect\citeauthoryear{{Mo} \& {White}}{{Mo} \&
  {White}}{1996}]{Mo1996a}
{Mo} H.~J.,  {White} S.~D.~M.,  1996, \mn@doi [\mnras]
  {10.1093/mnras/282.2.347}, \href
  {http://adsabs.harvard.edu/abs/1996MNRAS.282..347M} {282, 347}

\bibitem[\protect\citeauthoryear{{Mo}, {Jing}  \& {White}}{{Mo}
  et~al.}{1996}]{Mo1996b}
{Mo} H.~J.,  {Jing} Y.~P.,   {White} S.~D.~M.,  1996, \mn@doi [\mnras]
  {10.1093/mnras/282.3.1096}, \href
  {http://adsabs.harvard.edu/abs/1996MNRAS.282.1096M} {282, 1096}

\bibitem[\protect\citeauthoryear{{Moradinezhad Dizgah}, {Lee}, {Mu{\~n}oz}  \&
  {Dvorkin}}{{Moradinezhad Dizgah} et~al.}{2018}]{Moradinezhad2018}
{Moradinezhad Dizgah} A.,  {Lee} H.,  {Mu{\~n}oz} J.~B.,   {Dvorkin} C.,  2018,
  preprint, \href {http://adsabs.harvard.edu/abs/2018arXiv180107265M} {}
  (\mn@eprint {arXiv} {1801.07265})

\bibitem[\protect\citeauthoryear{{Newman}}{{Newman}}{2008}]{Newman2008}
{Newman} J.~A.,  2008, \mn@doi [\apj] {10.1086/589982}, \href
  {http://adsabs.harvard.edu/abs/2008ApJ...684...88N} {684, 88}

\bibitem[\protect\citeauthoryear{{Nikoloudakis}, {Shanks}  \&
  {Sawangwit}}{{Nikoloudakis} et~al.}{2013}]{Nikoloudakis2012}
{Nikoloudakis} N.,  {Shanks} T.,   {Sawangwit} U.,  2013, \mnras, \href
  {http://adsabs.harvard.edu/abs/2013MNRAS.429.2032N} {429, 2032}

\bibitem[\protect\citeauthoryear{{Orsi}, {Baugh}, {Lacey}, {Cimatti}, {Wang}
  \& {Zamorani}}{{Orsi} et~al.}{2010}]{Orsi2010}
{Orsi} A.,  {Baugh} C.~M.,  {Lacey} C.~G.,  {Cimatti} A.,  {Wang} Y.,
  {Zamorani} G.,  2010, \mn@doi [\mnras] {10.1111/j.1365-2966.2010.16585.x},
  \href {http://adsabs.harvard.edu/abs/2010MNRAS.405.1006O} {405, 1006}

\bibitem[\protect\citeauthoryear{{Padmanabhan} et~al.,}{{Padmanabhan}
  et~al.}{2007}]{Padmanabhan2007}
{Padmanabhan} N.,  et~al., 2007, \mnras, \href
  {http://adsabs.harvard.edu/abs/2007MNRAS.378..852P} {378, 852}

\bibitem[\protect\citeauthoryear{{Passaglia}, {Manzotti}  \&
  {Dodelson}}{{Passaglia} et~al.}{2017}]{Passaglia2017}
{Passaglia} S.,  {Manzotti} A.,   {Dodelson} S.,  2017, \mn@doi [\prd]
  {10.1103/PhysRevD.95.123508}, \href
  {http://adsabs.harvard.edu/abs/2017PhRvD..95l3508P} {95, 123508}

\bibitem[\protect\citeauthoryear{{Peacock} \& {Dodds}}{{Peacock} \&
  {Dodds}}{1994}]{Peacock1994}
{Peacock} J.~A.,  {Dodds} S.~J.,  1994, \mn@doi [\mnras]
  {10.1093/mnras/267.4.1020}, \href
  {http://adsabs.harvard.edu/abs/1994MNRAS.267.1020P} {267, 1020}

\bibitem[\protect\citeauthoryear{{Pietroni}}{{Pietroni}}{2008}]{Pietroni2008}
{Pietroni} M.,  2008, \mn@doi [\jcap] {10.1088/1475-7516/2008/10/036}, \href
  {http://adsabs.harvard.edu/abs/2008JCAP...10..036P} {10, 036}

\bibitem[\protect\citeauthoryear{{Planck Collaboration} et~al.,}{{Planck
  Collaboration} et~al.}{2016a}]{Planck2016_cosmopar}
{Planck Collaboration} et~al., 2016a, \mn@doi [\aap]
  {10.1051/0004-6361/201525830}, 594, A13

\bibitem[\protect\citeauthoryear{{Planck Collaboration} et~al.,}{{Planck
  Collaboration} et~al.}{2016b}]{Planck_PNG2016}
{Planck Collaboration} et~al., 2016b, \mn@doi [\aap]
  {10.1051/0004-6361/201525836}, \href
  {http://adsabs.harvard.edu/abs/2016A%26A...594A..17P} {594, A17}

\bibitem[\protect\citeauthoryear{{Raccanelli} et~al.,}{{Raccanelli}
  et~al.}{2012}]{Raccanelli2011}
{Raccanelli} A.,  et~al., 2012, \mn@doi [\mnras]
  {10.1111/j.1365-2966.2012.20634.x}, \href
  {http://adsabs.harvard.edu/abs/2012MNRAS.424..801R} {424, 801}

\bibitem[\protect\citeauthoryear{{Raccanelli}, {Montanari}, {Bertacca},
  {Dor{\'e}}  \& {Durrer}}{{Raccanelli} et~al.}{2016}]{Raccanelli2016}
{Raccanelli} A.,  {Montanari} F.,  {Bertacca} D.,  {Dor{\'e}} O.,   {Durrer}
  R.,  2016, \mn@doi [\jcap] {10.1088/1475-7516/2016/05/009}, \href
  {http://adsabs.harvard.edu/abs/2016JCAP...05..009R} {5, 009}

\bibitem[\protect\citeauthoryear{{Raccanelli}, {Shiraishi}, {Bartolo},
  {Bertacca}, {Liguori}, {Matarrese}, {Norris}  \& {Parkinson}}{{Raccanelli}
  et~al.}{2017}]{Raccanelli2017}
{Raccanelli} A.,  {Shiraishi} M.,  {Bartolo} N.,  {Bertacca} D.,  {Liguori} M.,
   {Matarrese} S.,  {Norris} R.~P.,   {Parkinson} D.,  2017, \mn@doi [Physics
  of the Dark Universe] {10.1016/j.dark.2016.10.006}, \href
  {http://adsabs.harvard.edu/abs/2017PDU....15...35R} {15, 35}

\bibitem[\protect\citeauthoryear{{Rahman}, {Mendez}, {M{\'e}nard}, {Scranton},
  {Schmidt}, {Morrison}  \& {Budav{\'a}ri}}{{Rahman} et~al.}{2016}]{Rahman2016}
{Rahman} M.,  {Mendez} A.~J.,  {M{\'e}nard} B.,  {Scranton} R.,  {Schmidt}
  S.~J.,  {Morrison} C.~B.,   {Budav{\'a}ri} T.,  2016, \mn@doi [\mnras]
  {10.1093/mnras/stw981}, \href
  {http://adsabs.harvard.edu/abs/2016MNRAS.460..163R} {460, 163}

\bibitem[\protect\citeauthoryear{{Salopek} \& {Bond}}{{Salopek} \&
  {Bond}}{1990}]{Salopek1990}
{Salopek} D.~S.,  {Bond} J.~R.,  1990, \mn@doi [\prd]
  {10.1103/PhysRevD.42.3936}, \href
  {http://adsabs.harvard.edu/abs/1990PhRvD..42.3936S} {42, 3936}

\bibitem[\protect\citeauthoryear{{Schmid} \& {Hui}}{{Schmid} \&
  {Hui}}{2013}]{Schmidt2012}
{Schmid} F.,  {Hui} L.,  2013, \mn@doi [Physical Review Letters]
  {10.1103/PhysRevLett.110.011301}, \href
  {http://adsabs.harvard.edu/abs/2013PhRvL.110a1301S} {110, 011301}

\bibitem[\protect\citeauthoryear{{Schmidt}}{{Schmidt}}{2016}]{Schmidt2015}
{Schmidt} F.,  2016, \mn@doi [\prd] {10.1103/PhysRevD.93.063512}, \href
  {http://adsabs.harvard.edu/abs/2016PhRvD..93f3512S} {93, 063512}

\bibitem[\protect\citeauthoryear{{Schmidt} \& {Kamionkowski}}{{Schmidt} \&
  {Kamionkowski}}{2010}]{Schmidt2010}
{Schmidt} F.,  {Kamionkowski} M.,  2010, \mn@doi [\prd]
  {10.1103/PhysRevD.82.103002}, \href
  {http://adsabs.harvard.edu/abs/2010PhRvD..82j3002S} {82, 103002}

\bibitem[\protect\citeauthoryear{{Schmidt}, {Jeong}  \& {Desjacques}}{{Schmidt}
  et~al.}{2013}]{Schmidt2013}
{Schmidt} F.,  {Jeong} D.,   {Desjacques} V.,  2013, \mn@doi [\prd]
  {10.1103/PhysRevD.88.023515}, \href
  {http://adsabs.harvard.edu/abs/2013PhRvD..88b3515S} {88, 023515}

\bibitem[\protect\citeauthoryear{{Schneider}, {Knox}, {Zhan}  \&
  {Connolly}}{{Schneider} et~al.}{2006}]{Schneider2006}
{Schneider} M.,  {Knox} L.,  {Zhan} H.,   {Connolly} A.,  2006, \mn@doi [\apj]
  {10.1086/507675}, \href {http://adsabs.harvard.edu/abs/2006ApJ...651...14S}
  {651, 14}

\bibitem[\protect\citeauthoryear{{Scoccimarro}, {Couchman}  \&
  {Frieman}}{{Scoccimarro} et~al.}{1999}]{Scoccimarro1999}
{Scoccimarro} R.,  {Couchman} H.~M.~P.,   {Frieman} J.~A.,  1999, \mn@doi
  [\apj] {10.1086/307220}, \href
  {http://adsabs.harvard.edu/abs/1999ApJ...517..531S} {517, 531}

\bibitem[\protect\citeauthoryear{{Scoccimarro}, {Sheth}, {Hui}  \&
  {Jain}}{{Scoccimarro} et~al.}{2001a}]{Scoccimarro2001}
{Scoccimarro} R.,  {Sheth} R.~K.,  {Hui} L.,   {Jain} B.,  2001a, \mn@doi
  [\apj] {10.1086/318261}, \href
  {http://adsabs.harvard.edu/abs/2001ApJ...546...20S} {546, 20}

\bibitem[\protect\citeauthoryear{{Scoccimarro}, {Feldman}, {Fry}  \&
  {Frieman}}{{Scoccimarro} et~al.}{2001b}]{Scoccimarro2001b}
{Scoccimarro} R.,  {Feldman} H.~A.,  {Fry} J.~N.,   {Frieman} J.~A.,  2001b,
  \mn@doi [\apj] {10.1086/318284}, \href
  {http://adsabs.harvard.edu/abs/2001ApJ...546..652S} {546, 652}

\bibitem[\protect\citeauthoryear{{Scoccimarro}, {Sefusatti}  \&
  {Zaldarriaga}}{{Scoccimarro} et~al.}{2004}]{Scoccimarro2003}
{Scoccimarro} R.,  {Sefusatti} E.,   {Zaldarriaga} M.,  2004, \mn@doi [\prd]
  {10.1103/PhysRevD.69.103513}, \href
  {http://adsabs.harvard.edu/abs/2004PhRvD..69j3513S} {69, 103513}

\bibitem[\protect\citeauthoryear{{Scoccimarro}, {Hui}, {Manera}  \&
  {Chan}}{{Scoccimarro} et~al.}{2012}]{Scoccimarro2011}
{Scoccimarro} R.,  {Hui} L.,  {Manera} M.,   {Chan} K.~C.,  2012, \mn@doi
  [\prd] {10.1103/PhysRevD.85.083002}, \href
  {http://adsabs.harvard.edu/abs/2012PhRvD..85h3002S} {85, 083002}

\bibitem[\protect\citeauthoryear{{Sefusatti}}{{Sefusatti}}{2009}]{Sefusatti2009}
{Sefusatti} E.,  2009, \mn@doi [\prd] {10.1103/PhysRevD.80.123002}, \href
  {http://adsabs.harvard.edu/abs/2009PhRvD..80l3002S} {80, 123002}

\bibitem[\protect\citeauthoryear{{Sefusatti} \& {Komatsu}}{{Sefusatti} \&
  {Komatsu}}{2007}]{Sefusatti2007}
{Sefusatti} E.,  {Komatsu} E.,  2007, \mn@doi [\prd]
  {10.1103/PhysRevD.76.083004}, \href
  {http://adsabs.harvard.edu/abs/2007PhRvD..76h3004S} {76, 083004}

\bibitem[\protect\citeauthoryear{{Sefusatti}, {Crocce}, {Pueblas}  \&
  {Scoccimarro}}{{Sefusatti} et~al.}{2006}]{Sefusatti2006}
{Sefusatti} E.,  {Crocce} M.,  {Pueblas} S.,   {Scoccimarro} R.,  2006, \mn@doi
  [\prd] {10.1103/PhysRevD.74.023522}, \href
  {http://adsabs.harvard.edu/abs/2006PhRvD..74b3522S} {74, 023522}

\bibitem[\protect\citeauthoryear{{Sefusatti}, {Crocce}  \&
  {Desjacques}}{{Sefusatti} et~al.}{2012}]{Sefusatti2012}
{Sefusatti} E.,  {Crocce} M.,   {Desjacques} V.,  2012, \mn@doi [\mnras]
  {10.1111/j.1365-2966.2012.21271.x}, \href
  {http://adsabs.harvard.edu/abs/2012MNRAS.425.2903S} {425, 2903}

\bibitem[\protect\citeauthoryear{{Seljak}}{{Seljak}}{2009}]{Seljak2009}
{Seljak} U.,  2009, \mn@doi [Physical Review Letters]
  {10.1103/PhysRevLett.102.021302}, \href
  {http://adsabs.harvard.edu/abs/2009PhRvL.102b1302S} {102, 021302}

\bibitem[\protect\citeauthoryear{{Senatore}}{{Senatore}}{2015}]{Senatore2014}
{Senatore} L.,  2015, \mn@doi [\jcap] {10.1088/1475-7516/2015/11/007}, \href
  {http://adsabs.harvard.edu/abs/2015JCAP...11..007S} {11, 007}

\bibitem[\protect\citeauthoryear{{Senatore}, {Smith}  \&
  {Zaldarriaga}}{{Senatore} et~al.}{2010}]{Senatore2009}
{Senatore} L.,  {Smith} K.~M.,   {Zaldarriaga} M.,  2010, \mn@doi [\jcap]
  {10.1088/1475-7516/2010/01/028}, \href
  {http://adsabs.harvard.edu/abs/2010JCAP...01..028S} {1, 028}

\bibitem[\protect\citeauthoryear{{Seo} \& {Eisenstein}}{{Seo} \&
  {Eisenstein}}{2003}]{Seo_2003}
{Seo} H.-J.,  {Eisenstein} D.~J.,  2003, \mn@doi [\apj] {10.1086/379122}, \href
  {http://adsabs.harvard.edu/abs/2003ApJ...598..720S} {598, 720}

\bibitem[\protect\citeauthoryear{{Sheth} \& {Tormen}}{{Sheth} \&
  {Tormen}}{1999}]{ShethTormen1999}
{Sheth} R.~K.,  {Tormen} G.,  1999, \mn@doi [\mnras]
  {10.1046/j.1365-8711.1999.02692.x}, \href
  {http://adsabs.harvard.edu/abs/1999MNRAS.308..119S} {308, 119}

\bibitem[\protect\citeauthoryear{{Slosar}, {Hirata}, {Seljak}, {Ho}  \&
  {Padmanabhan}}{{Slosar} et~al.}{2008}]{Slosar2008}
{Slosar} A.,  {Hirata} C.,  {Seljak} U.,  {Ho} S.,   {Padmanabhan} N.,  2008,
  \mn@doi [\jcap] {10.1088/1475-7516/2008/08/031}, \href
  {http://adsabs.harvard.edu/abs/2008JCAP...08..031S} {8, 031}

\bibitem[\protect\citeauthoryear{{Smith} et~al.,}{{Smith}
  et~al.}{2003}]{Smith2003}
{Smith} R.~E.,  et~al., 2003, \mn@doi [\mnras]
  {10.1046/j.1365-8711.2003.06503.x}, \href
  {http://adsabs.harvard.edu/abs/2003MNRAS.341.1311S} {341, 1311}

\bibitem[\protect\citeauthoryear{{Song}, {Taruya}  \& {Oka}}{{Song}
  et~al.}{2015}]{Song2015}
{Song} Y.-S.,  {Taruya} A.,   {Oka} A.,  2015, \mn@doi [\jcap]
  {10.1088/1475-7516/2015/08/007}, \href
  {http://adsabs.harvard.edu/abs/2015JCAP...08..007S} {8, 007}

\bibitem[\protect\citeauthoryear{{Takahashi}, {Sato}, {Nishimichi}, {Taruya}
  \& {Oguri}}{{Takahashi} et~al.}{2012}]{Takahashi2012}
{Takahashi} R.,  {Sato} M.,  {Nishimichi} T.,  {Taruya} A.,   {Oguri} M.,
  2012, \mn@doi [\apj] {10.1088/0004-637X/761/2/152}, \href
  {http://adsabs.harvard.edu/abs/2012ApJ...761..152T} {761, 152}

\bibitem[\protect\citeauthoryear{{Taruya} \& {Soda}}{{Taruya} \&
  {Soda}}{1999}]{Taruya1998}
{Taruya} A.,  {Soda} J.,  1999, \mn@doi [\apj] {10.1086/307612}, \href
  {http://adsabs.harvard.edu/abs/1999ApJ...522...46T} {522, 46}

\bibitem[\protect\citeauthoryear{{Tellarini}, {Ross}, {Tasinato}  \&
  {Wands}}{{Tellarini} et~al.}{2016}]{Tellarini2016}
{Tellarini} M.,  {Ross} A.~J.,  {Tasinato} G.,   {Wands} D.,  2016, \mn@doi
  [\jcap] {10.1088/1475-7516/2016/06/014}, \href
  {http://adsabs.harvard.edu/abs/2016JCAP...06..014T} {6, 014}

\bibitem[\protect\citeauthoryear{{Tinker}, {Weinberg}, {Zheng}  \&
  {Zehavi}}{{Tinker} et~al.}{2005}]{Tinker2005}
{Tinker} J.~L.,  {Weinberg} D.~H.,  {Zheng} Z.,   {Zehavi} I.,  2005, \mn@doi
  [\apj] {10.1086/432084}, \href
  {http://adsabs.harvard.edu/abs/2005ApJ...631...41T} {631, 41}

\bibitem[\protect\citeauthoryear{{Umeh}, {Jolicoeur}, {Maartens}  \&
  {Clarkson}}{{Umeh} et~al.}{2017}]{Umeh2017}
{Umeh} O.,  {Jolicoeur} S.,  {Maartens} R.,   {Clarkson} C.,  2017, \mn@doi
  [\jcap] {10.1088/1475-7516/2017/03/034}, \href
  {http://adsabs.harvard.edu/abs/2017JCAP...03..034U} {3, 034}

\bibitem[\protect\citeauthoryear{{Verde} \& {Matarrese}}{{Verde} \&
  {Matarrese}}{2009}]{Verde2009}
{Verde} L.,  {Matarrese} S.,  2009, \mn@doi [\apjl]
  {10.1088/0004-637X/706/1/L91}, \href
  {http://adsabs.harvard.edu/abs/2009ApJ...706L..91V} {706, L91}

\bibitem[\protect\citeauthoryear{{Verde}, {Heavens}, {Matarrese}  \&
  {Moscardini}}{{Verde} et~al.}{1998}]{Verde1998}
{Verde} L.,  {Heavens} A.~F.,  {Matarrese} S.,   {Moscardini} L.,  1998,
  \mn@doi [\mnras] {10.1046/j.1365-8711.1998.01937.x}, \href
  {http://adsabs.harvard.edu/abs/1998MNRAS.300..747V} {300, 747}

\bibitem[\protect\citeauthoryear{{Verde}, {Wang}, {Heavens}  \&
  {Kamionkowski}}{{Verde} et~al.}{2000a}]{Verde1999}
{Verde} L.,  {Wang} L.,  {Heavens} A.~F.,   {Kamionkowski} M.,  2000a, \mn@doi
  [\mnras] {10.1046/j.1365-8711.2000.03191.x}, \href
  {http://adsabs.harvard.edu/abs/2000MNRAS.313..141V} {313, 141}

\bibitem[\protect\citeauthoryear{{Verde}, {Heavens}  \& {Matarrese}}{{Verde}
  et~al.}{2000b}]{Verde2000}
{Verde} L.,  {Heavens} A.~F.,   {Matarrese} S.,  2000b, \mn@doi [\mnras]
  {10.1046/j.1365-8711.2000.03774.x}, \href
  {http://adsabs.harvard.edu/abs/2000MNRAS.318..584V} {318, 584}

\bibitem[\protect\citeauthoryear{{Verde}, {Heavens}, {Percival}  \&
  {Matarrese}}{{Verde} et~al.}{2002}]{Verde2002}
{Verde} L.,  {Heavens} A.~F.,  {Percival} W.~J.,   {Matarrese} S.,  2002, ArXiv
  Astrophysics e-prints, \href
  {http://adsabs.harvard.edu/abs/2002astro.ph.12311V} {}

\bibitem[\protect\citeauthoryear{{Wagner}, {Verde}  \& {Boubekeur}}{{Wagner}
  et~al.}{2010}]{Wagner2010}
{Wagner} C.,  {Verde} L.,   {Boubekeur} L.,  2010, \mn@doi [\jcap]
  {10.1088/1475-7516/2010/10/022}, \href
  {http://adsabs.harvard.edu/abs/2010JCAP...10..022W} {10, 022}

\bibitem[\protect\citeauthoryear{{Wagner}, {Schmidt}, {Chiang}  \&
  {Komatsu}}{{Wagner} et~al.}{2015}]{Wagner2015}
{Wagner} C.,  {Schmidt} F.,  {Chiang} C.-T.,   {Komatsu} E.,  2015, \mn@doi
  [\jcap] {10.1088/1475-7516/2015/08/042}, \href
  {http://adsabs.harvard.edu/abs/2015JCAP...08..042W} {8, 042}

\bibitem[\protect\citeauthoryear{{Wittman}, {Tyson}, {Kirkman}, {Dell'Antonio}
  \& {Bernstein}}{{Wittman} et~al.}{2000}]{Wittman2000}
{Wittman} D.~M.,  {Tyson} J.~A.,  {Kirkman} D.,  {Dell'Antonio} I.,
  {Bernstein} G.,  2000, \mn@doi [\nat] {10.1038/35012001}, \href
  {http://adsabs.harvard.edu/abs/2000Natur.405..143W} {405, 143}

\bibitem[\protect\citeauthoryear{{Xia}, {Bonaldi}, {Baccigalupi}, {De Zotti},
  {Matarrese}, {Verde}  \& {Viel}}{{Xia} et~al.}{2010a}]{Xia2010b}
{Xia} J.-Q.,  {Bonaldi} A.,  {Baccigalupi} C.,  {De Zotti} G.,  {Matarrese} S.,
   {Verde} L.,   {Viel} M.,  2010a, \mn@doi [\jcap]
  {10.1088/1475-7516/2010/08/013}, \href
  {http://adsabs.harvard.edu/abs/2010JCAP...08..013X} {8, 013}

\bibitem[\protect\citeauthoryear{{Xia}, {Viel}, {Baccigalupi}, {De Zotti},
  {Matarrese}  \& {Verde}}{{Xia} et~al.}{2010b}]{Xia2010a}
{Xia} J.-Q.,  {Viel} M.,  {Baccigalupi} C.,  {De Zotti} G.,  {Matarrese} S.,
  {Verde} L.,  2010b, \apjl, \href
  {http://adsabs.harvard.edu/abs/2010ApJ...717L..17X} {717, L17}

\bibitem[\protect\citeauthoryear{{Xia}, {Viel}, {Baccigalupi}, {De Zotti},
  {Matarrese}  \& {Verde}}{{Xia} et~al.}{2010c}]{Xia2010c}
{Xia} J.-Q.,  {Viel} M.,  {Baccigalupi} C.,  {De Zotti} G.,  {Matarrese} S.,
  {Verde} L.,  2010c, \mn@doi [\apjl] {10.1088/2041-8205/717/1/L17}, \href
  {http://adsabs.harvard.edu/abs/2010ApJ...717L..17X} {717, L17}

\bibitem[\protect\citeauthoryear{{Xia}, {Baccigalupi}, {Matarrese}, {Verde}  \&
  {Viel}}{{Xia} et~al.}{2011}]{Xia2011}
{Xia} J.-Q.,  {Baccigalupi} C.,  {Matarrese} S.,  {Verde} L.,   {Viel} M.,
  2011, \jcap, \href {http://adsabs.harvard.edu/abs/2011JCAP...08..033X} {8,
  33}

\bibitem[\protect\citeauthoryear{{Yamauchi}, {Yokoyama}  \&
  {Takahashi}}{{Yamauchi} et~al.}{2017}]{Yamauchi2017}
{Yamauchi} D.,  {Yokoyama} S.,   {Takahashi} K.,  2017, \mn@doi [\prd]
  {10.1103/PhysRevD.95.063530}, \href
  {http://adsabs.harvard.edu/abs/2017PhRvD..95f3530Y} {95, 063530}

\bibitem[\protect\citeauthoryear{{Zhan}}{{Zhan}}{2006}]{Zhan2006}
{Zhan} H.,  2006, \mn@doi [\jcap] {10.1088/1475-7516/2006/08/008}, \href
  {http://adsabs.harvard.edu/abs/2006JCAP...08..008Z} {8, 008}

\bibitem[\protect\citeauthoryear{{Zhan} \& {Tyson}}{{Zhan} \&
  {Tyson}}{2017}]{Zhan2017}
{Zhan} H.,  {Tyson} J.~A.,  2017, preprint, \href
  {http://adsabs.harvard.edu/abs/2017arXiv170706948Z} {} (\mn@eprint {arXiv}
  {1707.06948})

\makeatother
\end{thebibliography}

   \appendix
   
   \section{Standard Perturbation Theory} \label{StanPT}
   
   In this section we review the main results of standard perturbation theory (SPT). For a complete description of the subject the reader is referred to the review of \citet{Bernardeau2002} and references therein. The equations of motion (EOM) for the evolution of the cosmic fluid in an expanding FLRW metric are solved under the assumptions of a pressureless, self-gravitating, single component perfect fluid up to a quasi non-linear scale denoting the break down of SPT. The formulation of the EOM is done in terms of the matter overdensity field $\delta$ and the velocity field, characterized by its divergence $\theta=\nabla \cdot \mathbf v$, while the vorticity degree of freedom is neglected. Such an assumption is valid as long as the stress tensor $\sigma_{ij}\approx0$. This generally does not hold on small scales, where multi-streaming and shocks generate vorticity.
 
 The non-linear EOM in Fourier space are
 
 \begin{align}
  \frac{\partial\delta(\bk,\tau)}{\partial\tau}+\theta(\bk,\tau)&=-\int \frac{d^3q_{1}}{(2\pi)^3}\int \frac{d^3q_{2}}{(2\pi)^3} \delta_{D}(\bk -\bq_{12}) \nonumber \\  &\times\alpha(\bq_1,\bq_2)\theta(\bq_1,\tau)\delta(\bq_2,\tau) \, ,
 \end{align}
 
 \begin{align} 
  &\frac{\partial\theta(\bk,\tau)}{\partial\tau}+\mathcal{H}(\tau)\theta(\bk,\tau)+\frac{3}{2}\Omega_{m}\mathcal{H}^2(\tau)\delta(\bk,\tau)= \nonumber \\ &-\int \frac{d^3q_{1}}{(2\pi)^3} \int \frac{d^3q_{2}}{(2\pi)^3} \delta_{D}(\bk -\bq_{12})\beta(\bq_1,\bq_2)\theta(\bq_1,\tau)\theta(\bq_2,\tau)\, ,
 \end{align}

 \noindent where $\tau=\int dt/a$ is the conformal time and $\mathcal{H}=d\ln a/d\tau=H a$ is the conformal expansion rate, with $H$ being the Hubble constant. The coupling between different Fourier modes due to the non-linear evolution is encoded in the terms
 
 \begin{align}
 &\alpha(\bk_i,\bk_j)=\bk_{ij}\cdot\bk_i/k_i^2 \, , \\
 &\beta(\bk_i,\bk_j)=\bk_{ij}^2(\bk_i\cdot\bk_j)/(2k_i^2k_j^2) \, ,
 \end{align}

 \noindent where $\bk_{ij}=\bk_i+\bk_j$. They originate from the cosmological continuity and Euler equations in Fourier space. Such coupled differential equations can be solved by a perturbative solution of the form  
 \begin{align}
  \delta(\bk,\tau) &=\sum_{n=1}^{\infty}D_{n}(\tau)\delta^{(n)}(\bk,z)\, , \\
  \theta(\bk,\tau) &=-\mathcal{H}(\tau)f\sum_{n=1}^{\infty}D_{n}(\tau)\theta^{(n)}(\bk,z) \,.
 \end{align}

  \noindent Hereafter we drop the \textit{m} subscript from the matter density symbol. Each \textit{n}-th order term can be written in terms of the linear Gaussian fluctuation $\delta^{(1)}(\bk)$
  
  \begin{align}\label{eq:deltapt}
   &\delta(\bk,\tau)=\sum_{n=1}^{\infty}D^n(\tau) \int \frac{d^3k_{1}}{(2\pi)^3}\ldots\int \frac{d^3k_{n-1}}{(2\pi)^3}\int \frac{d^3k_{n}}{(2\pi)^3} \nonumber \\ &\times \delta_{D}(\bk-\bk_1-\ldots-\bk_{n})F_n(\bk_1,\ldots,\bk_n)\delta^{(1)}(\bk_1)\ldots\delta^{(1)}(\bk_n), \\
%
   &\theta(\bk,\tau)=-f\mathcal{H}\sum_{n=1}^{\infty}D^n(\tau) \int \frac{d^3k_{1}}{(2\pi)^3}\ldots\int \frac{d^3k_{n-1}}{(2\pi)^3}\int \frac{d^3k_{n}}{(2\pi)^3} \nonumber \\ &\times \delta_{D}(\bk-\bk_1-\ldots-\bk_{n})G_n(\bk_1,\ldots,\bk_n)\delta^{(1)}(\bk_1)\ldots\delta^{(1)}(\bk_n),
  \end{align}
  
  \noindent where $f=d\mathrm{ln}D/d\mathrm{ln}a$ is the growth rate, which can be approximated by $f\approx\Omega_{m}^{4/7}(z)$. The two symmetrized kernels $F_n^s(\bk_1,\ldots\bk_n)$ and $G_n^s(\bk_1,\ldots\bk_n)$ incorporate the non-linear Fourier mode coupling. The first-order solutions of the above equations are given by $\delta(\bk,\tau)=D(\tau)\delta^{(1)}(\bk)$ and $\theta^{(1)}(\bk,\tau)=-\mathcal{H}(\tau)fD(\tau)\delta^{(1)}(\bk)$ and hence they are completely characterized by the linear term $\delta^{(1)}(\bk)$. Below we list the symmetrised kernels for the density field up to the third perturbative order:

  \begin{align}
   &F_1(\bk)=1, \\
   &F_2^{(s)}(\bk_1,\bk_2)=\frac{5}{7}+\frac{1}{2}\frac{\bk_1\cdot\bk_2}{k_1k_2}\left(\frac{k_1}{k_2}+\frac{k_2}{k_1} \right)+\frac{2}{7}\left(\frac{\bk_1\cdot\bk_2}{k_1k_2}\right)^2, \label{eq:F2kernel} \\ 
   &F_3^{(s)}(\bk_1,\bk_2,\bk_3)=\frac{7}{54}\Big[F_2^{(s)}(\bk_1,\bk_2)\alpha(\bk_3,\bk_{12}) \nonumber \\
   &+F_2^{(s)}(\bk_2,\bk_3)\alpha(\bk_1,\bk_{23})+F_2^{(s)}(\bk_3,\bk_1)\alpha(\bk_2,\bk_{31}) \nonumber \\
   &+G_2^{(s)}(\bk_1,\bk_2)\alpha(\bk_{12},\bk_3)+G_2^{(s)}(\bk_2,\bk_3)\alpha(\bk_{23},\bk_1) \nonumber \\
   &+G_2^{(s)}(\bk_3,\bk_1)\alpha(\bk_{31},\bk_2)\Big]+\frac{2}{27}\Big[G_2^{(s)}(\bk_1,\bk_2)\beta(\bk_{12},\bk_3) \nonumber \\
   &+G_2^{(s)}(\bk_2,\bk_3)\beta(\bk_{23},\bk_1)+G_2^{(s)}(\bk_3,\bk_1)\beta(\bk_{31},\bk_2)\Big], \\ \label{eq:F3kernel}
   \nonumber \\
   &F_4^{(s)}(\bk_1,\bk_2,\bk_3,\bk_4)=\frac{1}{792} \Big[2 G_2^{(s)}(\bk_1,\bk_4) \nonumber \\
   &\big[18F_2^{(s)}(\bk_2,\bk_3)\alpha(\bk_1+\bk_4,\bk_2+\bk_3) \nonumber \\
   &+8G_2^{(s)}(\bk_2,\bk_3)\beta(\bk_2+\bk_3,\bk_1+\bk_4)\big]\nonumber \\
   &+2G_2^{(s)}(\bk_1,\bk_3)\big[18F_2^{(s)}(\bk_2,\bk_4)\alpha(\bk_1+\bk_3,\bk_2+\bk_4) \nonumber \\
   &+8G_2^{(s)}(\bk_2,\bk_4)\beta(\bk_1+\bk_3,\bk_2+\bk_4)\big]\nonumber \\
   &+2G_2^{(s)}(\bk_1,\bk_2)\big[18F_2^{(s)}(\bk_3,\bk_4)\alpha(\bk_1+\bk_2,\bk_3+\bk_4) \nonumber \\
   &+8G_2^{(s)}(\bk_3,\bk_4)\beta(\bk_1+\bk_2,\bk_3+\bk_4)\big]\nonumber \\
   &+6G_3^{(s)}(\bk_1,\bk_2,\bk_3) \big[9\alpha(\bk_1+\bk_2+\bk_3,\bk_4) \nonumber \\
   &+4\beta(\bk_1+\bk_2+\bk_3,\bk_4)\big]\nonumber \\
   &+6G_3^{(s)}(\bk_1,\bk_2,\bk_4) \big[9\alpha(\bk_1+\bk_2+\bk_4,\bk_3) \nonumber \\
   &+4\beta(\bk_3,\bk_1+\bk_2+\bk_4)\big]\nonumber \\
   &+6G_3^{(s)}(\bk_1,\bk_3,\bk_4) \big[9\alpha(\bk_1+\bk_3+\bk_4,\bk_2)\nonumber \\
   &+4\beta(\bk_2,\bk_1+\bk_3+\bk_4)\big]\nonumber \\
   &+6G_3^{(s)}(\bk_2,\bk_3,\bk_4) \big[9\alpha(\bk_2+\bk_3+\bk_4,\bk_1)\nonumber \\
   &+4\beta(\bk_1,\bk_2+\bk_3+\bk_4)\big]\nonumber \\
   &+36F_2^{(s)}(\bk_1,\bk_4) G_2^{(s)}(\bk_2,\bk_3)\alpha(\bk_2+\bk_3,\bk_1+\bk_4)\nonumber \\
   &+36F_2^{(s)}(\bk_1,\bk_3) G_2^{(s)}(\bk_2,\bk_4)\alpha(\bk_2+\bk_4,\bk_1+\bk_3)\nonumber \\
   &+36F_2^{(s)}(\bk_1,\bk_2) G_2^{(s)}(\bk_3,\bk_4)\alpha(\bk_3+\bk_4,\bk_1+\bk_2)\nonumber \\
   &+54F_3^{(s)}(\bk_1,\bk_2,\bk_3)\alpha(\bk_4,\bk_1+\bk_2+\bk_3) \nonumber \\
   &+54F_3^{(s)}(\bk_1,\bk_2,\bk_4)\alpha(\bk_3,\bk_1+\bk_2+\bk_4)\nonumber \\
   &+54F_3^{(s)}(\bk_1,\bk_3,\bk_4)\alpha(\bk_2,\bk_1+\bk_3+\bk_4) \nonumber \\
   &+54F_3^{(s)}(\bk_2,\bk_3,\bk_4)\alpha(\bk_1,\bk_2+\bk_3+\bk_4)\Big].
  \end{align}
  
  \noindent For the velocity divergence field the SPT kernels are:
  
  \begin{align}
   &G_1(\bk)=1, \\
   &G_2^{(s)}(\bk_1,\bk_2)=\frac{3}{7}+\frac{1}{2}\frac{\bk_1\cdot\bk_2}{k_1k_2}\left(\frac{k_1}{k_2}+\frac{k_2}{k_1} \right)+\frac{4}{7}\left(\frac{\bk_1\cdot\bk_2}{k_1k_2}\right)^2, \label{eq:G2kernel} \\ 
   &G_3^{(s)}(\bk_1,\bk_2,\bk_3)=\frac{1}{18}\Big[F_2^{(s)}(\bk_1,\bk_2)\alpha(\bk_3,\bk_{12}) \nonumber \\
   &+F_2^{(s)}(\bk_2,\bk_3)\alpha(\bk_1,\bk_{23})+F_2^{(s)}(\bk_3,\bk_1)\alpha(\bk_2,\bk_{31}) \nonumber \\
   &+G_2^{(s)}(\bk_1,\bk_2)\alpha(\bk_{12},\bk_3)+G_2^{(s)}(\bk_2,\bk_3)\alpha(\bk_{23},\bk_1) \nonumber \\
   &+G_2^{(s)}(\bk_3,\bk_1)\alpha(\bk_{31},\bk_2)\Big]+\frac{2}{9}\Big[G_2^{(s)}(\bk_1,\bk_2)\beta(\bk_{12},\bk_3) \nonumber \\
   &+G_2^{(s)}(\bk_2,\bk_3)\beta(\bk_{23},\bk_1)+G_2^{(s)}(\bk_3,\bk_1)\beta(\bk_{31},\bk_2)\Big], \\ \label{eq:G3kernel}
   \nonumber \\
   &G_4^{(s)}(\bk_1,\bk_2,\bk_3,\bk_4)=\frac{1}{792} \Big[2 G_2^{(s)}(\bk_1,\bk_4)\nonumber \\ 
   &\big[6F_2^{(s)}(\bk_2,\bk_3)\alpha(\bk_1+\bk_4,\bk_2+\bk_3)\nonumber \\
   &+32G_2^{(s)}(\bk_2,\bk_3)\beta(\bk_2+\bk_3,\bk_1+\bk_4)\big]\nonumber \\
   &+2G_2^{(s)}(\bk_1,\bk_3) \big[6F_2^{(s)}(\bk_2,\bk_4)\alpha(\bk_1+\bk_3,\bk_2+\bk_4)\nonumber \\
   &+32G_2^{(s)}(\bk_2,\bk_4)\beta(\bk_1+\bk_3,\bk_2+\bk_4)\big]\nonumber \\
   &+2G_2^{(s)}(\bk_1,\bk_2) \big[6F_2^{(s)}(\bk_3,\bk_4)\alpha(\bk_1+\bk_2,\bk_3+\bk_4)\nonumber \\
   &+32G_2^{(s)}(\bk_3,\bk_4)\beta(\bk_1+\bk_2,\bk_3+\bk_4)\big]\nonumber \\
   &+6G_3^{(s)}(\bk_1,\bk_2,\bk_3) (3\alpha(\bk_1+\bk_2+\bk_3,\bk_4)+16\beta(\bk_1+\bk_2+\bk_3,\bk_4))\nonumber \\
   &+6G_3^{(s)}(\bk_1,\bk_2,\bk_4) (3\alpha(\bk_1+\bk_2+\bk_4,\bk_3)+16\beta(\bk_3,\bk_1+\bk_2+\bk_4))\nonumber \\
   &+6G_3^{(s)}(\bk_1,\bk_3,\bk_4) (3\alpha(\bk_1+\bk_3+\bk_4,\bk_2)+16\beta(\bk_2,\bk_1+\bk_3+\bk_4))\nonumber \\
   &+6G_3^{(s)}(\bk_2,\bk_3,\bk_4) (3\alpha(\bk_2+\bk_3+\bk_4,\bk_1)+16\beta(\bk_1,\bk_2+\bk_3+\bk_4))\nonumber \\
   &+12F_2^{(s)}(\bk_1,\bk_4) G_2^{(s)}(\bk_2,\bk_3)\alpha(\bk_2+\bk_3,\bk_1+\bk_4)\nonumber \\
   &+12F_2^{(s)}(\bk_1,\bk_3) G_2^{(s)}(\bk_2,\bk_4)\alpha(\bk_2+\bk_4,\bk_1+\bk_3)\nonumber \\
   &+12F_2^{(s)}(\bk_1,\bk_2) G_2^{(s)}(\bk_3,\bk_4)\alpha(\bk_3+\bk_4,\bk_1+\bk_2)\nonumber \\
   &+18F_3^{(s)}(\bk_1,\bk_2,\bk_3)\alpha(\bk_4,\bk_1+\bk_2+\bk_3)\nonumber \\
   &+18F_3^{(s)}(\bk_1,\bk_2,\bk_4)\alpha(\bk_3,\bk_1+\bk_2+\bk_4)\nonumber \\
   &+18F_3^{(s)}(\bk_1,\bk_3,\bk_4)\alpha(\bk_2,\bk_1+\bk_3+\bk_4)\nonumber \\
   &+18F_3^{(s)}(\bk_2,\bk_3,\bk_4)\alpha(\bk_1,\bk_2+\bk_3+\bk_4)\Big].
  \end{align}

   \section{MPTbreeze Formalism}
  \label{app:MPT}
  We start by introducing a more general perturbation theory, the Renormalised Perturbation Theory.
By defining $\eta=\log a$ and then a vector
\begin{equation}
\Psi\left(\textbf{k},\eta\right)=\left(\delta\left(\textbf{k},\eta\right),-\theta\left(\textbf{k},\eta\right)/\mathcal{H}\right) \, ,
\end{equation}
the usual fluid equations for can be recast in matrix notation as
\begin{multline}
\partial_\eta \Psi_a\left(\textbf{k},\eta\right)+\Omega_{ab}\left(\textbf{k},\eta\right)=  \\
\gamma_{abc}^{(s)}\left(\textbf{k},\textbf{k}_1,\textbf{k}_2\right)\Psi_b\left(\textbf{k}_1,\eta\right)\Psi_c\left(\textbf{k},\eta\right) \, ,
\end{multline}
where $\gamma_{abc}^{(s)}$ is a symmetrised vertex matrix and
\begin{equation}
\Omega_{ab}=\left(\begin{array}{cc}
                   0 & -1/2\\
                   -3/2 & 1/2
                  \end{array}
\right) \, .
\end{equation}
The above equation has solutions
\begin{multline}
\label{eqpsi}
\Psi_a\left(\textbf{k},\eta\right)=g_{ab}\left(\eta\right)\phi\left(\textbf{k}\right)+\int_0^{\eta}d\eta'g_{ab}\left(\eta-\eta'\right) \\ \times 
\gamma_{bcd}^{(s)}\left(\textbf{k},\textbf{k}_1,\textbf{k}_2\right)\Psi_c\left(\textbf{k}_1,\eta'\right)\Psi_d\left(\textbf{k}_2,\eta'\right) \, ,
\end{multline}
where $g_{ab}$ is the linear propagator, which is non-zero only for for positive $\eta$:
\begin{align}
g_{ab}\left(\eta\right)=\frac{e^{\eta}}{5}\left(\begin{array}{cc}
                   3 & 2\\
                   3 & 2
                  \end{array}\right)-\frac{e^{-3\eta/2}}{5}\left(\begin{array}{cc}
                   -2 & -2\\
                   3 & -3
                  \end{array}\right) \, .
\end{align}

\noindent Analogously to SPT, Eq.~(\ref{eqpsi}) can be solved by a series expansion
\begin{equation}
\Psi_a\left(\textbf{k},\eta\right)=\sum_{n=1}^{\infty}\Psi_a^{(n)}\left(\textbf{k},\eta\right) \, ,
\label{sumpsi}
\end{equation}
with
\begin{multline}
\Psi_a^{(n)}\left(\textbf{k},\eta\right)= \int \delta_D\left(\textbf{k}-\textbf{k}_{1 \cdots n}\right)\mathcal{F}^{(n)}_{aa_1 \cdots a_n}\left(\textbf{k}_1, \cdots, \textbf{k}_n; \eta\right)  \\
\times \phi(\textbf{k}_1) \cdots \phi(\textbf{k}_n)
\end{multline}
where $\mathcal{F}^{(n)}$ are kernels and $\textbf{k}_{1 \cdots n}=\textbf{k}_1+ \cdots \textbf{k}_n$. Non-linearities modify both the propagator and the vertex functions. The non-linear propagator is defined by
\begin{equation}
G_{ab}\left(k,\eta\right)\delta_D\left(\textbf{k}-\textbf{k}'\right)=\left\langle\frac{\delta\Psi_a(\textbf{k},\eta)}{\delta \phi_b(\textbf{k}')}\right\rangle \, 
\end{equation}
and it can be expressed as an infinite series using Eq.~\ (\ref{sumpsi}),
\begin{equation}
G_{ab}\left(k,\eta\right)=g_{ab}\left(k,\eta\right)+\sum_{n=2}^{\infty}\left\langle\frac{\delta\Psi_a^{(n)}(\textbf{k},\eta)}{\delta \phi_b(\textbf{k}')}\right\rangle \, .
\end{equation}
The full vertex functions $\Gamma$ are defined in terms of the fully non-linear propagator,
\begin{multline}
\left\langle\frac{\delta^2\Psi_a(\textbf{k},\eta)}{\delta \phi_e(\textbf{k}_1) \delta \phi_f(\textbf{k}_2)}\right\rangle = 2 \int_0^{\eta}ds\int_0^s ds_1 \int_0^s ds_2 G_{ab}\left(\eta-s\right)  \\
\times \Gamma_{bcd}^{(s)}\left(\textbf{k},s;\textbf{k}_1,s_1;\textbf{k}_2,s_2\right)G_{ce}(s_1)G_{df}(s_2) \, .
\end{multline}
By using the Feynman diagram formalism, one can see that the non-linear propagator satisfies Dyson's formula:
\begin{multline}
G_{ab}\left(\textbf{k},\eta\right)=g_{ab}\left(\eta\right)+\int_0^{\eta}ds_1 \int_0^{s_1}ds_2 g_ac\left(\eta-s_1\right) \\ \times 
\Sigma_{cd}\left(\textbf{k},s_1,s_2\right)G_{db}\left(\textbf{k},s_2,\eta'\right) \, ,
\end{multline}
where $\Sigma$ represents the sum of the principal path irreducible diagrams.

With this formalism, one can calculate the \emph{n}-point correlation function in RPT for an arbitrary number of loops, but the actual computations are difficult because they involve solving numerically a set of integro-differential equations. Nevertheless, this method provides a well-defined perturbative expansion in the non-linear regime, which is not the case in SPT.

A simplification of this model,\MPT, was developed in \citet{Bernardeau2008} and \citet{Crocce2012} that only requires the late-time propagator. Hence, in this new theory no time integrations are required. The non-linear propagator can be generalised to an arbitrary number of points; the $(p+1)$-point propagator $\Gamma^{(p)}$ is defined as
\begin{multline}
\frac{1}{p!}\left\langle \frac{\delta \Psi_a^p\left(\textbf{k},a\right)}{\delta \phi_{b_1}(\textbf{k}_1) \cdots \delta \phi_{b_p}(\textbf{k}_p)}\right\rangle \\
= \delta_D\left(\textbf{k}-\textbf{k}_{1 \cdots p}\right) \Gamma_{ab_1 \cdots b_p}^{(p)}\left(\textbf{k}_1,\cdots,\textbf{k}_p,a\right) \, .
\end{multline}
Then the power spectrum becomes
\begin{multline}
P\left(k,z\right)=\sum_{r \geq 1} r! \int\delta_D\left(\textbf{k}-\textbf{q}_{1 \cdots r}\right)\left[\Gamma^{(r)}\left(\textbf{q}_1, \cdots, \textbf{q}_r,z \right)\right]^2  \\
\times P_{\text{lin}}(q_1) \cdots P_{\text{lin}}(q_r) d^3q_1 \cdots d^3q_r \, .
\end{multline}
where the propagator takes the following simple form
\begin{multline}
\Gamma_\delta^{(n)}\left(\textbf{k}_1, \cdots, \textbf{k}_n;z\right) =  \\
D^n\left(z\right)F_n^{(s)}\left(\textbf{k}_1, \cdots, \textbf{k}_n\right) \exp\left[f(k)D^2(z)\right] \, .
\label{gamman}
\end{multline}
The function $f$ can be expressed in terms of an integral over the power spectrum today
\begin{multline} \label{eq:fk}
f\left(k\right)=\int \frac{d^3q}{(2\pi)^3}\frac{P_{\text{lin}}\left(q,z=0\right)}{504k^3q^5} \left[ 6k^7q-79k^5q^3 +50q^5k^3 \right.  \\
\left.- 21kq^7 + \frac{3}{4}\left(k^2-q^2\right)^3\left(2k^2+7q^2\right)\log \frac{|k-q|^2}{|k+q|^2} \right] \, .
\end{multline}

Up to one loop, the power spectrum and bispectrum up to one loop take the form:
\begin{align}
\MoveEqLeft[3] P_{\text{linear}}^{\text{MPTbreeze}}\left(k,z\right)=\exp \left[2 f(k)D^2(z)\right]P_{\text{lin}}\left(k\right) \label{eq:MPT_pk} \\
\MoveEqLeft[3] P_{\text{1-loop}}^{\text{MPTbreeze}}\left(k,z\right)=\exp\left[2 f(k)D^2(z)\right]P_{\text{22}}\left(k\right)  \\
\MoveEqLeft[3] B^{\text{MPTbreeze}}_{\text{tree}}\left(k_1,k_2,k_3,z\right)= B_{\text{tree}}^{\mathrm{SPT}}\left(k_1,k_2,k_3,z\right)  \nonumber \\
&\times \exp\left[\left(f(k_1)+f(k_2)+f(k_3)\right)D^2(z)\right] \label{eq:MPT_Bk} \\
\MoveEqLeft[3] B^{\text{MPTbreeze}}_{\text{1-loop}}\left(k_1,k_2,k_3,z\right)= \nonumber \\
&\left(B_{222}+B_{321}^I\right)\left(k_1,k_2,k_3,z\right) \nonumber \\
&\times \exp\left[\left(f(k_1)+f(k_2)+f(k_3)\right)D^2(z)\right] \, ,
\end{align}
where Eq.~(24) from \citet{2012PhRvD..85l3519B} has been used for the bispectrum expansion.

    \section{Derivation of Bias Parameters with the Peak-Background Split Method}\label{PBSbias}
  
  The peak-background split approach relates the bias parameters to the mean tracer abundance, where the density fluctuation field is decomposed into a low-amplitude long-wavelength linear fluctuation and a noisy short wavelength one, $\delta_h(\mathbf{x})=\delta_{h,l}(\mathbf{x})+\delta_{h,s}(\mathbf{x})$. The short-wavelength fluctuations ride on top of the large-wavelength ones, where we treat the matter perturbations as a superposition of small and large modes separated by a cut-off wavenumber. The short wavelength modes are the sources of the dark matter halos, while the long-wavelength modes increase or decrease the background density in large patches of the sky. In this picture the peaks of the small modes that are located over the peaks of long-wavelength modes will be clustered more than the average ones, as well as being the first to collapse forming galaxy clusters. This explains why galaxy clusters are more clustered than the galaxies themselves.  
  
  The presence of the long-wavelength linear fluctuation ($\delta_l=\delta^{(1)}$) will modify the background density and alter the height of the peaks to an effective value
  
  \begin{equation}\label{eq:neff}
   \nu\rightarrow\nu_{\text{eff}}=\frac{\delta_c-\delta_l}{\sigma_R},
  \end{equation}
  The overdensity field of halos within a cell of volume $V$ in the Eulerian frame is given by
  
  \begin{align}
   \delta_h^E(M|M_1,V)&=\frac{\mathcal{N}(M|M_1,V,z)}{n_h(M,z)V}-1 \nonumber \\
   &=\frac{n_h(M|M_1,V_0,z)}{n_h(M,z)}(1+\delta)-1,
  \end{align}
  
  \noindent where $\mathcal{N}(M|M_1,V,z)$ is the number of subhalos of mass $M$, corresponding to the small wavelength peaks ready to collapse on top of the long mode, above some mass $M_1$ defined by the ``background'' (\ie long wavelength) mode and $n_h(M,z)$ is the mean number of halos above mass $M$ as given by the halo mass function. The transformation rule, in the spherical collapse framework, between the Lagrangian and Eulerian frames is \citep{Mo1996a}
  
  \begin{equation}\label{eq:ELmap}
   \delta_h^E(M|M_1,V)=(1+\delta_h^L(M|M_1,V_0))(1+\delta)-1.
  \end{equation}
  
  \noindent From that we can derive the halo number density in Lagrangian coordinates as 
  
  \begin{equation}
   \delta_h^L(M|M_1,V_0)=\frac{n_h(M|M_1,V_0,z)}{n_h(M,z)}-1.
  \end{equation}
  
  \noindent The initial volume here is $V_0=M/\mean{\rho}=V(1+\delta)$. The Lagrangian bias can be identified from a matching between a Taylor expansion of the above equation and the Lagrangian local-in-matter bias relation as
  
  \begin{align}\label{eq:dhalo}
   \delta_h^L &=\sum_{n=0}^{\infty}\frac{1}{n!}b_n^L(M)[\delta^{(1)}]^n=b_1^h(M)\delta^{(1)}+\frac{b_2^h(M)}{2}[\delta^{(1)}]^2+\ldots \nonumber \\
   &=\frac{1}{n_h(M,z)}\sum_{N=0}^{\infty}\frac{1}{N!}\frac{\partial^Nn_h(M|M_1,V_0,z)}{\partial\delta_l^N}\bigg| _{\delta_l=0}\delta_l^N.
  \end{align}
  The bias parameters can be then identified by the coefficients of the above expansion as  
  
  \begin{align}\label{eq:PBS_arg}
   b_N^L(M,z)&=\frac{1}{n_h(M,z)}\frac{\partial^Nn_h(M,z)}{\partial\delta_l^N}\bigg|_{\delta_l=0}=\frac{(-\nu)^N}{\delta_c^Nf(\nu)}\frac{d^Nf(\nu)}{d\nu^N}\,.
  \end{align}
  This result is general and can be used to derive the Lagrangian halo bias for any type of universal mass function. Here we will present the Lagrangian bias in the case of Sheth-Tormen mass function. For the first four local-in-matter halo bias parameters we have \citep{Mo1996a,Mo1996b,Scoccimarro2001}

  \begin{align}
   &b_1^L(M,z)=\frac{q\nu^2-1}{\delta_c}+\frac{2p}{\delta_c(1+(q\nu^2)^p)}, \label{eq:b1h} \\
   &b_2^L(M,z)=\frac{q\nu^2(q\nu^2-3)}{\delta_c^2}\nonumber \\ &+\left(\frac{1+2p}{\delta_c}+\frac{2(q\nu^2-1)}{\delta_c}\right)\frac{2p}{\delta_c(1+(q\nu^2)^p)},\label{eq:b2h} \\
   &b_3^L(M,z)=\frac{q\nu^2}{\delta_c^3}(q^2\nu^4 - 6q\nu^2 + 3) \nonumber \\
   &+\left[\frac{4(p^2-1) + 6pq\nu^2}{\delta_c^2} + 3\left(\frac{q\nu^2-1}{\delta_c}\right)^2\right]\frac{2p}{\delta_c(1+(q\nu^2)^p)}, \label{eq:b3h} \\
   &b_4^L(M,z)=\left(\frac{q\nu^2}{\delta_c^2}\right)^2(q^2\nu^4 - 10q\nu^2 + 15) \nonumber \\
   &+\frac{2p}{\delta_c(1+(q\nu^2)^p)}\bigg[\frac{2q\nu^2}{\delta_c^2}\left(2\frac{q^2\nu^4}{\delta_c} - 15\frac{q\nu^2-1}{\delta_c}\right) \nonumber \\
   &+2\frac{(1 + p)}{\delta_c^2}\left(\frac{4(p^2-1) + 8(p-1)q\nu^2 + 3}{\delta_c} +6q\nu^2\frac{q\nu^2-1}{\delta_c}\right)\bigg]. \label{eq:b4h}
  \end{align}  
   
  \noindent where $q=0.707$ and $p=0.3$.
  
    The halo density field in \eref{eq:dhalo}) needs to be transformed to the Eulerian frame in order to take into account the dynamics of halos. To define the mapping between Eulerian and Lagrangian bias, usually one expands in real space the initial linear Lagrangian density field $\delta^{(1)}$ in the spherical collapse approximation with respect to the Eulerian non-linear overdensity as 
  
  \begin{equation}\label{eq:d0expa}
   \delta^{(1)}=\sum_{i=1}^{\infty}a_i\delta^i=\alpha_1\delta+\alpha_2\delta^2+\alpha_3\delta^3+\ldots\; .
  \end{equation}
  
  \noindent The first four coefficients are, $\alpha_1=1$, $\alpha_2=-17/21$, $\alpha_3=2815/3969$ and $\alpha_4=-590725/916839$. They are derived after calculating the spherical average of the kernels in the perturbative expansion of SPT. Coefficients $\alpha_3$ and $\alpha_4$ of the expansion in \eref{eq:d0expa}) are taken from \citet{Wagner2015}, where results of  \citet{Mo1996b} are corrected. Finally, the relations for the bias factors between the two frames are \citep{Mo1996a,Mo1996b}
  
  \begin{align}
   b_1^E&=1+b_1^L, \label{eq:b1hE} \\
   b_2^E&=b_2^L+2(\alpha_1+\alpha_2)b_1^L=\frac{8}{21}(b_1^E-1)+b_2^L, \label{eq:b2hE} \\
   b_3^E&=6(\alpha_2 + \alpha_3)b_1^L+ 3(1 + 2\alpha_2)b_2^L +b_3^L \nonumber \\
   &=\frac{20}{198}(b_1^E-1)-\frac{13}{7}b_2^E+b_3^L, \label{eq:b3hE} \\
   b_4^E&=24(\alpha_3 + \alpha_4)b_1^L+ 12(\alpha_2^2 + 2(\alpha_2 + \alpha_3))b_2^L \nonumber \\
   &+4(1 + 3\alpha_2)b_3^L +b_4^L \nonumber \\
   &=\frac{3680}{43659}(b_1^E-1)-\frac{6820}{1323}b_2^E-\frac{40}{7}b_3^E+b_4^L. \label{eq:b4hE}
  \end{align}
  
\begin{figure}
    \centering
    \resizebox{\hsize}{!}{\includegraphics{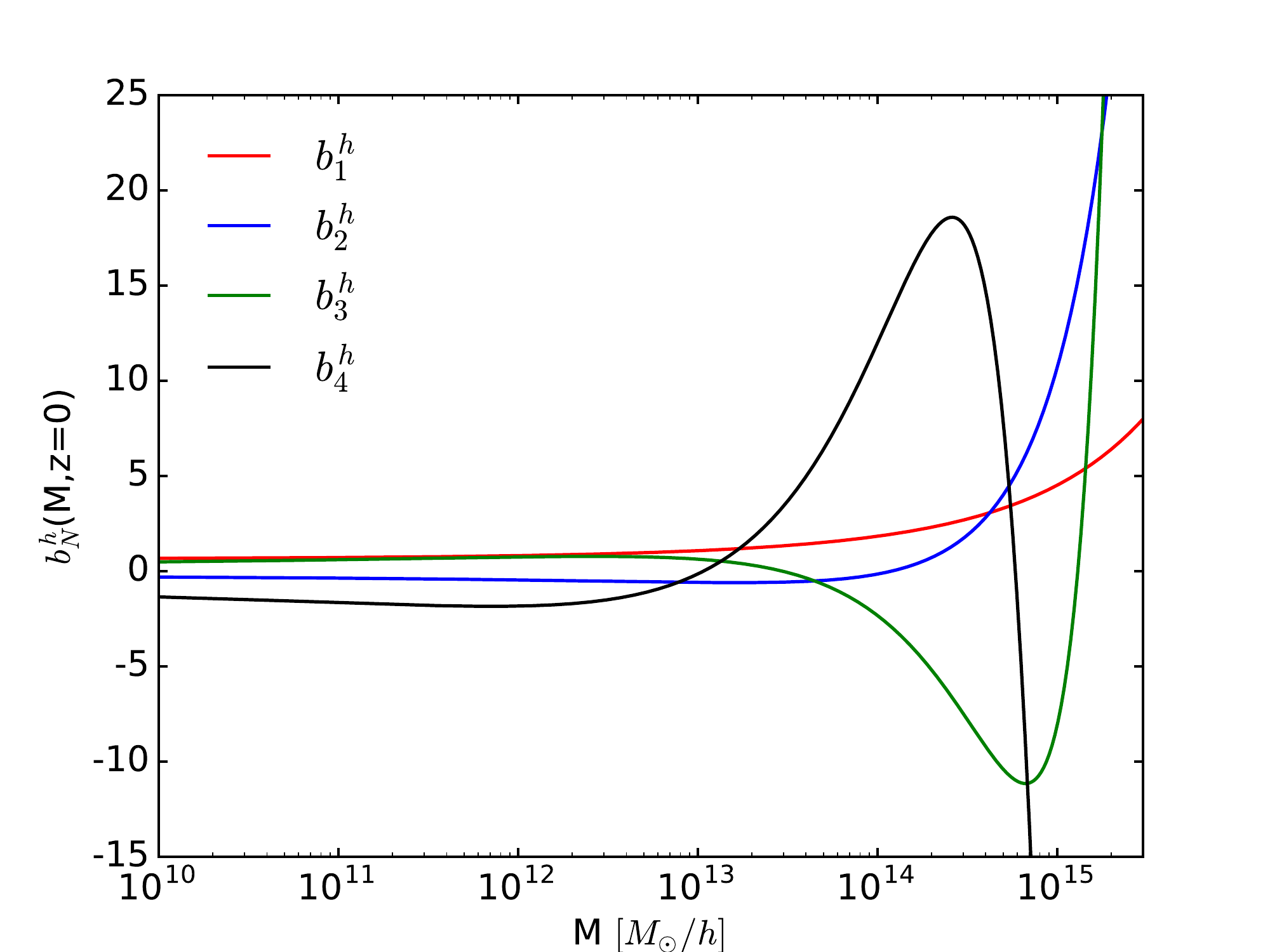}}
    \caption{The Eulerian halo bias parameters up to fourth order redshift $z=0$ as a function of the halo mass.}
    \label{fig:bias_halo}
    \end{figure}

  \noindent The Eulerian bias parameters listed above are plotted in Fig. \ref{fig:bias_halo} as a function of the halo mass.

  In the case of non-Gaussianity the peak-background split treatment of the halo bias must be generalized to consider the response of the number density with respect to a rescaling in the initial density fluctuations. What changes from the Gaussian case is that now the long and short wavelength modes are coupled to each other. For the local case, after splitting the Gaussian part of the primordial gravitational potential into long and short wavelength fluctuations, $\Phi_{G}=\phi_l+\phi_s$ and substituting into the local expansion of the non-Gaussian potential field $\Phi$ we get
  
  \begin{equation}
   \Phi=\phi_l+f_{NL}\phi_l^2+(1+2f_{NL}\phi_l)\phi_s+f_{NL}\phi_s.
  \end{equation}
  
  \noindent The most important term here is the coupling term, $(1+2f_{NL}\phi_l)\phi_s$, between long and short modes, since the long wavelength linear fluctuations will introduce a scale dependence rescaling in the amplitude of the short modes. In the case of a general non-local non-Gaussianity this rescaling of long-wavelength part of the initial density fluctuations can be parametrized through

  \begin{equation}
   \delta_l(\bk)\rightarrow[1+2\eps k^{-\alpha}]\delta_l(\bk),
  \end{equation}
  
  \noindent where the $\eps$ is an infinitesimal parameter, which becomes $\eps=\fnl\phi_l$ for local non-Gaussianity. The modulation in the primordial large wavelength density mode will affect the variance of the small scale modes introducing additional dependences in the number of collapsed objects. The short wavelength variance will transform at the lowest order to

  \begin{equation}\label{eq:varNG}
   \sigma_R\rightarrow \sigma_R\left[1+2\eps\frac{\sigma_{R,-\alpha}^2}{\sigma_R^2}\right]\, .
  \end{equation}
  
  \noindent The Jacobian ,$J\equiv\left|\frac{\mathrm{dln}\nu}{\mathrm{dln}M}\right|$, will also be transformed for the non-local case into \citep{Desjacques2011a,Desjacques2011b}

  \begin{equation}\label{eq:jacNG}
   J\rightarrow J\left[1+4\eps\frac{\sigma_{R,-\alpha}^2}{\sigma_R^2}\left(\frac{d\ln\sigma_{R,-\alpha}^2}{d\ln\sigma_{R}^2}-1\right)\right] \, .
  \end{equation}

  \noindent The peak-background split argument of \eref{eq:PBS_arg} can be easily generalized to

  \begin{equation}\label{eq:PBSNG}
   b_{\Psi\delta^N}^L=\frac{1}{n_h(M,z)}\frac{\partial^{N+1}n_h(M,z)}{\partial\delta_l^N\partial\eps}\bigg|_{\delta_l=0,\eps=0},
  \end{equation}
  
  \noindent where the average halo number density $n_h$ is taken after substituting in \eref{eq:neff}, \eref{eq:varNG} and \eref{eq:jacNG}. The leading non-Gaussian bias $b_{\Psi}$ can be derived from the above relation as a special case for $N=0$. The local non-Gaussian result can be then calculated from it for $\alpha=0$, where for a mass function with a universal form we get

  \begin{align}
   b_{\Phi}^L(M,z)&\equiv b_{\Psi}^L(\alpha=0)=\frac{1}{n_h(M,z)}\frac{\partial n_h(M,z)}{\partial\eps}\bigg|_{\eps=0} \nonumber \\ &=-2\frac{\nu}{\delta_cf(\nu)}\frac{df(\nu)}{d\nu}=2\fnll\delta_cb_1^L \, ,
  \end{align}
  which yields the well-known formula for the scale-dependent bias, after being plugged into \eref{eq:sc_dep_stand}.
    
  \noindent The first higher order non-Gaussian bias parameter $b_{\Psi\delta}$ can be easily derived through \eref{eq:PBSNG} for $N=1$ as
  
  \begin{align}\label{eq:bpsid}
  b_{\Psi\delta}^L(M,z)&=\frac{1}{n_h(M,z)}\frac{\partial^2 n_h(M,z)}{\partial\delta_l\partial\eps}\bigg|_{\delta_l=0,\eps=0} \nonumber \\ &=A\fnl^X\left[2\delta_cb_2^L+b_1^L\left(4\frac{d\ln\sigma_{R,-\alpha}^2}{d\ln\sigma_R^2}-6\right)\frac{\sigma_{R,-\alpha}^2}{\sigma_R^2}\right], 
  \end{align}
  
  \noindent which reduces for the local case to \citep{Giannantonio2010}
  
  \begin{equation}
   b_{\Phi\delta}^L(M,z)=2\fnll(\delta_cb_2^L-b_1^L) \, .
  \end{equation}
  
  \noindent Following the same steps as for the local-in-matter bias parameters, we relate the Lagrangian bias to the desired Eulerian one throughout \eref{eq:d0expa} as \citep{Giannantonio2010,Baldauf2011}
  
  \begin{align}
   b_{\Psi}^E&=b_{\Psi}^L \, , \\
   b_{\Psi\delta}^E&=b_{\Psi\delta}^L+b_{\Psi}^L\, . 
  \end{align}

  \noindent Combining the above equations with \eref{eq:bpsid} and the Eulerian results for the local-in-matter bias parameters we can derive \eref{eq:bpsidE}.
  
  The non-Gaussian mass function considered here is the LoVerde  mass function, where the fractional correction $R_{NG}$ will introduce a scale-independent offset in the bias parameters originating from the partial derivative in \eref{eq:PBSNG}. These terms will depend on $f_{NL}$ due to the presence of a non-zero skewness $S_3$ in the Edgeworth expansion of the LV mass function. These corrections will be:

  \begin{align}
   &\delta b_{1,NG}^E(f_{NL})=\delta b_{1,NG}^L=-\frac{1}{\delta_c}\frac{\nu}{R_{NG}}\frac{\partial R_{NG}}{\partial\nu}\nonumber \\ &=-\frac{\nu}{6\delta_c}\frac{f(\nu,0)}{f(\nu,f_{NL})}\biggl[3\sigma_RS_3(\nu^2-1)-\frac{d^2S_3}{d\ln\nu^2}\left(1-\frac{1}{\nu^2}\right)\nonumber \\ &+\frac{dS_3}{d\ln\nu}\left(\nu^2-4-\frac{1}{\nu^2}\right)\biggl], 
    \end{align}
    
    \noindent and
   
   \begin{align}
   &\delta b_{2,NG}^E(f_{NL})=\delta b_{2,NG}^L+\frac{8}{21}\delta b_{1,NG}^E\nonumber \\ &=\frac{\nu^2}{\delta_c^2R_{NG}}\frac{\partial^2R_{NG}}{\partial\nu^2}+2\nu\left(b_1^E-\frac{17}{21}\right)\delta b_{1,NG}^E \nonumber \\ 
   &=\frac{\nu^2}{6\delta_c}\frac{f(\nu,0)}{f(\nu,f_{NL})}\biggl[6\sigma_RS_3\nu+\frac{dS_3}{d\ln\nu}\left(5\nu-\frac{3}{\nu}+\frac{2}{\nu^3}\right)\nonumber \\ & +\frac{d^2S_3}{d\ln\nu^2}\left(\nu-\frac{4}{\nu}-\frac{3}{\nu^3}\right)-\frac{d^3S_3}{d\ln\nu^3}\left(\frac{1}{\nu}-\frac{1}{\nu^3}\right)\biggl] \nonumber \\ & +2\nu\left(b_1^E-\frac{17}{21}\right)\delta b_{1,NG}^E.
  \end{align}
  
  \noindent Here we consider only the first order correction in $f_{NL}$ to the Gaussian bias and therefore we set $f(\nu,\fnl)=f(\nu,0)$ in the above expressions.


  \section{Redshift space kernels}\label{app:RSD}

  The peculiar velocities of galaxies distort their spatial distribution in redshift space due to the change in their individual redshift values caused by Doppler shift, which leads to inaccurate distance measurements. This must be taken into account in order to properly model the power spectrum and bispectrum of galaxies in a survey. The Fourier transformation of the galaxy density field in redshift space up to second order is

  \begin{align}
   &\delta_g^s(\bk_i)=\delta_g(\bk_i)+f\mu_i^2\theta(\bk_i) \nonumber \\ 
   &+\inte{q_1}\inte{q_2}[\delta_g(\bq_1)+f\mu_1^2\theta(\bq_1)]f\mu_{12}q_{12}\frac{\mu_2}{q_2}\theta(\bq_2), \label{eq:rsd2}
  \end{align}
  where $f$ is the linear growth rate, $\mu_i=\bk_i\cdot\hat{z}/k_i$ is the cosine of the angle between the wavevector $\bk_i$ and the line-of-sight $\hat{z}$, $\mu_{ij}=(\mu_ik_i+\mu_jk_j)/k_{ij}$ and $k_{ij}^2=(\bk_i+\bk_j)^2$. In addition $G_2(\bk_i,\bk_j)$ is the second order velocity kernel of SPT (\eref{eq:G2kernel} ).
  In order to derive the galaxy statistics including the effect of RSD we can generalise the kernel formalism of SPT to include the bias terms of the expansion in \eref{eq:dg}, by plugging it in \eref{eq:rsd2}. The galaxy overdensity in redshift space can be written as \citep{Verde1998}
  
  \begin{align}
   \delta_g^s(\bk,z)&=\sum_{n=1}^{\infty}D^n(z)\int \frac{d^3q_1}{(2\pi)^3}\ldots \frac{d^3q_n}{(2\pi)^3}\delta_D(\bk-\bq_1\ldots-\bq_n)\\ \nonumber
   &\times Z_n(\bq_1,\ldots,\bq_n)\delta^{(1)}(\bk_1)\ldots\delta^{(1)}(\bk_n),
  \end{align}
  where $Z_n$ are the $n^{\text{th}}$ order redshift space galaxy kernels. In the case of a general quadratic PNG the redshift space kernels up to second order are
  
  \begin{align}
   &Z_1(\bk_i)=b_1+f\mu_i^2+\frac{b_{\Psi}k_i^{\alpha}}{M(k_i,z)}, \label{eq:Z1}\\
   &Z_2(\bk_i,\bk_j)=b_1F_2(\bk_i,\bk_j)+f\mu_{ij}^2G_2(\bk_i,\bk_j)+\frac{b_2}{2} \nonumber \\ 
   &+b_{s^2}S_2(\bk_i,\bk_j)+\frac{f\mu_{ij}k_{ij}}{2}\left[\frac{\mu_i}{k_i}Z_1(\bk_j)+\frac{\mu_j}{k_j}Z_1(\bk_i)\right] \nonumber \\ 
   &+\frac{1}{2}\left(\frac{(b_{\Psi\delta}-b_{\Psi}N_2(\bk_j,\bk_i))k_i^{\alpha}}{M(k_i,z)}+\frac{(b_{\Psi\delta}-b_{\Psi}N_2(\bk_i,\bk_j))k_j^{\alpha}}{M(k_j,z)}\right) \label{eq:Z2},
  \end{align}
  where we keep up to $\order{\fnl}$ terms. The bias parameters in the above expressions are those in Eulerian framework (observed time slice).

\end{document}